\documentclass[aps, prl, reprint, nofootinbib]{revtex4-1}
 
\usepackage{graphicx}
\usepackage{dcolumn}
\usepackage{bm}
\usepackage[hidelinks, bookmarksopen=true, bookmarks=true]{hyperref}
    \hypersetup{colorlinks = true, citecolor = [rgb]{0.8,0,0}, urlcolor = [rgb]{0,0.2,0.7}, linkcolor = [rgb]{0,0.2,0.7}}
\usepackage{amsmath}
\usepackage{amssymb}
\usepackage{xfrac}
\usepackage{natbib}
\usepackage{titletoc}
\usepackage{bookmark}

\begin{document}
 
\pdfbookmark[0]{Bound states in the continuum of gravitational waves}{Title}
\title{Bound states in the continuum of gravitational waves}
\author{Rodrigo Bert\'{e}}
\affiliation{Instituto de F\'{i}sica de S\~{a}o Carlos, Universidade de S\~{a}o Paulo, Caixa Postal 369, 13560-970, S\~{a}o Carlos, SP, Brazil}
\email{rodrigo.berte@alumni.usp.br}

\pdfbookmark[1]{Abstract}{Abstract}
\begin{abstract}
\hypersetup{bookmarksdepth=-1}
Bound states in the continuum (BICs) are
ubiquitous wave phenomena, but have not yet been demonstrated for gravitational waves (GWs). 
Here, in-plane periodic
perturbations, exponentially localized at the plane $z = 0$, are shown to lead to distributional surface energy tensors at this plane and to be regular vacuum solutions ($T_{\mu \nu} = 0$) of the linearized Einstein field equations 
outside of it. These are achieved by explicitly calculating the Ricci tensor components and the Ricci scalar from the metric perturbation tensor. To fulfill each vacuum solution 
($R_{{\sigma \nu}_{(+, \times)}} = 0$ 
and 
$R_{{}_{(+, \times)}} = 0$), 
different surface polariton-like dispersions are required.
These bound
perturbations decay exponentially to a flat metric
($h_{{BIC}_{(+,\times)}} \propto e^{- k_z |z|}$), and
each localized metric has a correspondence to a different planar GW polarization ($+,\times$). 
The Lorenz gauge-fulfilling solutions exist at the $\Gamma$ point in momentum space, dwelling within the continuum of wavevectors of propagating GWs.
The strains in the unit cell of the periodic 
perturbations have opposite parities relative to the corresponding planar GWs under 
a $C_2$ in-plane rotation, making them incompatible by symmetry with their propagating counterpart, 
indicating localization via symmetry protection.
\end{abstract}
    \hypersetup{bookmarksdepth}
    ﻿
    \hypersetup{bookmarksdepth=-1}
    \maketitle
    \hypersetup{bookmarksdepth}
    							
    \pdfbookmark[section]{Introduction}{Introduction}
    The detection of gravitational waves (GWs) has become a new paradigm to test gravitational effects and the limits of their best description, \mbox{general relativity (GR) \cite{Abbott2016}.} GW detections can now complement electromagnetic and neutrino-based observations in the new era of \mbox{multi-messenger astronomy \cite{AbbottAJ2017}.} However, GWs may allow us to peer into conditions where light is absent, such as in black hole \mbox{mergers \cite{Abbott2016}} and the early stages of the \mbox{universe \cite{Krauss2010}.} Localizing GWs could better resolve ringdown nonlinearities \cite{Mitman2023}, determine the equation of state of neutron stars \cite{Annala2018}, and clarify more elusive  aspects of gravity, by improving the detection of single gravitons \cite{Tobar2024} and enhancing the interconversion between GWs and photons \cite{Gertsenshtein1962}.
    
     The localization of waves in space is counterintuitive given that we usually perceive them (light, sound, waves in fluids and solids) as transient excitations that propagate and quickly dissipate. Therefore, the spatial localization of waves in bound states in the continuum (BICs) is quite unfamiliar and intriguing phenomena that only arises under certain special conditions \cite{Hsu2016}, among them symmetry incompatibility between propagating and localized modes. Equivalently, the localized mode is \textit{protected} (from leaking) by the \textit{symmetry} of the local excitation, whose effects cancel out in the far-field. In photonic systems these symmetry-protected bound states are usually probed as \textit{quasi}-BICs (\textit{q}BICs) in finite platforms, for instance, through a deliberate symmetry-breaking in discrete arrays \cite{Fedotov2007, Koshelev2018, Berte2023} or in \mbox{continuous media \cite{Berte2025}.}
    	
    As BICs
    have been investigated in quantum \cite{NeumannWigner1929}, hydrodynamic \cite{Ursell1951}, elastic \cite{Lim1969}, acoustic \cite{Parker1966} and, especially, optical \cite{Fedotov2007, Plotnik2011, Koshelev2018, Berte2023, Berte2025} systems, it is reasonable to inquire if the same bound states could be achieved for propagating perturbations of the flat spacetime metric, i.e., GWs.
    In other words, regarding symmetry-protected BICs: are there flat spacetime disturbances that might be exponentially localized at a plane, 
    and which are prevented to leak out from it by the symmetry of the excitation, akin to electric/magnetic fields in photonic BICs \cite{Fedotov2007, Hsu2016, Koshelev2018, Berte2023}? 
    Could the parities of the bound modes be different than that of a propagating GW (in-phase or \textit{even} under a 180$^{\circ}$ rotation, and out-of-phase or \textit{odd} under a 90$^{\circ}$ rotation) when rotated in the plane in which they are localized?
    Here, two metrics corresponding to localized perturbations $h_{{\mu \nu}_{(+,\times)}}$ of the flat (Minkowski) spacetime metric $\eta_{\mu \nu}$ 
    are shown to
    lead to distributional stress energy tensors $^{(\Sigma)}T^{\mu \nu}$ at the localization plane (hypersurface $\Sigma$ or $z = 0$) and to regular
    vacuum solutions of the Einstein field equations ($T_{\mu \nu} = 0$)
    in the linearized (weak) gravity regime
    for all other values of $z$.
    
    These consequences of perturbations which are exponentially localized at the plane $z = 0$ ($h_{{BIC}_{(+,\times)}} \propto e^{- k_z |z|}$), are demonstrated by explicitly calculating the Ricci curvature tensor components ($R_{{\sigma \nu}_{(+, \times)}}$) and the Ricci scalar ($R_{{}_{(+,\times)}}$), obtaining the dispersions by which these are nullified outside the localization plane.
    These solutions are unbounded in the $xy$ plane, fulfilling the idealized condition of being infinite in at least one spatial dimension, as required by symmetry-protected BICs \cite{Hsu2016, Plotnik2011}. The mode is an infinite mathematical construct as an edge would constitute a translational symmetry breaking, leading from a theoretically infinite quality factor ($Q$) to a finite one. Nevertheless, in photonic systems, symmetry-protected BIC signatures were excited for short times in equivalent finite symmetric metasurfaces \cite{vanHoof2021, Dong2022}. 
    The components of the Riemann 
    curvature tensor ($R_{{\sigma \mu \nu}_{(+, \times)}}^\rho$) 
    are also calculated for each bound solution. 
    It is also demonstrated that the proposed metrics fulfill the Lorenz gauge ($\partial^{\mu}\overline{h}_{\mu \nu_{(+, \times)}}$)
    and that the components of their trace-reversed perturbation tensor ($\overline{h}_{{\mu \nu}_{(+, \times)}}$) are solutions of the homogeneous wave equation for the same surface-polariton-like dispersions mentioned above outside the localization plane. Energy-momentum conservation is also demonstrated, regardless of the distributional nature of the stress-energy tensor at the hypersurface $\Sigma$. The second order Kretschmann scalar invariant and the energy-momentum flux via the Isaacson stress-energy tensor are also computed, yielding pathology-free regular spacetimes outside the localization plane and no energy-momentum flux from the bound modes, respectively.

    \pdfbookmark[section]{Results}{Results}
    \textit{Results.---} 
    A ($+,-,-,-$) metric signature is employed and a null cosmological constant ($\Lambda = 0$) is considered for the calculations. 
    The following solutions require
    modifications in the $\eta_{tt}$ and $\eta_{zz}$ terms of the flat spacetime metric, similarly to other 
    linear and non-linear solutions of GWs \cite{Halilsoy1988, Bondi2000, Stephani2003, Bondi2004, Golat2020, Glod2022}. 
    Although these degrees of freedom in $h_{\mu \nu}$ (i.e. $h_{tt}$ and $h_{zz}$) are redundant for planar GWs in the TT-gauge of linearized gravity, leading to $R_{tt}$ and $R_{zz}$ terms which become zero for the same linear dispersion ($\omega^2 = c^2 k_z^2$) required to nullify $R_{xx} = R_{yy} = R_{xy} = R_{yx} = 0$, for the bound metrics here proposed they are crucial to generate the vacuum solutions outside the localization plane ($z = 0$). 

    \begin{figure*}[t]
    	\includegraphics[scale=1.2]{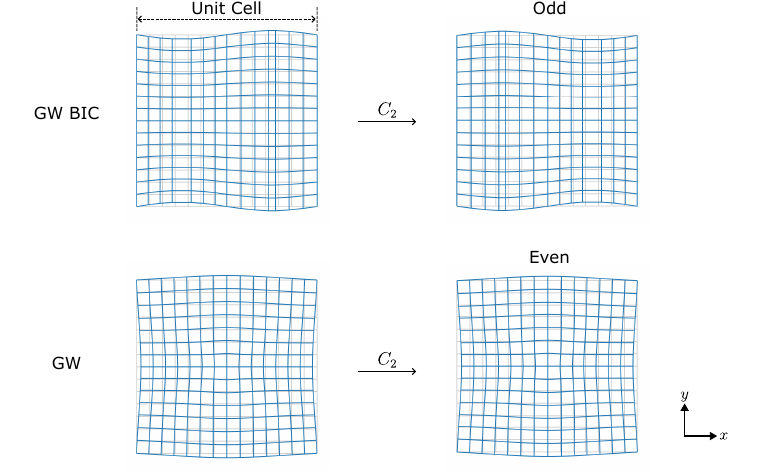}
    	\caption{\textbf{Symmetry incompatibility between bound and ($+$)-polarized propagating GWs.} (top) The strain 
        in the unit cell (periodic in $x$) 
        of the bound spacetime perturbation 
        is anti-symmetric (odd) under a $C_2$ transformation (180$^{\circ}$ in-plane rotation). 
        (bottom) A propagating GW is symmetric (even) under the same transformation. The symmetry incompatibility suggests localization via symmetry protection for the bound metric. The unperturbed (Minkowski) spacetime is shown in light grey.
    	}
      	\label{fig1}
    \end{figure*}
    \subsection{Ricci tensor and Ricci scalar}
    \subsubsection{$h_{xx} , h_{yy}$ perturbation}
    We first consider a perturbation involving the diagonal terms ($h_{xx} , h_{yy}$) of the metric tensor. These terms correspond to a plus-polarized ($+$) planar GW. 
    For the ansatz, we may take inspiration in photonic BICs in semiconductor 1D gratings, localized at the $\Gamma$ point in reciprocal space, and in which the electric field varies sinusoidally along the thin film/grating, while decaying exponentially away from it \cite{Berte2023, Berte2025}.
    In cartesian coordinates, the localized metric can be written as:
    \begin{equation} \label{eq_bic_+}
     \begin{split}
        ds^2 =  (1 + h_{{BIC}_+}) ~ c^2 dt^2 - (1 + h_{{BIC}_+}) ~ dx^2 \\ 
    	- ~ (1 - h_{{BIC}_+}) ~ dy^2 - (1 + h_{{BIC}_+}) ~ dz^2
     \end{split}
    \end{equation} 
    being the standing strain perturbation
    \begin{equation} \label{eq_hbic_+}
        h_{{BIC}_+}(ct,x,z) = A ~ cos(k_x x) ~ cos(\omega ct/c) ~ e^{- k_z |z|}
    \end{equation}
    where $A$ is the amplitude, $k_x$ the in-plane wavevector (defining the in-plane periodicity), $k_z$ is the spatial decay constant and $\omega$ is the angular frequency of the flat spacetime metric perturbation. 
    Thus, the metric perturbation decays exponentially from (thus it is localized at) \mbox{the plane $z = 0$.} 
    Differently from a planar ($+$)-polarized GW in the TT-gauge, this solution requires the additional perturbation of the $\eta_{tt}$ and $\eta_{zz}$ metric components.
    It must be clarified that while the bound metrics ($\propto e^{-k_z |z|}$) are continuous everywhere, their first derivative with respect to $z$ is discontinuous at $z = 0$, $\partial_z (h_{\mu \nu_{(+, \times)}}) \propto k_z ~ sign(z)$, termed a weak derivative. This will lead to Dirac deltas ($\delta (z)$) in the curvature from the second derivative with respect to $z$ and to distributional sources, as shown below. However, the proposed metrics fulfill all the requirements of a regular metric associated with a distributional curvature tensor: locally bound $g_{\mu \nu}$, existence of locally bound inverse $g^{\mu \nu}$, and also existence of locally square-integrable weak first derivative (i.e.~finite $\int{|sign(z)~e^{-k_z |z|}|^2dz}$) \cite{Geroch1987}.
This metric perturbation yields the following non-zero Ricci tensor components 
$R_{{\sigma \nu}_{+}}$
in linearized gravity:
\begin{equation}
    \begin{split}
        & R_{{tt}_+} = 
        - R_{{xx}_+} =
        R_{{yy}_+} =
        - R_{{zz}_+} \\ 
        & = 
         \frac{h_{{BIC}_+}}{2} \big(
        - k_x^2 + k_z( k_z ~ sign^2(z) - 2\delta (z))  
        + \omega^2/c^2 \big)
        \\
        \therefore
        & =
        \frac{h_{{BIC}_+}}{2} (
        - k_x^2 + k_z^2 
        + \omega^2/c^2
        ) \quad for ~ z \neq 0
    \end{split}
\end{equation}

Therefore, for $z \neq 0$, all non-zero components of the Ricci tensor $R_{{\sigma \nu}_{+}}$ (see Supplemental Material, SM, Eqs.~\ref{rtt+} - \ref{rzz+} for detailed calculations)
vanish for the following 
surface polariton-like dispersion (being a surface polariton an electromagnetic wave exponentially bound to an interface between media of positive and negative real permittivities) \cite{AgranovichBook}:
\begin{equation} \label{eq_disp_+}
    \omega^2 = c^2 (k_x^2 - k_z^2)
\end{equation}
yielding a real frequency if $|k_x| > |k_z|$.
For a given frequency $\omega$, a smaller in-plane periodicity ($\propto 1/k_x$) can be obtained by increasing the out-of-plane confinement (larger $k_z$), without apparent upper bounds, as long as the difference in their squares remains constant. While there is no lower bound for $k_z$, in the limit of no confinement ($ k_z \rightarrow 0$), the lower bound of $k_x$ is achieved, corresponding to the linear dispersion of a plane wave ($\omega = k_x c$).
It results for the Ricci scalar $R_{{}_+}$:
\begin{equation}
\begin{split}
    R_{{}_+}
    & = h_{{BIC}_+} \big(
    \omega^2/c^2 - k_x^2
    + k_z( k_z ~ sign^2(z) - 2\delta (z))  
    \big) 
\\ 
    \therefore
        & = 
        h_{{BIC}_+} (
         \omega^2/c^2
        - k_x^2 + k_z^2
        ) \quad for ~ z \neq 0
\end{split}
\end{equation}
    implying that the same dispersion above (Eq.~\ref{eq_disp_+}) leads to $R_{{}_+} = 0$ (SM Eq.~\ref{ricci_scalar_+}) outside the localization plane $z = 0$.
    These results imply a vacuum solution (i.e. $R_{{\sigma \nu}_+} - R_{{}_+}\eta_{\sigma \nu}/2 = (8\pi G/c^4) T_{{\sigma \nu}_+} = 0$) for $z \neq 0$ in the weak gravity regime \footnote{An equivalent solution can be obtained when replacing the in-plane modulation in $x$ by an in-plane modulation in $y$ ($cos(k_y y)$). For that one must also change the signs of $h_{tt}$ and $h_{zz}$ in the metric tensor, i.e. $g_{tt} = 1 - h_{BIC}$ and $g_{zz} = -1 + h_{BIC}$}. 
    The combination of $R_{{\sigma \nu}_+}$ and $R_{{}_+}$ also implies a distributional stress-energy tensor $^{(\Sigma)}T_{\sigma \nu_{+}}$ for the $yy$ component at $z = 0$.
    
    The corresponding unit cell strain of the GW BIC is shown in Fig.~\ref{fig1}. While a planar propagating GW has an even (invariant) symmetry under a 180$^{\circ}$ ($C_2$) rotation \cite{MisnerBook}, 
    the bound GW is anti-symmetric (odd)
    under the same transformation, making them incompatible by symmetry and indicating localization via symmetry protection \cite{Hsu2016}.
    The non-zero components of $R_{{\sigma \mu \nu}_+}^\rho$ 
    are shown in Eqs.~\ref{lg_eq_RCTc_BIC_hxxhyy}
    of the SM.
    
    \subsubsection{$h_{xy} , h_{yx}$ perturbation}
    As for planar GWs, we might consider BICs that would exist regarding perturbations in the elements of the metric that correspond to an orthogonal (cross or $\times$) polarization, in case, involving the off-diagonal elements $h_{xy}$ and $ h_{yx}$. 
    The perturbation in the off-diagonal elements $h_{xy}$ and $ h_{yx}$ has further requirements relative to the previous case to generate a solution of the linearized EFEs, as discussed below.
    \begin{figure*}[t]
    	\includegraphics[scale=1.2]{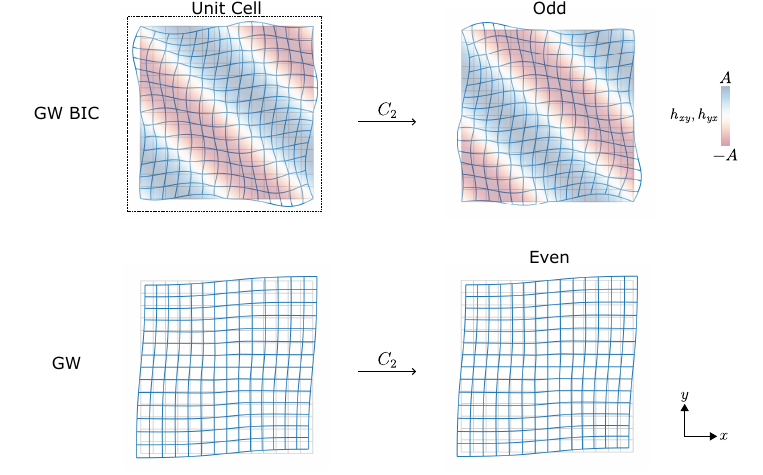}
    	\caption{\textbf{Symmetry incompatibility between bound and ($\times$)-polarized propagating GWs.} (top) The strain 
        in the unit cell (periodic at 45$^\circ$ relative to the $x$ and $y$ directions) 
        of the bound spacetime perturbation 
        is anti-symmetric (odd) under a $C_2$ transformation (180$^{\circ}$ in-plane rotation). The unit cell of the flat metric perturbation $h_{xy},h_{yx}$ (periodic in $x$ and in $y$) is shown in red and blue colors.  
        (bottom) A ($\times$)-polarized propagating GW is symmetric (even) under the same transformation, indicating 
        localization via symmetry protection. 
        The unperturbed (Minkowski) spacetime is shown in light grey.
    	}
      	\label{fig2}
    \end{figure*}
    The line element of this bound metric can be written as:     
    \begin{equation} \label{eq_bic_x}
     \begin{split}
        ds^2 =  (1 + h_{{BIC}_{\times}}) & ~ c^2 dt^2 - dx^2 - dy^2 \\ - & 2h_{{BIC}_{\times}} ~ dxdy 
    	 - (1 + h_{{BIC}_{\times}}) ~ dz^2
     \end{split}
    \end{equation} 
    
    in which the strain perturbation is:
    \begin{equation}
    \begin{split}
        h_{{BIC}_{\times}}(ct,x,y,z) = A ~ \big(& cos(k_x x)  sin(k_yy) ~ + \\ & sin(k_xx) cos(k_yy)\big) ~ cos(\omega ct/c) ~ 
        e^{- k_z |z|}
    \end{split}
    \end{equation}
    being the perturbation $h_{{BIC}_{\times}}$ modulated both in $x$ and in $y$ directions. As for the previous bound perturbation, this solution requires modifications in the $\eta_{tt}$ and $\eta_{zz}$ flat metric components.
    The non-zero coefficients of the Ricci tensor $R_{{\sigma \nu}_{\times}}$ are (SM Eqs.~\ref{rttx} - \ref{rzzx} for detailed calculations):
\begin{equation}
\begin{split}
     R_{{tt}_{\times}} 
    = - R_{{zz}_{\times}} =
    & \frac{h_{{BIC}_{\times}}}{2} \big(
        - k_x^2
        - k_y^2
        \\ &
        + k_z( k_z ~ sign^2(z) - 2\delta (z)) 
        + \omega^2/c^2
        \big)
\end{split}
\end{equation}
\begin{equation}
\begin{split}
    R_{{tx}_{\times}} 
    & = R_{{xt}_{\times}}
    = - R_{{ty}_{\times}}
    = - R_{{yt}_{\times}} \\
    & = \frac{A \omega}{2 c} (
    k_y
    - k_x
    ) 
    ~ \big(sin(k_x x) sin(k_y y) \\ 
    & \quad ~ - cos(k_x x) cos(k_y y)\big) ~ sin( \omega ct/c) ~ 
    e^{- k_z |z|}
\end{split}
\end{equation}
\begin{equation}
    R_{{xx}_{\times}} 
    = h_{{BIC}_{\times}} k_x (
    - k_y
    + k_x
    )
\end{equation}
\begin{equation}
\begin{split}
    R_{{xy}_{\times}}
    = R_{{yx}_{\times}}
    & = \frac{h_{{BIC}_{\times}}}{2} (
    - \omega^2/c^2
    \\
    & - k_z( k_z ~ sign^2(z) - 2\delta (z))
    + k_y k_x 
    + k_y k_x
    )
\end{split}
\end{equation}
\begin{equation}
\begin{split}
    R_{{xz}_{\times}}
    & = R_{{zx}_{\times}}
    = - R_{{yz}_{\times}}
    = - R_{{zy}_{\times}} \\
    & = \frac{A k_z ~ sign(z)}{2} (
    k_y
    - k_x
    ) 
    ~ \big(sin(k_x x) sin(k_y y) \\ 
    & \quad ~  - cos(k_x x) cos(k_y y)\big) ~ cos( \omega ct/c) ~
    e^{- k_z |z|}
\end{split}
\end{equation}
\begin{equation}
     R_{{yy}_{\times}} 
     = h_{{BIC}_{\times}} k_y (
    - k_x
    + k_y
    )
\end{equation}
It can be seen that if $k_x = k_y$ (i.e. in case of a square lattice perturbation) all of the following $R_{{\sigma \nu}_{\times}}$ coefficients become zero:
\begin{equation}
\begin{split}
    R_{{tx}_{\times}} = R_{{xt}_{\times}}
    = R_{{ty}_{\times}} = R_{{yt}_{\times}}
    = R_{{xx}_{\times}} \\
    = R_{{xz}_{\times}} = R_{{zx}_{\times}}
    = R_{{yy}_{\times}}
    = R_{{yz}_{\times}} = R_{{zy}_{\times}}
    = 0
\end{split}
\end{equation}
while the $R_{{xy}_{\times}}$ and $R_{{yx}_{\times}}$ coefficients may be rewritten as:
\begin{equation}
\begin{split}
    R_{{xy}_{\times}}
    = R_{{yx}_{\times}}
    & = \frac{h_{{BIC}_{\times}}}{2} (
        - \omega^2/c^2
        \\
       & - k_z( k_z ~ sign^2(z) - 2\delta (z)) 
        + k_x^2
        + k_y^2
        )
\end{split}
\end{equation}
Thus, all remaining non-zero terms ($R_{{tt}_{\times}},  R_{{xy}_{\times}}, R_{{yx}_{\times}}$ and $R_{{zz}_{\times}}$) can be nullified outside the localization plane for a similar surface polariton-like dispersion:
\begin{equation} \label{eq_dispersion_x}
    \omega^2 = c^2(k_x^2 + k_y^2 - k_z^2)
\end{equation}
Similar arguments for the ($+$)-polarized BIC applies here, regarding a real frequency ($k_x^2 + k_y^2 > k_z^2$), the decrease in in-plane unit cell periodicity ($\propto 1/k_x, 1/k_y$) with an increase in the out-of-plane confinement without wavevector upper bounds, and a linear dispersion in the limit of no confinement ($\omega = \sqrt{2}k_x c$, $k_x = k_y$). For both modes, the excessive confinement for a given in-plane periodicity ($k_z^2 > k_x^2 + k_y^2$ for the ($\times$)-BIC) leads to imaginary frequencies. This is expected to generate instabilities in the form of exponential growth as function of the time coordinate ($cos(\omega t) \rightarrow cosh(\omega t) \propto e^{\omega t}$), for which the assumption of small perturbations ($h_{\mu \nu_{(+,\times)}} \propto e^{\omega t} \ll 1$) of the Minkowsky metric by the bound mode breaks down.
The ($\times$)-polarized bound metric perturbation yields the following Ricci scalar $R_{{}_\times}$ (SM, Eq.~\ref{ricci_scalar_x}) in the weak gravity regime:
\begin{equation}
\begin{split}
    R_{{}_{\times}} 
    & = h_{{BIC}_{\times}} (
    - k_x^2
    - k_y^2
    + k_z( k_z ~ sign^2(z) - 2\delta (z))
    + \omega^2/c^2
    ) \\
    & \quad ~  - h_{{BIC}_{\times}} k_x (
    - k_y
    + k_x
    ) 
    - h_{{BIC}_{\times}} k_y (
    - k_x
    + k_y
    )
\end{split}
\end{equation}
which also becomes null for $z \neq 0$ if $k_x = k_y$ and for the dispersion in Eq.~\ref{eq_dispersion_x}. Given that 
$R_{{\sigma \nu}_{\times}} = 0$ for all spacetime indices and $R_{{}_{\times}} = 0$ we may conclude that the metric given by Eq.~\ref{eq_bic_x} is a vacuum solution ($T_{\sigma \nu} = 0$) of the linearized EFEs outside the localization plane. At $z = 0$, $R_{{\sigma \nu}_\times}$ and $R_{{}_\times}$ lead to a distributional stress-energy tensor $^{(\Sigma)}T_{\sigma \nu_{\times}}$ with 4 non-zero components ($xx, xy, yx, yy$).

The corresponding unit cell strain of the \mbox{($\times$)-polarized} GW BIC is illustrated in Fig.~\ref{fig2}. Identically to a \mbox{($+$)-polarized} planar GW, its orthogonal \mbox{($\times$)-polarization} (Fig.~\ref{fig2}, bottom) is symmetric, or even under a $C_2$ rotation in the $xy$ plane. Meanwhile, the bound strain is odd under the same rotation, thus incompatible by symmetry with the propagating mode. The non-zero components of $R_{{\sigma \mu \nu}_\times}^\rho$ are shown in Eqs.~\ref{lg_eq_RCTc_BIC}
of the SM.

\subsection{Non-trivial scalar invariants}
Further insight into the proposed bound spacetime perturbations as physical solutions can be gained by computing non-trivial quadratic, or second order, scalar invariants. Here, the Kretschmann scalar ($K$) is calculated for both solutions. Being defined as \cite{Cherubini2003}:
\begin{equation}
    K_{_{(+,\times)}} = R_{\rho \sigma \mu \nu} R^{\rho \sigma \mu \nu}
\end{equation}
As shown in the SM, the non-trivial $K_{_{(+,\times)}}$ are ill-defined at the hypersurface $\Sigma$, corresponding to the localization plane $z = 0$ due to the product of distributions ($\delta^2(z)$).
Nevertheless, we may split them into two parts, an ill-defined component ($K_{\Sigma_{(+,\times)}}$) which is zero everywhere except at $z = 0$, and a smooth ($C^\infty$), well-behave function ($K_{S_{(+,\times)}}$) outside of $\Sigma$ where the EFEs yield vacuum solutions, with which we proceed with our analysis:
\begin{equation}
    K_{_{(+,\times)}} = K_{S_{(+,\times)}} + K_{\Sigma_{(+,\times)}}
\end{equation}
The well-behaved, smooth second order scalars have the same structure for both plus and cross bound metrics (see SM for detailed calculations). For the $h_{xx},h_{yy}$ perturbation:
\begin{equation}
\begin{split}
     K_{S_{+}} =  4 A^2 
     \big( & k_x^4 cos^2(\omega ct/c)
     + \frac{\omega^4}{c^4}  cos^2(k_x x)
     \\ &
     - \frac{k_x^2 \omega^2}{c^2} \big)
     e^{- 2 k_z |z|}
\end{split}
\end{equation}
While for the $h_{xy},h_{yx}$ perturbation:
\begin{equation}
\begin{split}
    K_{S_{\times}} = ~ 4A^2 
    & \big[ 
    (k_x^2 + k_y^2)^2 cos^2(\omega ct/c)
    \\ &+ \frac{\omega^4}{c^4}
    \big(cos(k_x x)sin(k_yy) + sin(k_xx)cos(k_yy)\big)^2
    \\
    & - \frac{(k_x^2 + k_y^2) \omega^2}{c^2}
    \big] e^{- 2 k_z |z|}
\end{split}
\end{equation}
Thus, both $K_{S_{+}}$ and $K_{S_{\times}}$ are finite, corresponding to regular spacetimes
without divergences/singularities and free of coordinate pathologies outside the localization plane, exponentially decaying away from $z = 0$. Note that, while mostly positive, 
there are 
values of the time-like coordinate $ct$ (i.e. for $cos^2(\omega ct/c) \approx 0$)
for which the curvature is negative \cite{Henry2000}, albeit orders of magnitude smaller compared with its maximum positive value (assuming, e.g~$k_x^4 \gg k_x^2 \omega^2/c^2$). Given that $R_{{\sigma \nu}_{(+, \times)}} = 0$ and $R_{{}_{(+, \times)}} = 0$ outside $\Sigma$, the Weyl scalars ($C_{\rho \sigma \mu \nu} C^{\rho \sigma \mu \nu}$) for both solutions are identical to the calculated Kretschmann second-order invariants.

\subsection{Lorenz gauge}
The perturbed metrics proposed in Eq.~\ref{eq_bic_+} and in Eq.~\ref{eq_bic_x} were shown to lead to distributional expressions via the linearized EFEs by explicitly calculating the Ricci tensors and scalar from the perturbation tensors $h_{{\mu \nu}_{(+, \times)}}$, which involves dealing with coupled differential equations. A simpler and equivalent approach might be taken by demonstrating that the metric perturbations fulfill the Lorenz gauge:
\begin{equation}
    \partial^{\mu}\overline{h}_{\mu \nu_{(+, \times)}}
    = 0
\end{equation}
where $\overline{h}_{\mu \nu_{(+, \times)}}$ are the respective trace-reversed perturbation tensors, defined as:
\begin{equation} \label{eq_trpt}
    \overline{h}_{{\mu \nu}_{(+, \times)}} = h_{{\mu \nu}_{(+, \times)}} - \frac{1}{2} \eta_{\mu \nu} h_{{}_{(+, \times)}}
\end{equation}
Once fulfilled the Lorenz gauge, the distributional stress-energy tensor components can be obtained directly from the wave equation
($\Box \overline{h}_{{\mu \nu}_{(+, \times)}} 
$, being $\Box = \partial_t\partial_t - \bigtriangleup$ the d'Alambert operator).
As shown in the SM, the proposed perturbations $h_{\mu \nu_{(+,\times)}}$ are not a mere gauge field configuration, i.e., they cannot be set to zero via a gauge transformation by a vector field $\xi_\mu$:
\begin{equation}
    \tilde{h}_{\mu \nu{(+,\times)}} = h_{\mu \nu{(+,\times)}} + \partial_\mu \xi_\nu + \partial_\nu \xi_\mu
\end{equation}
Setting $\tilde{h}_{\mu \nu{(+,\times)}} = 0$ leads to spurious equalities between functions with different spacetime coordinate dependencies. Thus, there are no $\xi_\mu$ that fulfill simultaneously all conditions imposed by the spacetime indices of the gauge symmetry of linearized gravity. Note also that a pure gauge configuration would lead to completely null Riemann curvature tensors, which is not the case for the proposed bound metrics.

\subsubsection{$h_{xx} , h_{yy}$ perturbation}
The trace of the perturbation tensor $h_{{\mu \nu}_+}$ is (SM Eq.~\ref{trace_bic_+}):
\begin{equation}
    h_{{}_+} = 2 h_{{BIC}_+}
\end{equation}
yielding, through Eq.~\ref{eq_trpt}, only one non-zero trace-reversed perturbation tensor coefficient (SM, Eq.~\ref{trace_reversed_bic_+}):
\begin{equation}
    \overline{h}_{yy} = 2 h_{{BIC}_+}
\end{equation}
Finally, the Lorenz gauge condition can be verified (SM, Eq.~\ref{lor_gauge_bic_+}):
\begin{equation}
    \partial^{\mu}\overline{h}_{y \mu}
    = \partial^{y}\overline{h}_{yy}
    = - \partial_y (2 h_{{BIC}_+}) 
    = 0
\end{equation}
as the perturbation $h_{{BIC}_+}$ is independent of $y$ (Eq.~\ref{eq_hbic_+}), fulfilling the condition of the Lorenz gauge. By applying the d'Alambertian to the trace-reversed perturbation $\bar{h}_{{\mu \nu}_+}$:
\begin{equation}
\begin{split}
    \Box \overline{h}_{{\mu \nu}_+}
    = & ~ \Box (2 h_{{BIC}_+}) \\
    =  & ~ \big[-\omega^2/c^2 + k_x^2 
     - k_z( k_z ~ sign^2(z) - 2\delta (z)) \big]
     \\
     & \times (2 h_{{BIC}_+})
\end{split}
\end{equation}
leading to a distributional $yy$ component of $^{(\Sigma)}T_{\mu \nu_+}$ at the localization plane and to a regular vacuum solution outside it 
for an identical dispersion as shown in Eq.~\ref{eq_disp_+}. 
\subsubsection{$h_{xy} , h_{yx}$ perturbation}
Similarly, for the trace of $h_{{\mu \nu}_\times}$ (SM, Eq.~\ref{trace_bic_x}):
\begin{equation}
    h_{{}_\times} = 2 h_{{BIC}_{\times}}
\end{equation}
The non-zero coefficients of $\overline{h}_{{\mu \nu}_ \times}$ are (SM, Eq.~\ref{trace_reversed_bic_x}):
\begin{equation} \label{tracerev_x_components}
    \overline{h}_{xx} = 
    - \overline{h}_{xy} = 
    - \overline{h}_{yx}
    = \overline{h}_{yy}
    = h_{{BIC}_\times}
\end{equation}
implying for the Lorenz gauge (SM, Eq.~\ref{lor_gauge_bic_x}):
\begin{equation}
\begin{split}
   \partial^{\mu}\overline{h}_{x \mu}
   & = 
   \partial^{x}\overline{h}_{xx}
   + \partial^{y}\overline{h}_{xy}
   = - \partial_x (h_{{BIC}_\times})
   - \partial_y (-h_{{BIC}_\times})
   \\
   & = (k_x - k_y) 
   ~ (sin(k_x x) sin(k_y y) 
   \\ & \quad ~ - cos(k_x x) cos(k_y y)) ~ cos( \omega ct/c) ~ 
   e^{- k_z |z|}
\end{split}
\end{equation}
\begin{equation}
\begin{split}
   \partial^{\mu}\overline{h}_{y \mu}
   & = 
   \partial^{x}\overline{h}_{yx}
   + \partial^{y}\overline{h}_{yy}
   = - \partial_x ( - h_{{BIC}_\times})
   - \partial_y (h_{{BIC}_\times})
   \\
   & = (-k_x + k_y) 
   ~ (sin(k_x x) sin(k_y y) 
   \\ & \quad ~  - cos(k_x x) cos(k_y y)) ~ cos( \omega ct/c) ~ 
   e^{- k_z |z|}
\end{split}
\end{equation}
Both sums become null, fulfilling the Lorenz gauge, for a square lattice (i.e.~$k_x = k_y$), an identical condition required when solving the linearized EFEs via the Ricci tensor and scalar above. Applying the d'Alambertian: 
\begin{equation}
\begin{split}
    \Box \overline{h}_{{\mu \nu}_\times} = &  
        ~ \Box (\pm h_{{BIC}_\times})
        \\
        = & ~ \big[-\omega^2/c^2 + k_x^2 + k_y^2 
        \\
        & - k_z( k_z ~ sign^2(z) - 2\delta (z))
        \big](\pm h_{{BIC}_\times})
\end{split}
\end{equation}
implying again a distributional stress-energy tensor with same non-zero indices as the trace-reversed perturbation above (Eq.~\ref{tracerev_x_components}). Outside $z = 0$, the metric is a regular vacuum solution
as long as the bound mode obeys the surface polariton-like dispersion shown in Eq.~\ref{eq_dispersion_x}. 
\subsection{Energy-momentum conservation}
It can be demonstrated that up to first order the distributional stress-energy momentum tensors obey local energy-momentum conservation:
\begin{equation} \label{lemconservation}
    \partial_\mu T^{\mu \nu} = 0
\end{equation}
regardless of their distributional nature at $z = 0$. The $T_{\mu \nu}$ components can be obtained directly from the trace-reversed perturbation tensor (given that the Lorenz gauge is fulfilled) by:
\begin{equation}
    \Box \bar{h}_{\mu \nu_{(+,\times)}} = -\frac{16 \pi G}{c^4} T_{\mu \nu_{(+,\times)}}
\end{equation}
\subsubsection{$h_{xx} , h_{yy}$ perturbation}
For this bound polarization the only non-zero stress-energy tensor component is $T_{yy} = T^{yy}$. So, Eq.~\ref{lemconservation} is evidently zero for $\nu = (t,x,z)$. Otherwise it implies (SM Eq.~\ref{lemconservation_+}):
\begin{equation}
\begin{split}
    \partial_\mu T^{\mu \nu} & = 
    \partial_y T^{yy}\\ 
    &  =
    ~ \partial_y \bigg[ \frac{c^4}{8 \pi G}  \bigg(\frac{\omega^2}{c^2} - k_x^2 + k_z \big(k_z~sign^2(z) -2\delta (z)\big) \bigg) 
    \\
    & \times A ~ cos(k_x x) ~ cos(\omega ct/c) ~
    e^{- k_z |z|} \bigg]
    = 0
\end{split}
\end{equation}
as the stress-energy tensor is not a function of the spacetime coordinate $y$, yielding a divergence-free tensor.
\subsubsection{$h_{xy}, h_{yx}$ perturbation}
The ($\times$)-polarized BIC has the following non-zero $T_{\mu \nu}$ components (SM, Eq.~\ref{SET_x}):
\begin{equation}
\begin{split}
    T_{xx} = &
    - T_{xy} = 
    - T_{yx} =
    T_{yy} = 
    \frac{c^4}{16 \pi G}  
    \\
    & \times \bigg(\frac{\omega^2}{c^2} - k_x^2 - k_y^2 + k_z \big(k_z~sign^2(z) -2\delta (z)\big) \bigg) h_{{BIC}_\times}
\end{split}
\end{equation}
being identical to the respective contravariant $T^{\mu \nu}$ components. Eq.~\ref{lemconservation} leads to (SM, Eq.~\ref{lemconservation_SM_x}):
\begin{equation}
\begin{split}    
    \partial_\mu T^{\mu x} = & ~ 
    \partial_x T^{xx} + \partial_y T^{yx} 
    \\
    = & ~
    (-k_x + k_y) 
    \\ & \times \frac{Ac^4}{16 \pi G}  \bigg(\frac{\omega^2}{c^2} - k_x^2 - k_y^2 + k_z \big(k_z~sign^2(z) -2\delta (z)\bigg) \\ & \times
    (cos(k_x x) cos(k_y y) - sin(k_y y) sin(k_x x))
    \\ & \times ~ cos(\omega ct/c) ~ e^{- k_z |z|}
\end{split}
\end{equation}
\begin{equation}
\begin{split}   
    \partial_\mu T^{\mu y} = & ~ 
    \partial_x T^{xy} + \partial_y T^{yy} 
    \\
    = & ~
    (k_x - k_y) 
    \\ & \times \frac{Ac^4}{16 \pi G}  \bigg(\frac{\omega^2}{c^2} - k_x^2 - k_y^2 + k_z \big(k_z~sign^2(z) -2\delta (z)\bigg) 
    \\ & \times
    (cos(k_x x) cos(k_y y) - sin(k_y y) sin(k_x x)) 
    \\ & \times cos(\omega ct/c) ~ e^{- k_z |z|}
\end{split}
\end{equation}
both being nullified for a square lattice perturbation ($k_x = k_y$). For $\nu = (t,z)$ the divergence is evidently zero. Thus, both bound metrics obey local energy-momentum conservation, a condition which is in general not fulfilled by regular metrics \cite{Geroch1987}.
\subsection{Isaacson stress-energy tensor}
The energy-momentum density and flux of the GW-BICs can be calculated using a second-order effective averaging tensor, known as the Isaacson stress-energy tensor $t_{\mu \nu}^{BIC_{_{(+,\times)}}}$ \cite{Isaacson1968}.
In the Lorenz gauge, the tensor can be written as:
\begin{equation} \label{isaacsonset}
    t_{\mu \nu}^{BIC_{_{(+,\times)}}} = \frac{c^4}{32 \pi G} 
    \langle 
    \partial_\mu \bar{h}_{\alpha \beta_{_{(+,\times)}}} ~ 
    \partial_\nu \bar{h}^{\alpha \beta}_{_{(+,\times)}}
    -\frac{1}{2}
    \partial_\mu \bar{h}_{_{(+,\times)}} ~ \partial_\nu \bar{h}_{_{(+,\times)}}
    \rangle
\end{equation}
where $\bar{h}_{_{(+,\times)}}$ is the trace of the trace-reversed perturbation and $\langle \cdot \cdot \cdot \rangle$ is the spacetime average of the high-frequency terms in the expression in brackets, usually performed for a region comprising a few wavelengths of the GW.  
The components for the ($+$)-polarized BIC are (SM, Eqs.~\ref{isaacson_tt+} - \ref{isaacson_zz+}):
\begin{equation}
    t_{\mu \nu}^{BIC_{+}} = e^{- 2k_z |z|}
    \begin{pmatrix}
			\frac{A^2 \omega^2 c^2}{64 \pi G} 	& 0 		& 0 		& 0 \\ 
			0 	& \frac{A^2 k_x^2 c^4}{64 \pi G}  	& 0 		& 0 \\ 
			0 	& 0 		& 0 	& 0 \\ 
			0 	& 0 		& 0 		& \frac{A^2 k_z^2 c^4}{64 \pi G} sign^2(z)
    \end{pmatrix}
\end{equation}
The non-zero localized components are associated with energy density ($t_{tt}$) and pressure ($t_{xx}, t_{zz}$). Interestingly, the $t_{zz}$ becomes zero ($sign^2(0) = 0$) at the localization plane. For the ($\times$)-polarized BIC (SM, Eqs.~\ref{isaacson_ttx} - \ref{isaacson_zzx}):
\begin{equation}
\begin{split}
    t_{\mu \nu}^{BIC_{\times}} & = e^{- 2 k_z |z|} \\
    &  \times \begin{pmatrix}
			\frac{A^2 \omega^2 c^2}{64 \pi G} 	& 0 		& 0 		& 0 \\ 
			0 	& \frac{A^2 k_x^2 c^4}{64 \pi G}  	& \frac{A^2 k_x k_y c^4}{64 \pi G} 		& 0 \\ 
			0 	& \frac{A^2 k_x k_y c^4}{64 \pi G} 		& \frac{A^2 k_y^2 c^4}{64 \pi G} 	& 0 \\ 
			0 	& 0 		& 0 		& \frac{A^2 k_z^2 c^4}{64 \pi G} sign^2(z)
    \end{pmatrix}
\end{split}
\end{equation}
in which again $t_{tt} = 0$ at $z = 0$. Here an additional pressure component ($t_{yy}$) and two shear components ($t_{xy}, t_{yx}$) are present compared to the Isaacson tensor of the first metric. Regardless, all coefficients for both bound metrics decay exponentially to zero away from $z = 0$. Similarly to a planar propagating GW, the $tt$ and $zz$ terms are non-zero. 
Conversely to $z$-propagating planar GWs, there is no energy flux/momentum density for the localized modes ($t_{tk}^{BIC_{(+,\times)}} = t_{kt}^{BIC_{(+,\times)}} = 0$) along this, or any other spatial direction.

The mode energy can be calculated by integrating the energy density $t_{tt}^{BIC_{(+,\times)}}$, which is identical for both BIC polarizations, over the spatial coordinates in the unit cell:
\begin{equation} \label{modeenergy}
    E_{BIC_{(+,\times)}} = \int_{Unit~cell}{t_{tt}^{BIC_{(+,\times)}} d^3x}
\end{equation}
Identically to photonic BICs in 1D gratings \cite{Berte2023, Berte2025}, a sensible finite result for the mode energy requires the truncation of the unit cell in the direction at which the mode is invariant (see SM for details), here applied to the ($+$)-polarized BIC (invariant in the $y$-direction). By truncating the unit cell for this polarization within the interval $[0, 2 \pi/k_y]$, an identical finite mode energy per unit cell is obtained for both polarizations:
\begin{equation}
    E_{BIC_{(+,\times)}} = \frac{A^2 \omega^2 c^2 \pi}{16 G k_x k_y k_z}
\end{equation}
\begin{figure}[t]
\includegraphics[scale=1.2]{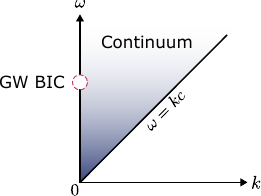}
\caption{\textbf{Scheme of GW BIC within the continuum of propagating wavevectors.} The continuum of propagating planar GWs lies between the wavevectors $k = 0$ and \mbox{$k = +\sqrt{\omega^2/c^2}$} (considering the linear dispersion), shaded area in blue. Regular non-propagating bound states exist below the light line ($\omega = kc$, i.e. for $k > \omega/c$). The symmetry-protected GW BICs (dashed red circle) solutions dwell at the $\Gamma$ point ($k = 0$), thus are spatially localized or bound, even though they are embedded in the continuum of propagating modes.
}
\label{fig3}
\end{figure} 

    \pdfbookmark[section]{Discussion}{Discussion}		
    \textit{Discussion.---}
    The bound states of GWs demonstrated here correspond to symmetry-protected BICs - which live at high-symmetry points in the reciprocal (momentum) space. In this case, as there is no phase variation between the edges of the unit cell, the BICs dwell at the $\Gamma$ point of the reciprocal lattice (i.e., at $k = 0$ - not to be confused with the in-plane modulation $k_x, k_y$ of the flat spacetime themselves, see Fig. 3). 
    The \textit{continuum}, defined by the set of wavevectors of propagating modes might be visualized by planes that intersect a planar GW. For a plane perpendicular to the propagation of the GW, it intersects the wave at an invariant phase, corresponding to $k = 0$ in reciprocal space. If the plane is parallel to the propagation of the GW it experiences the highest possible phase variation in space, i.e., given by the linear dispersion. 
    Thus, being at the $\Gamma$ point, these bound solutions are embedded in the \textit{continuum} of propagating wavevectors between $k = 0$ and $k = +\sqrt{\omega^2/c^2}$, 
    while regular bound states dwell below the light line ($\omega = kc$). Indeed, 
    GWs that do not propagate and decay exponentially from a given source (compact binaries) have been recently demonstrated \cite{Golat2020}. These non-propagating terms were shown to be dominant near their finite sources, but have not been considered for inifintely periodic systems as in here. As for photonic BICs \cite{Hsu2016}, there might also be GW BICs which are localized outside these high-symmetry points - termed accidental BICs - albeit with a non-intuitive spacetime strain in the unit cell.
    
The existence of symmetry-protected BICs as self-sustaining oscillations, set as an initial condition of an idealized system cannot be ruled out. However the strict requirements of such system (infinite in at least one spatial dimension, lack of absorption losses and symmetry breaking features) and infinite energy for a non-vanishingly small amplitude of oscillation $A$ make this scenario unlikely. Nevertheless, the excitation of the photonic analogues, modes corresponding to symmetry-protected BIC in lossy, finite metasurfaces has been demonstrated via near-field coupling \cite{vanHoof2021} and via electron energy loss spectroscopy \cite{Dong2022}.

The linearized equations are globally well-defined in the unit cell in the sense of distributional sources. However, the second order scalar invariants $K_{_{(+,\times)}}$ are ill-defined at the hypersurface $z = 0$ due to the product of distributions, being smooth and well-behaved though for all other values of $z$. No secularly accumulating effects are observed in the metric nor in its derived first and second order quantities, nor any coordinate pathologies in the form of event horizons/singularities are present. The distributions at the localization plane $\Sigma$ are thus true divergences in the curvature and not coordinate artifacts. The extraction of the energy from the mode in a realistic scenario would involve the analysis of the interaction between a \textit{quasi}-BIC with the kinetic and internal degrees of freedom of a massive body, which goes beyond the scope of this manuscript. The bound metrics also yield spacetimes which are not locally flat (given the exponentially decaying, non-zero Riemann curvature tensor components) outside the localization plane,
differently from other distributional sources in the form of domain walls
\cite{Vilenkin1983}.

    Given the demonstration of bound, in-plane periodic, solution of the linearized EFEs, which are also incompatible by symmetry with a planar GW, 
    it remains to be determined if corresponding GW \textit{quasi}-BIC 
    resonances
    could be excited in periodic $C_2$-asymmetric
    distributions of matter by a propagating GW (analogously to the excitation of a photonic \textit{q}BIC by a planar electromagnetic wave in optics). Additionally, if significant strain enhancements (relative to the incident GW amplitude) can be obtained at densities available with common materials (similarly to the electric/magnetic field enhancements in photonic \textit{q}BICs \cite{Seok2011}). Symmetry-protected \textit{q}BICs are quite compelling because localized states of longer lifetimes (i.e.~of higher $Q$-factors) can be excited with smaller asymmetries in the system \cite{Koshelev2018, Berte2023}.    
    Obviously, for practical devices (metasurfaces), these would need to be of feasible sizes. Smaller devices could be aimed at hypothetical GWs of higher frequencies (MHz and GHz) \cite{Aggarwal2021}. Laser-generated GWs at PHz range \cite{Atonga2024} are also an alternative for even smaller metasurfaces.
    Strain enhancements could provide direct measurements of GWs from inspiraling binaries at early times \cite{Taylor1982} and from other nearly continuous wave sources, such as asymmetric pulsars \cite{Christensen2022}, besides more stringent limits on tests of GR and \mbox{alternative models \cite{Christensen2022}.}
	
Eventual GW \textit{q}BICs of high $Q$-factors (e.g.  controlled by the degree of matter geometrical/density asymmetry \cite{Koshelev2018, Berte2023, Berte2025}) could have interesting effects on quantum systems, e.g., via mediating entanglement or in wavefunction collapse \cite{Carney2019}.
Phase-matching conditions for the interconversion between GWs and photons \cite{Gertsenshtein1962} (i.e. the coherence requirement between photons and gravitons in strong magnetic fields \cite{Aggarwal2021, Dyson2013}) could be relaxed in eventual GW \textit{q}BIC platforms, in which the conversion efficiency may be defined by the induced GW resonance, akin to non-linear generation in nanophotonic systems \cite{Berte2025, AgioBook}.
Future works will address the localization of GWs in the full non-linear GR theory. 

\pdfbookmark[section]{Ackowledgments}{Ackowledgments}	
\textit{Acknowledgments.---} 
I 
thank 
Prof. Daniel Vanzella
for revising this manuscript and for his invaluable comments and suggestions.

\pdfbookmark[section]{Data availability statement}{Data availability statement}	
\textit{Data availability statement.---} 
All data that support the findings of this study are included within the article and the  Supplemental Material.


    \onecolumngrid

\newpage

    \pdfbookmark[0]{Supplemental Material for Bound states in the continuum of gravitational waves}{Title_SM}
    \begin{center}
	\large{\textbf{Supplemental Material for \\ Bound states in the continuum of gravitational waves}}
	\vspace{4mm}
	\\
	\normalsize{Rodrigo Bert\'{e}}
    \\
    \textit{\small{Instituto de F\'{i}sica de S\~{a}o Carlos, Universidade de S\~{a}o Paulo \\ Caixa Postal 369, 13560-970, S\~{a}o Carlos, SP, Brazil}}
\end{center}

This Supplemental Material is organized as follows: it starts with a calculation of the Riemann and Ricci curvature tensor components, followed by the Ricci scalar for each of the proposed solutions; the non-trivial Kretschmann scalar invariant is then calculated for both polarizations; the proposed metrics are shown not to be a pure gauge transformation of flat space, to fulfill the Lorenz gauge condition and their trace-reversed perturbation tensors to be solutions of the homogeneous wave equation for suitable dispersions outside the localization hypersurface $\Sigma ~ (z = 0)$, followed by the demonstration of energy-momentum conservation by the bound metrics; lastly, the energy-momentum flux is obtained via the second order Isaacson stress-energy tensor.

We will assume a localized perturbation of the Minkowski flat spacetime metric ($\eta_{\mu \nu}$) such that the metric tensors $g_{{\mu \nu}_{(+,\times)}}$ are:
\begin{equation}
	g_{{\mu \nu}_{(+,\times)}} = 
    \eta_{\mu \nu} + h_{{\mu \nu}_{(+,\times)}}
    , \quad |h_{{\mu \nu}_{(+,\times)}}| \ll 1
\end{equation}

The fact we assume $|h_{{\mu \nu}_{(+,\times)}}| \ll 1$ implies that we are dealing with a linearized (weak) gravity regime. The subscripts ($+,\times$) refer to different non-zero coefficients in the perturbation tensor, as explained below. Here a ($+,-,-,-$) metric signature will be employed, such that:
\begin{equation}
  \eta_{\mu \nu} = 
  \eta^{\mu \nu} =
  \begin{pmatrix}
			1 	& 0 		& 0 		& 0 \\ 
			0 	& -1 	& 0 		& 0 \\ 
			0 	& 0 		& -1 	& 0 \\ 
			0 	& 0 		& 0 		& -1
 \end{pmatrix}
\end{equation}

A null cosmological constant ($\Lambda = 0$) is considered for the calculations. For a gravitational wave BIC protected by symmetry in which the $h_{xx}$ and $h_{yy}$ coefficients are non-zero, the proposed localized metric tensor $g_{{\mu \nu}_+}$ is:
\begin{equation} \label{eq_BIC_+}
  g_{{\mu \nu}_+} = 
  \begin{pmatrix}
			1 + h_{{BIC}_+} 	& 0 		& 0 		& 0 \\ 
			0 	& -1 - h_{{BIC}_+} 	& 0 		& 0 \\ 
			0 	& 0 		& -1 + h_{{BIC}_+} 	& 0 \\ 
			0 	& 0 		& 0 		& -1 - h_{{BIC}_+}
 \end{pmatrix}
\end{equation}
The ($+$) subscript in the metric tensor is chosen because the $h_{xx}$ and $h_{yy}$ perturbation tensor coefficients are the ones present for a ($+$)-polarized planar GW.
For this metric, the localized, standing wave-like, perturbation strain is:
\begin{equation} \label{eq_hbic_plus}
	h_{{BIC}_+}(ct, x, z) = A ~ cos(k_x x) ~ cos(\omega ct/c) ~
    e^{- k_z |z|}
\end{equation}
The metric perturbation decays exponentially from, or equivalently, is localized at, the plane $z = 0$. 
It also implies that we are dealing with a flat spacetime in the $z$ far-field, while the periodic perturbation is infinite in the $xy$ plane. 
Please note that, as we are dealing with exponentially-localized perturbations at $z = 0$, these do not correspond to regular standing waves arising from the interference of counter-propaging planar GWs.

On the other hand, when the perturbation involves the off-diagonal coefficients $h_{xy}$ and $h_{yx}$, the metric $g_{{\mu \nu}_{\times}}$ is written as:

\begin{equation} \label{eq_BIC_x}
  g_{{\mu \nu}_{\times}} = 
  \begin{pmatrix}
			1 + h_{{BIC}_{\times}} 	& 0 		& 0 		& 0 \\ 
			0 	& -1 	& - h_{{BIC}_{\times}} 		& 0 \\ 
			0 	& - h_{{BIC}_{\times}} 		& -1 	& 0 \\ 
			0 	& 0 		& 0 		& -1 - h_{{BIC}_{\times}}
 \end{pmatrix}
\end{equation}
Analogously, the ($\times$) subscript is chosen due to the cross-polarized planar GW being associated with these perturbation coefficients. The perturbation strain for this solution is: 
\begin{equation} \label{eq_hbic_x}
	h_{{BIC}_{\times}}(ct,x,y,z) = A ~ \big( cos(k_x x) sin(k_yy) + sin(k_xx) cos(k_yy)\big) ~ cos(\omega ct/c) ~ 
    e^{- k_z |z|}
\end{equation}
Differently from the strain in eq.~\ref{eq_hbic_plus}, now the perturbation is periodic in both $x$ and $y$ directions, which is essential for a vacuum solution outside the localization plane $z = 0$, as demonstrated below.

\subsection{Riemann curvature tensor} \label{sec:Riemann}
In this limit of weak gravitational fields, or linearized gravity, we may start by calculating the Riemann curvature tensor (RCT) components as:
\begin{equation} \label{lg_eq_RCTc}
	R_{\sigma \mu \nu}^\rho = 
		\frac{1}{2} \eta^{\rho\alpha}(	
		\partial_\mu \partial_\sigma h_{\alpha \nu}
		- \partial_\mu \partial_\alpha h_{\nu \sigma} 
		- \partial_\nu \partial_\sigma h_{\alpha \mu} 
		+ \partial_\nu \partial_\alpha h_{\mu \sigma}
		)
\end{equation}
\vspace{-2.5 em}
\subsubsection{$h_{xx}, h_{yy}$ BIC}
\vspace{-0.5 em}
Henceforth, the $ct$ time-like coordinate will be simply denoted as $t$ in the tensor indices and derivative operators. Note, however, that the derivatives in the time-like coordinate are performed with respect to $ct$. The subscripts $(+, \times)$ will also be omitted from the RCT components for clarity. The non-zero RCT tensor components for the $h_{{\mu \nu}_+}$ perturbation tensor are:
\begin{equation} \label{lg_eq_RCTc_BIC_hxxhyy}
 \begin{split}
	R_{xtx}^t = - R_{xxt}^t = 
	R_{ttx}^x = - R_{txt}^x =
	& ~ \frac{A}{2} (k_x^2 - \frac{\omega^2}{c^2}) ~ cos(k_x x) ~ cos( \omega ct/c) ~ e^{- k_z |z|} \\ \\
	R_{xtz}^t = - R_{xzt}^t = 
    R_{ztx}^t = - R_{zxt}^t =
    R_{ttz}^x = - R_{tzt}^x =
    R_{yyz}^x = - R_{yzy}^x = \quad \quad & \\
    - R_{xyz}^y = R_{xzy}^y =
    R_{zxy}^y = - R_{zyx}^y = 
    R_{ttx}^z = - R_{txt}^z = 
    - R_{yxy}^z = R_{yyx}^z = 
    & ~ - \frac{A}{2} k_x k_z ~ sign(z) ~ sin(k_x x) ~ cos(\omega ct/c) ~ 
    e^{- k_z |z|} \\ \\
    R_{xxz}^t = - R_{xzx}^t =
    - R_{yyz}^t = R_{yzy}^t =
    R_{txz}^x = - R_{tzx}^x =
    - R_{ztx}^x = R_{zxt}^x = \quad \quad & \\  
    - R_{tyz}^y = R_{tzy}^y = 
    R_{zty}^y = - R_{zyt}^y =
    R_{xtx}^z = - R_{xxt}^z =
    - R_{yty}^z = R_{yyt}^z =
    & ~ - \frac{A}{2} k_z \frac{\omega}{c} ~ sign(z) ~ cos( k_x x) ~ sin(\omega ct/c) ~ 
    e^{- k_z |z|} \\ \\
    R_{yty}^t = - R_{yyt}^t =
    R_{tty}^y = - R_{tyt}^y = 
    & ~ \frac{A}{2} \frac{\omega^2}{c^2} ~ cos(k_x x) ~ cos(\omega ct/c) ~ e^{- k_z |z|} \\ \\
    R_{yxy}^t = - R_{yyx}^t = 
    - R_{zxz}^t = R_{zzx}^t =
    - R_{yty}^x = R_{yyt}^x =
    R_{ztz}^x = - R_{zzt}^x = \quad \quad & \\  
    R_{txy}^y = - R_{tyx}^y =
    R_{xty}^y = - R_{xyt}^y =
    - R_{txz}^z = R_{tzx}^z =
    - R_{xtz}^z = R_{xzt}^z =
    & ~ - \frac{A}{2} k_x \frac{\omega}{c} ~ sin(k_x x) ~ sin(\omega ct/c) ~ e^{- k_z |z|} \\ \\
	R_{ztz}^t = - R_{zzt}^t =
    R_{ttz}^z = - R_{tzt}^z =
	& ~ - \frac{A}{2} \big(k_z (k_z ~ sign^2(z) - 2\delta (z))  + \frac{\omega^2}{c^2} \big) \\ ~
    & \times cos(k_x x) ~ cos(\omega ct/c) ~ e^{- k_z |z|} \\ \\
    - R_{yxy}^x = R_{yyx}^x =
    R_{xxy}^y = - R_{xyx}^y =
    & ~ \frac{A}{2} k_x^2 ~ cos(k_x x) ~ cos(\omega ct/c) ~ 
    e^{- k_z |z|} \\ \\
	R_{zxz}^x = - R_{zzx}^x =
    - R_{xxz}^z = R_{xzx}^z =
	&  ~ \frac{A}{2} \big(k_x^2 - k_z (k_z ~ sign^2(z) - 2\delta (z)) \big) 
    \\ & \times cos(k_x x) ~ cos(\omega ct/c) ~ e^{- k_z |z|} \\ \\
    R_{zyz}^y = - R_{zzy}^y = 
    - R_{yyz}^z = R_{yzy}^z = 
    & ~ \frac{A}{2} k_z(k_z ~ sign^2(z) - 2\delta (z)) ~ 
    \\ & \times cos(k_x x) ~ cos(\omega ct/c) ~ e^{- k_z |z|} \\ \\
 \end{split}
\end{equation}
where $sign(z)$ is the sign function of $z$ and $\delta(z)$ is the Dirac delta-function.
Note that the proposed bound metric leads to discontinuities at $z = 0$ in some of the components of the RCT (those that involve derivatives in $z$, yielding the $sign(z)$ and $\delta (z)$ terms). 

\subsubsection{$h_{xy}, h_{yx}$ BIC}
For the $h_{{\mu \nu}_{\times}}$ perturbation tensor, the non-zero RCT tensor components are:
\begin{equation} \label{lg_eq_RCTc_BIC}
 \begin{split}
    R_{xtx}^t = - R_{xxt}^t = 
	R_{ttx}^x = - R_{txt}^x =
    R_{zxz}^x = - R_{zzx}^x =
    - R_{xxz}^z = R_{xzx}^z =
	& ~ \frac{A}{2} k_x^2 ~ (sin(k_x x) cos(k_y y) ~ + ~ \\ & sin(k_y y) cos(k_x x)) ~ cos( \omega ct/c) ~ 
    e^{- k_z |z|}
    \\ \\
	R_{xty}^t = - R_{xyt}^t = 
	R_{ytx}^t = - R_{yxt}^t = 
    R_{tty}^x = - R_{tyt}^x = ~
	R_{ttx}^y = - R_{txt}^y = 
	& ~ \frac{A}{2} (k_x k_y - \frac{\omega^2}{c^2}) ~ (sin(k_x x) cos(k_y y) ~ + ~ \\ & sin(k_y y) cos(k_x x)) ~ cos( \omega ct/c) ~ 
    e^{- k_z |z|} \\ \\	
	R_{xtz}^t = - R_{xzt}^t = 
    R_{ztx}^t = - R_{zxt}^t =
    R_{ttz}^x = - R_{tzt}^x =
    - R_{yxz}^x = R_{yzx}^x = \quad \quad & \\
    - R_{zxy}^x = R_{zyx}^x = 
    R_{xxz}^y = - R_{xzx}^y =
    R_{ttx}^z = - R_{txt}^z =
    R_{xxy}^z = - R_{xyx}^z =
    & ~ - \frac{A}{2} k_x k_z ~ sign(z) ~ (sin(k_x x) sin(k_y y) ~ \\ - & cos(k_x x) cos(k_y y)) ~ cos( \omega ct/c) ~ 
    e^{- k_z |z|} \\ \\
    R_{yty}^t = - R_{yyt}^t =
    R_{tty}^y = - R_{tyt}^y =
    R_{zyz}^y = - R_{zzy}^y =
    - R_{yyz}^z = R_{yzy}^z =
    & ~ \frac{A}{2} k_y^2 ~ (sin(k_x x) cos(k_y y) ~ + ~ \\ & sin(k_y y) cos(k_x x)) ~ cos( \omega ct/c) ~ 
    e^{- k_z |z|} 
    \\ \\    			
    R_{ytz}^t = - R_{yzt}^t =
    R_{zty}^t = - R_{zyt}^t =
    R_{yyz}^x = - R_{yzy}^x =
    R_{ttz}^y = - R_{tzt}^y = \quad \quad & \\
    - R_{xyz}^y = R_{xzy}^y = 
    R_{zxy}^y = - R_{zyx}^y =
    R_{tty}^z = - R_{tyt}^z = 
    - R_{yxy}^z = R_{yyx}^z = 
    & ~ - \frac{A}{2} k_y k_z ~ sign(z) ~ (sin(k_x x) sin(k_y y) ~ \\ - & cos(k_x x) cos(k_y y)) ~ cos(\omega ct/c) ~ 
    e^{- k_z |z|} \\ \\
    - R_{xxy}^t = R_{xyx}^t = 
    - R_{zxz}^t = R_{zzx}^t =
    - R_{txy}^x = R_{tyx}^x =
    R_{ytx}^x = - R_{yxt}^x = \quad \quad & \\  
    R_{ztz}^x = - R_{zzt}^x =
    - R_{xtx}^y = R_{xxt}^y =
    - R_{txz}^z = R_{tzx}^z = 
    - R_{xtz}^z = R_{xzt}^z =
    & ~ - \frac{A}{2} k_x \frac{\omega}{c} ~ (sin(k_x x) sin(k_y y) ~ - ~ \\ & cos(k_x x) cos(k_y y)) ~ sin( \omega ct/c) ~ 
    e^{- k_z |z|} \\ \\
    R_{yxy}^t = - R_{yyx}^t = 
    - R_{zyz}^t = R_{zzy}^t =
    - R_{yty}^x = R_{yyt}^x =
    R_{txy}^y = - R_{tyx}^y = \quad \quad & \\
    R_{xty}^y = - R_{xyt}^y =
    R_{ztz}^y = - R_{zzt}^y =
    - R_{tyz}^z = R_{tzy}^z =
   - R_{ytz}^z = R_{yzt}^z =
    & ~ - \frac{A}{2} k_y \frac{\omega}{c} ~ (sin(k_x x) sin(k_y y) ~ - ~ \\ & cos(k_x x) cos(k_y y)) ~ sin(\omega ct/c) ~ 
    e^{- k_z |z|} \\ \\
    R_{xyz}^t = - R_{xzy}^t = 
    R_{yxz}^t = - R_{yzx}^t = 
    R_{tyz}^x = - R_{tzy}^x = 
    - R_{zty}^x = R_{zyt}^x = \quad \quad & \\
    R_{txz}^y = - R_{tzx}^y =  
    - R_{ztx}^y = R_{zxt}^y = 
    R_{xty}^z = - R_{xyt}^z = 
    R_{ytx}^z = - R_{yxt}^z = 
    & ~ - \frac{A}{2} k_z ~ sign(z) \frac{\omega}{c} ~ (sin(k_x x) cos(k_y y) ~ ~ \\ + & sin(k_y y) cos(k_x x)) ~ sin(\omega ct/c) ~ e^{- k_z |z|} \\ \\
	R_{ztz}^t = - R_{zzt}^t =
    R_{ttz}^z = - R_{tzt}^z =
	& ~ - \frac{A}{2} (k_z (k_z ~ sign^2(z) - 2\delta (z)) + \frac{\omega^2}{c^2}) \\ \times & (sin(k_x x) cos(k_y y) + sin(k_y y) cos(k_x x)) \\ & \times cos( \omega ct/c) ~ 
    e^{- k_z |z|} \\ \\
 \end{split}
\end{equation}

\begin{align*}
 \begin{split}
    - R_{yxy}^x = R_{yyx}^x =
    R_{xxy}^y = - R_{xyx}^y =
    & ~ A k_x k_y ~ (sin(k_x x) cos(k_y y) ~ + ~ \\ & sin(k_y y) cos(k_x x)) ~ cos( \omega ct/c) ~ 
    e^{- k_z |z|} \\ \\
    R_{zyz}^x = - R_{zzy}^x = 
    R_{zxz}^y = - R_{zzx}^y = 
    - R_{xyz}^z = R_{xzy}^z = 
    - R_{yxz}^z = R_{yzx}^z = 
    & ~ \frac{A}{2} (k_x k_y - k_z (k_z ~ sign^2(z) - 2\delta (z))) \\ \times & (sin(k_x x) cos(k_y y) + sin(k_y y) cos(k_x x)) \\ \times & cos( \omega ct/c) ~ 
    e^{- k_z |z|} \\ \\
 \end{split}
\end{align*}

Similarly to the previous BIC perturbation, some components of the RCT are discontinuous at $z = 0$ (those containing $sign(z)$ or $\delta (z)$ terms).

			
\subsection*{Ricci tensor} \label{RicciTensor}
In the same weak gravity limit, we may calculate the Ricci tensor components $R_{{\sigma \nu}_{(+, \times)}}$ as:

\begin{equation} \label{lg_eq_RT_GW}
	R_{{\sigma \nu}_{(+, \times)}} = 
		\frac{1}{2} (\partial_\mu \partial_\sigma h_{~\nu}^\mu 
		+ \partial_\nu \partial_\alpha h_{~ \sigma}^\alpha
		- \Box h_{\nu \sigma}
		- \partial_\nu \partial_\sigma h)
\end{equation}

in which $h = h_{~\mu}^\mu = \eta^{\mu \alpha} h_{\alpha \mu}$ and $\Box = \eta^{\mu \alpha} \partial_\mu \partial_\alpha = \partial_\mu \partial^\mu$ is the d'Alembert operator. The $(+, \times)$ subscripts have and will be omitted from the individual perturbation tensor coefficients for clarity.

\subsubsection{$h_{xx}, h_{yy}$ BIC}
The $R_{{\sigma \nu}_+}$ components that result from the $h_{{\mu \nu}_+}$ perturbation tensor (Eq.~\ref{eq_BIC_+}, Eq.~\ref{eq_hbic_plus}) are:
\begin{equation} \label{rtt+}
\begin{split}
    R_{{tt}_+} 
    & = \frac{1}{2} \bigg(\partial_\mu \partial_t h_{~t} ^\mu 
		+ \partial_t \partial_\alpha h_{~ t}^\alpha
		- \Box h_{tt}
		- \partial_t \partial_t h
        \bigg) \\
    & = \frac{1}{2} \bigg(\partial_t^2 \eta^{tt} h_{tt}
    + \partial_t^2 \eta^{tt} h_{tt}
    - \eta^{tt} \partial_t^2 h_{tt}
    - \eta^{xx} \partial_x^2 h_{tt}
    - \eta^{zz} \partial_z^2 h_{tt}
    - \partial_t^2 (
    \eta^{tt} h_{tt} + \eta^{xx} h_{xx} + \eta^{yy} h_{yy} + \eta^{zz} h_{zz})
    \bigg) \\
    & = \frac{1}{2} \bigg(
    \partial_x^2 h_{tt} + \partial_z^2 h_{tt}
    + \partial_t^2 h_{xx} + \partial_t^2 h_{yy} + \partial_t^2 h_{zz}
    \bigg) \\
    & = \frac{1}{2} \bigg(
    \partial_x^2 h_{{BIC}_+} + \partial_z^2 h_{{BIC}_+} 
    - \partial_t^2 h_{{BIC}_+} + \partial_t^2 h_{{BIC}_+} - \partial_t^2 h_{{BIC}_+}
    \bigg) \\
    & = \frac{h_{{BIC}_+}}{2} \bigg(
    - k_x^2 + k_z( k_z ~ sign^2(z) - 2\delta (z))  
    + (\omega/c)^2
    \bigg) \\
    \therefore
    & = \frac{h_{{BIC}_+}}{2} \bigg(
    - k_x^2 + k_z^2 
    + \omega^2/c^2
    \bigg) ~ for ~ z \neq 0 ~ ; ~ 
    = \frac{cos(k_x x) ~ cos(\omega ct/c)}{2} \bigg(
    - k_x^2 - 2 k_z \delta (0)  
    + \omega^2/c^2
    \bigg) ~ for ~ z = 0
\end{split}        
\end{equation}

\begin{equation}
\begin{split}
    R_{{tx}_+}
    = R_{{xt}_+}
    & = \frac{1}{2} \bigg(
        \partial_\mu \partial_t h_{~ x} ^\mu 
		+ \partial_x \partial_\alpha h_{~ t}^\alpha
		- \Box h_{xt}
		- \partial_x \partial_t h
        \bigg) \\
    & = \frac{1}{2} \bigg(
        \partial_x \partial_t \eta^{xx} h_{xx}
        + \partial_x \partial_t \eta^{tt} h_{tt}
        - \partial_x \partial_t (
        \eta^{tt} h_{tt} + \eta^{xx} h_{xx} + \eta^{yy} h_{yy} + \eta^{zz} h_{zz})
        \bigg) \\
    & = \frac{1}{2} \bigg(
        - \partial_x \partial_t h_{xx}
        + \partial_x \partial_t h_{tt}
        - \partial_x \partial_t h_{tt} 
        + \partial_x \partial_t h_{xx} 
        + \partial_x \partial_t h_{yy} 
        + \partial_x \partial_t h_{zz}
        \bigg) \\
    & = \frac{1}{2} \bigg(
        \partial_x \partial_t h_{yy} 
        + \partial_x \partial_t h_{zz}
        \bigg) \\
    & = \frac{1}{2} \bigg(
        \partial_x \partial_t h_{{BIC}_+}
        - \partial_x \partial_t h_{{BIC}_+}
        \bigg) \\
    & = 0
\end{split}        
\end{equation}

\begin{equation}
\begin{split}
    R_{{ty}_+}
    = R_{{yt}_+}
    & = \frac{1}{2} \bigg(
        \partial_\mu \partial_t h_{~ y} ^\mu 
		+ \partial_y \partial_\alpha h_{~ t}^\alpha
		- \Box h_{yt}
		- \partial_y \partial_t h
        \bigg) \\
    & = \frac{1}{2} \bigg(
        \partial_t^2 h_{~ y} ^t
        + \partial_x \partial_t h_{~ y} ^x 
        + \partial_y \partial_t h_{~ y} ^y
        + \partial_z \partial_t h_{~ y} ^z 
        \bigg) \\
    & = 0
\end{split}        
\end{equation}

\begin{equation}
\begin{split}
    R_{{tz}_+}
    = R_{{zt}_+}
    & = \frac{1}{2} \bigg(
        \partial_\mu \partial_t h_{~ z} ^\mu 
		+ \partial_z \partial_\alpha h_{~ t}^\alpha
		- \Box h_{zt}
		- \partial_z \partial_t h
        \bigg) \\
    & = \frac{1}{2} \bigg(
        \partial_z \partial_t \eta^{zz} h_{zz}
        + \partial_z \partial_t \eta^{tt} h_{tt}
        - \partial_z \partial_t (
        \eta^{tt} h_{tt} + \eta^{xx} h_{xx} + \eta^{yy} h_{yy} + \eta^{zz} h_{zz})
        \bigg) \\
    & = \frac{1}{2} \bigg(
        - \partial_z \partial_t h_{zz}
        + \partial_z \partial_t h_{tt}
        - \partial_z \partial_t h_{tt} 
        + \partial_z \partial_t h_{xx} 
        + \partial_z \partial_t h_{yy} 
        + \partial_z \partial_t h_{zz}
        \bigg) \\
    & = \frac{1}{2} \bigg(
         \partial_z \partial_t h_{{BIC}_+}
        + \partial_z \partial_t h_{{BIC}_+}
        - \partial_z \partial_t h_{{BIC}_+} 
        - \partial_z \partial_t h_{{BIC}_+} 
        + \partial_z \partial_t h_{{BIC}_+}
        - \partial_z \partial_t h_{{BIC}_+}
        \bigg) \\
    & = 0
\end{split}        
\end{equation}

\begin{equation} \label{rxx+}
\begin{split}
    R_{{xx}_+} 
    & = \frac{1}{2} \bigg(\partial_\mu \partial_x h_{~ x} ^\mu 
		+ \partial_x \partial_\alpha h_{~ x}^\alpha
		- \Box h_{xx}
		- \partial_x \partial_x h
        \bigg) \\
    & = \frac{1}{2} \bigg(\partial_x^2 \eta^{xx} h_{xx}
    + \partial_x^2 \eta^{xx} h_{xx}
    - \eta^{tt} \partial_t^2 h_{xx}
    - \eta^{xx} \partial_x^2 h_{xx}
    - \eta^{zz} \partial_z^2 h_{xx}
    - \partial_x^2 (
    \eta^{tt} h_{tt} + \eta^{xx} h_{xx} + \eta^{yy} h_{yy} + \eta^{zz} h_{zz})
    \bigg) \\
    & = \frac{1}{2} \bigg(- \partial_x^2 h_{xx}
    - \partial_x^2 h_{xx}
    - \partial_t^2 h_{xx}
    + \partial_x^2 h_{xx}
    + \partial_z^2 h_{xx}
    - \partial_x^2 h_{tt} 
    + \partial_x^2 h_{xx} 
    + \partial_x^2 h_{yy} 
    + \partial_x^2 h_{zz}
    \bigg) \\
    & = \frac{1}{2} \bigg(\partial_x^2 h_{{BIC}_+}
    + \partial_t^2 h_{{BIC}_+}
    - \partial_z^2 h_{{BIC}_+}
    - \partial_x^2 h_{{BIC}_+}
    - \partial_x^2 h_{{BIC}_+} 
    + \partial_x^2 h_{{BIC}_+} 
    - \partial_x^2 h_{{BIC}_+}
    \bigg) \\
    & = \frac{h_{{BIC}_+}}{2} \bigg(
    - (\omega/c)^2 - k_z( k_z ~ sign^2(z) - 2\delta (z))  
    + k_x^2
    \bigg) \\
    \therefore
    & = \frac{h_{{BIC}_+}}{2} \bigg(
    - \omega^2/c^2 - k_z^2 
    + k_x^2
    \bigg) ~ for ~ z \neq 0 ~ ; ~ 
    = \frac{cos(k_x x) ~ cos(\omega ct/c)}{2} \bigg(
    - \omega^2/c^2 + 2 k_z \delta (0)  
    + k_x^2
    \bigg) ~ for ~ z = 0
\end{split}        
\end{equation}

\begin{equation}
\begin{split}
    R_{{xy}_+}
    = R_{{yx}_+}
    & = \frac{1}{2} \bigg(
        \partial_\mu \partial_x h_{~ y} ^\mu 
		+ \partial_y \partial_\alpha h_{~ x}^\alpha
		- \Box h_{yx}
		- \partial_y \partial_x h
        \bigg) \\
    & = \frac{1}{2} \bigg(
        \partial_t \partial_x h_{~ y} ^t
        + \partial_x^2 h_{~ y} ^x 
        + \partial_y \partial_x h_{~ y} ^y
        + \partial_z \partial_x h_{~ y} ^z 
        \bigg) \\
    & = 0
\end{split}        
\end{equation}

\begin{equation}
\begin{split}
    R_{{xz}_+}
    = R_{{zx}_+}
    & = \frac{1}{2} \bigg(
        \partial_\mu \partial_x h_{~ z} ^\mu 
		+ \partial_z \partial_\alpha h_{~ x}^\alpha
		- \Box h_{zx}
		- \partial_z \partial_x h
        \bigg) \\
    & = \frac{1}{2} \bigg(
        \partial_z \partial_x \eta^{zz} h_{zz}
        + \partial_z \partial_x \eta^{xx} h_{xx}
        - \partial_z \partial_x (
        \eta^{tt} h_{tt} + \eta^{xx} h_{xx} + \eta^{yy} h_{yy} + \eta^{zz} h_{zz})
        \bigg) \\
    & = \frac{1}{2} \bigg(
        - \partial_z \partial_x h_{zz}
        - \partial_z \partial_x h_{xx}
        - \partial_z \partial_x h_{tt} 
        + \partial_z \partial_x h_{xx} 
        + \partial_z \partial_x h_{yy} 
        + \partial_z \partial_x h_{zz}
        \bigg) \\
    & = \frac{1}{2} \bigg(
         \partial_z \partial_x h_{{BIC}_+}
        + \partial_z \partial_x h_{{BIC}_+}
        - \partial_z \partial_x h_{{BIC}_+} 
        - \partial_z \partial_x h_{{BIC}_+} 
        + \partial_z \partial_x h_{{BIC}_+}
        - \partial_z \partial_x h_{{BIC}_+}
        \bigg) \\
    & = 0
\end{split}        
\end{equation}

\begin{equation} \label{ryy+}
\begin{split}
    R_{{yy}_+} 
    & = \frac{1}{2} \bigg(
        \partial_\mu \partial_y h_{~ y} ^\mu 
		+ \partial_y \partial_\alpha h_{~ y}^\alpha
		- \Box h_{yy}
		- \partial_y \partial_y h
        \bigg) \\
    & = -\frac{1}{2} \bigg(
    \eta^{tt} \partial_t^2 h_{yy}
    + \eta^{xx} \partial_x^2 h_{yy}
    + \eta^{zz} \partial_z^2 h_{yy}
    \bigg) \\
    & = -\frac{1}{2} \bigg(
    \partial_t^2 h_{yy}
    - \partial_x^2 h_{yy}
    - \partial_z^2 h_{yy}
    \bigg) \\
    & = -\frac{1}{2} \bigg(
    \partial_t^2 h_{{BIC}_+}
    - \partial_x^2 h_{{BIC}_+}
    - \partial_z^2 h_{{BIC}_+}
    \bigg) \\
    & = - \frac{h_{{BIC}_+}}{2} \bigg(
    - (\omega/c)^2 + k_x^2
    - k_z( k_z ~ sign^2(z) - 2\delta (z))  
    \bigg) 
    \\
    \therefore
    & = - \frac{h_{{BIC}_+}}{2} \bigg(
    - \omega^2/c^2 + k_x^2 - k_z^2 
    \bigg) ~ for ~ z \neq 0 ~ ; ~ 
    = - \frac{cos(k_x x) ~ cos(\omega ct/c)}{2} \bigg(
    - \omega^2/c^2 + 2 k_z \delta (0)  
    + k_x^2
    \bigg) ~ for ~ z = 0
\end{split}        
\end{equation}

\begin{equation}
\begin{split}
    R_{{yz}_+}
    = R_{{zy}_+}
    & = \frac{1}{2} \bigg(
        \partial_\mu \partial_y h_{~ z} ^\mu 
		+ \partial_z \partial_\alpha h_{~ y}^\alpha
		- \Box h_{zy}
		- \partial_z \partial_y h
        \bigg) \\
    & = \frac{1}{2} \bigg(
        \partial_z \partial_t h_{~ y} ^t
        + \partial_z \partial_x h_{~ y} ^x 
        + \partial_z \partial_y h_{~ y} ^y
        + \partial_z^2 h_{~ y} ^z 
        \bigg) \\
    & = 0
\end{split}        
\end{equation}

\begin{equation} \label{rzz+}
\begin{split}
    R_{{zz}_+} 
    & = \frac{1}{2} \bigg(
        \partial_\mu \partial_z h_{~ z} ^\mu 
		+ \partial_z \partial_\alpha h_{~ z}^\alpha
		- \Box h_{zz}
		- \partial_z \partial_z h
        \bigg) \\
    & = \frac{1}{2} \bigg(
    \partial_z^2 \eta^{zz} h_{zz}
    + \partial_z^2 \eta^{zz} h_{zz}
    - \eta^{tt} \partial_t^2 h_{zz}
    - \eta^{xx} \partial_x^2 h_{zz}
    - \eta^{zz} \partial_z^2 h_{zz}
    - \partial_z^2 (
    \eta^{tt} h_{tt} + \eta^{xx} h_{xx} + \eta^{yy} h_{yy} + \eta^{zz} h_{zz})
    \bigg) \\
    & = \frac{1}{2} \bigg(
    - \partial_z^2 h_{zz}
    - \partial_z^2 h_{zz}
    - \partial_t^2 h_{zz}
    + \partial_x^2 h_{zz}
    + \partial_z^2 h_{zz}
    - \partial_z^2 h_{tt} 
    + \partial_z^2 h_{xx} 
    + \partial_z^2 h_{yy} 
    + \partial_z^2 h_{zz}
    \bigg) \\
    & = \frac{1}{2} \bigg(
    - \partial_t^2 h_{zz}
    + \partial_x^2 h_{zz}
    - \partial_z^2 h_{tt} 
    + \partial_z^2 h_{xx} 
    + \partial_z^2 h_{yy} 
    \bigg) \\
    & = \frac{1}{2} \bigg(
    \partial_t^2 h_{{BIC}_+}
    - \partial_x^2 h_{{BIC}_+}
    - \partial_z^2 h_{{BIC}_+} 
    - \partial_z^2 h_{{BIC}_+} 
    + \partial_z^2 h_{{BIC}_+} 
    \bigg) \\
    & = \frac{1}{2} \bigg(
    \partial_t^2 h_{{BIC}_+}
    - \partial_x^2 h_{{BIC}_+}
    - \partial_z^2 h_{{BIC}_+}
    \bigg) \\
    & = \frac{h_{{BIC}_+}}{2} \bigg(
    - (\omega/c)^2 + k_x^2
    - k_z( k_z ~ sign^2(z) - 2\delta (z))  
    \bigg) 
    \\
    \therefore
    & = \frac{h_{{BIC}_+}}{2} \bigg(
    - \omega^2/c^2 + k_x^2 - k_z^2 
    \bigg) ~ for ~ z \neq 0 ~ ; ~ 
    = \frac{cos(k_x x) ~ cos(\omega ct/c)}{2} \bigg(
    - \omega^2/c^2 + 2 k_z \delta (0)  
    + k_x^2
    \bigg) ~ for ~ z = 0
\end{split}        
\end{equation}

Except for the diagonal terms $R_{{tt}_+}$, $R_{{xx}_+}$, $R_{{yy}_+}$ and $R_{{zz}_+}$, all other Ricci tensor elements are zero. In addition, except for $z = 0$ where the terms imply a distributional stress-energy tensor $^{(\Sigma)}T^{\mu \nu}_{_{+}}$, all these non-zero coefficients become zero if the following dispersion is obeyed by the bound wave:
\begin{equation} \label{eq_dispersion_+}
        \omega^2 = c^2(k_x^2 - k_z^2)
\end{equation}
Interestingly, this dispersion relation is identical (apart from the permittivity term) to that of a surface polariton (an electromagnetic wave bound at the interface between two media of opposite real parts in their permittivity) \cite{AgranovichBook}. In essence, being the surface polariton also a mode that varies sinusoidally in the plane while decaying exponentially away from the interface.

\subsubsection{$h_{xy}, h_{yx}$ BIC}

Similarly, we may calculate the $R_{{\sigma \nu}_{\times}}$ components from the $h_{{\mu \nu}_{\times}}$ perturbation tensor (Eq.~\ref{eq_BIC_x}, Eq.~\ref{eq_hbic_x}) as:

\begin{equation} \label{rttx}
\begin{split}
    R_{{tt}_{\times}} 
    & = \frac{1}{2} \bigg(
        \partial_\mu \partial_t h_{~t} ^\mu 
		+ \partial_t \partial_\alpha h_{~ t}^\alpha
		- \Box h_{tt}
		- \partial_t \partial_t h
        \bigg) \\
    & = \frac{1}{2} \bigg(
    \partial_t^2 \eta^{tt} h_{tt}
    + \partial_t^2 \eta^{tt} h_{tt}
    - \eta^{tt} \partial_t^2 h_{tt}
    - \eta^{xx} \partial_x^2 h_{tt}
    - \eta^{yy} \partial_y^2 h_{tt}
    - \eta^{zz} \partial_z^2 h_{tt}
    - \partial_t^2 (
    \eta^{tt} h_{tt} + \eta^{zz} h_{zz})
    \bigg) \\
    & = \frac{1}{2} \bigg(
    \partial_x^2 h_{tt}
    + \partial_y^2 h_{tt}
    + \partial_z^2 h_{tt}
    + \partial_t^2 h_{zz}
    \bigg) \\
    & = \frac{1}{2} \bigg(
    \partial_x^2 h_{{BIC}_{\times}}
    + \partial_y^2 h_{{BIC}_{\times}}
    + \partial_z^2 h_{{BIC}_{\times}}
    - \partial_t^2 h_{{BIC}_{\times}}
    \bigg) \\
    & = \frac{h_{{BIC}_{\times}}}{2} \bigg(
        - k_x^2
        - k_y^2
        + k_z( k_z ~ sign^2(z) - 2\delta (z)) 
        + \omega^2/c^2
        \bigg) 
    \\
    \therefore
    & = \frac{h_{{BIC}_{\times}}}{2} \bigg(
        - k_x^2 - k_y^2 + k_z^2 + \omega^2/c^2  
        \bigg) ~ for ~ z \neq 0 ~ ; ~ 
        \\
    & = \frac{(cos(k_x x)sin(k_y y) + sin(k_x x)cos(k_y y)) ~ cos(\omega ct/c)}{2} \bigg(
        - k_x^2 - k_y^2
         + 2 k_z \delta (0)
         - \omega^2/c^2
        \bigg) ~ for ~ z = 0
\end{split}        
\end{equation}

\begin{equation}
\begin{split}
    R_{{tx}_{\times}}
    = R_{{xt}_{\times}}
    & = \frac{1}{2} \bigg(
        \partial_\mu \partial_t h_{~ x} ^\mu 
		+ \partial_x \partial_\alpha h_{~ t}^\alpha
		- \Box h_{xt}
		- \partial_x \partial_t h
        \bigg) \\
    & = \frac{1}{2} \bigg(
        \partial_y \partial_t \eta^{yy} h_{yx}
        + \partial_x \partial_t \eta^{tt} h_{tt}
        - \partial_x \partial_t (
        \eta^{tt} h_{tt} + \eta^{zz} h_{zz})
        \bigg) \\
    & = \frac{1}{2} \bigg(
        - \partial_y \partial_t h_{yx}
        + \partial_x \partial_t h_{zz}
        \bigg) \\
    & = \frac{1}{2} \bigg(
        \partial_y \partial_t h_{{BIC}_{\times}}
        - \partial_x \partial_t h_{{BIC}_{\times}}
        \bigg) \\
    & = \frac{A \omega}{2 c} \bigg(
        k_y
        - k_x
        \bigg) 
        ~ (sin(k_x x) sin(k_y y) - cos(k_x x) cos(k_y y)) ~ sin( \omega ct/c) ~ 
        e^{- k_z |z|}\\
\end{split}        
\end{equation}

\begin{equation}
\begin{split}
    R_{{ty}_{\times}}
    = R_{{yt}_{\times}}
    & = \frac{1}{2} \bigg(
        \partial_\mu \partial_t h_{~ y} ^\mu 
		+ \partial_y \partial_\alpha h_{~ t}^\alpha
		- \Box h_{yt}
		- \partial_y \partial_t h
        \bigg) \\
    & = \frac{1}{2} \bigg(
        \partial_x \partial_t \eta^{xx} h_{xy}
        + \partial_y \partial_t \eta^{tt} h_{tt}
        - \partial_y \partial_t (
        \eta^{tt} h_{tt} + \eta^{zz} h_{zz})
        \bigg) \\
    & = \frac{1}{2} \bigg(
        - \partial_x \partial_t h_{xy}
        + \partial_y \partial_t h_{zz}
        \bigg) \\
    & = \frac{1}{2} \bigg(
         \partial_x \partial_t h_{{BIC}_{\times}}
        - \partial_y \partial_t h_{{BIC}_{\times}}
        \bigg) \\
    & = \frac{A \omega}{2 c} \bigg(
        k_x
        - k_y
        \bigg) 
        ~ (sin(k_x x) sin(k_y y) - cos(k_x x) cos(k_y y)) ~ sin( \omega ct/c) ~ 
        e^{- k_z |z|} \\
\end{split}        
\end{equation}

\begin{equation}
\begin{split}
    R_{{tz}_{\times}}
    = R_{{zt}_{\times}}
    & = \frac{1}{2} \bigg(
        \partial_\mu \partial_t h_{~ z} ^\mu 
		+ \partial_z \partial_\alpha h_{~ t}^\alpha
		- \Box h_{zt}
		- \partial_z \partial_t h
        \bigg) \\
    & = \frac{1}{2} \bigg(
        \partial_z \partial_t \eta^{zz} h_{zz}
        + \partial_z \partial_t \eta^{tt} h_{tt}
        - \partial_z \partial_t (
        \eta^{tt} h_{tt} + \eta^{zz} h_{zz})
        \bigg) \\
    & = 0
\end{split}        
\end{equation}

\begin{equation} \label{rxxx}
\begin{split}
    R_{{xx}_{\times}} 
    & = \frac{1}{2} \bigg(
        \partial_\mu \partial_x h_{~ x} ^\mu 
		+ \partial_x \partial_\alpha h_{~ x}^\alpha
		- \Box h_{xx}
		- \partial_x \partial_x h
        \bigg) \\
    & = \frac{1}{2} \bigg(
    \partial_y \partial_x \eta^{yy} h_{yx}
    + \partial_x \partial_y \eta^{yy} h_{yx}
    - \partial_x^2 (
    \eta^{tt} h_{tt} + \eta^{zz} h_{zz})
    \bigg) \\
    & = \frac{1}{2} \bigg(
    - \partial_y \partial_x h_{yx}
    - \partial_x \partial_y h_{yx}
    - \partial_x^2 h_{tt} 
    + \partial_x^2 h_{zz}
    \bigg) \\
    & = \frac{1}{2} \bigg(
    \partial_y \partial_x h_{{BIC}_{\times}}
    + \partial_x \partial_y h_{{BIC}_{\times}}
    - \partial_x^2 h_{{BIC}_{\times}} 
    - \partial_x^2 h_{{BIC}_{\times}}
    \bigg) \\
    & = \frac{h_{{BIC}_{\times}}}{2} \bigg(
    - k_y k_x
    - k_x k_y
    + k_x^2 
    + k_x^2
    \bigg) \\
    & = h_{{BIC}_{\times}} k_x \bigg(
    - k_y
    + k_x
    \bigg) \\
\end{split}        
\end{equation}

\begin{equation}
\begin{split}
    R_{{xy}_{\times}}
    = R_{{yx}_{\times}}
    & = \frac{1}{2} \bigg(
        \partial_\mu \partial_x h_{~ y} ^\mu 
		+ \partial_y \partial_\alpha h_{~ x}^\alpha
		- \Box h_{yx}
		- \partial_y \partial_x h
        \bigg) \\
    & = \frac{1}{2} \bigg(
        \partial_x^2 \eta^{xx} h_{xy} 
        + \partial_y^2 \eta^{yy} h_{yx} 
        - \eta^{tt} \partial_t^2 h_{yx}
        - \eta^{xx} \partial_x^2 h_{yx}
        - \eta^{yy} \partial_y^2 h_{yx}
        - \eta^{zz} \partial_z^2 h_{yx}
        - \partial_y \partial_x (
        \eta^{tt} h_{tt} + \eta^{zz} h_{zz})
        \bigg) \\
    & = \frac{1}{2} \bigg(
        - \partial_t^2 h_{yx}
        + \partial_z^2 h_{yx}
        - \partial_y \partial_x h_{tt} 
        + \partial_y \partial_x h_{zz}
        \bigg) \\
    & = \frac{1}{2} \bigg(
        \partial_t^2 h_{{BIC}_{\times}} 
        - \partial_z^2 h_{{BIC}_{\times}} 
        - \partial_y \partial_x h_{{BIC}_{\times}}  
        - \partial_y \partial_x h_{{BIC}_{\times}} 
        \bigg) \\
    & = \frac{h_{{BIC}_{\times}}}{2} \bigg(
        - \omega^2/c^2
        - k_z( k_z ~ sign^2(z) - 2\delta (z)) 
        + k_y k_x 
        + k_y k_x
        \bigg) 
    \\
    \therefore
    & = \frac{h_{{BIC}_{\times}}}{2} \bigg(
        - \omega^2/c^2
        - k_z^2
        + k_y k_x 
        + k_y k_x
        \bigg) ~ for ~ z \neq 0 ~ ; ~ 
        \\
    & = \frac{(cos(k_x x)sin(k_y y) + sin(k_x x)cos(k_y y)) ~ cos(\omega ct/c)}{2} \bigg(
         - \omega^2/c^2
         + 2 k_z \delta (0)
        + k_y k_x 
        + k_y k_x
        \bigg) ~ for ~ z = 0
\end{split}        
\end{equation}

\begin{equation}
\begin{split}
    R_{{xz}_{\times}}
    = R_{{zx}_{\times}}
    & = \frac{1}{2} \bigg(
        \partial_\mu \partial_x h_{~ z} ^\mu 
		+ \partial_z \partial_\alpha h_{~ x}^\alpha
		- \Box h_{zx}
		- \partial_z \partial_x h
        \bigg) \\
    & = \frac{1}{2} \bigg(
        \partial_z \partial_x \eta^{zz} h_{zz}
        + \partial_z \partial_y \eta^{yy} h_{yx}
        - \partial_z \partial_x (
        \eta^{tt} h_{tt} + \eta^{zz} h_{zz})
        \bigg) \\
    & = \frac{1}{2} \bigg(
        - \partial_z \partial_y h_{yx}
        - \partial_z \partial_x h_{tt}
        \bigg) \\
    & = \frac{1}{2} \bigg(
        \partial_z \partial_y h_{{BIC}_{\times}}
        - \partial_z \partial_x h_{{BIC}_{\times}}
        \bigg) \\
    & = \frac{A k_z ~ sign(z)}{2} \bigg(
        k_y
        - k_x
        \bigg) 
        ~ (sin(k_x x) sin(k_y y) - cos(k_x x) cos(k_y y)) ~ cos( \omega ct/c) ~ 
        e^{- k_z |z|} \\
\end{split}        
\end{equation}

\begin{equation} \label{ryyx}
\begin{split}
    R_{{yy}_{\times}} 
    & = \frac{1}{2} \bigg(
        \partial_\mu \partial_y h_{~ y} ^\mu 
		+ \partial_y \partial_\alpha h_{~ y}^\alpha
		- \Box h_{yy}
		- \partial_y \partial_y h
        \bigg) \\
    & = \frac{1}{2} \bigg(
    \partial_x \partial_y \eta^{xx} h_{xy}
    + \partial_y \partial_x \eta^{xx} h_{xy}
    - \partial_y^2 (
    \eta^{tt} h_{tt} + \eta^{zz} h_{zz})
    \bigg) \\
    & = \frac{1}{2} \bigg(
    - \partial_x \partial_y h_{xy}
    - \partial_y \partial_x h_{xy}
    - \partial_y^2 h_{tt} 
    + \partial_y^2 h_{zz}
    \bigg) \\
    & = \frac{1}{2} \bigg(
    \partial_x \partial_y h_{{BIC}_{\times}}
    + \partial_y \partial_x h_{{BIC}_{\times}}
    - \partial_y^2 h_{{BIC}_{\times}} 
    - \partial_y^2 h_{{BIC}_{\times}}
    \bigg) \\
    & = \frac{h_{{BIC}_{\times}}}{2} \bigg(
    - k_x k_y
    - k_y k_x
    + k_y^2
    + k_y^2
    \bigg) \\
    & = h_{{BIC}_{\times}} k_y \bigg(
    - k_x
    + k_y
    \bigg) \\
\end{split}        
\end{equation}

\begin{equation}
\begin{split}
    R_{{yz}_{\times}}
    = R_{{zy}_{\times}}
    & = \frac{1}{2} \bigg(
        \partial_\mu \partial_y h_{~ z} ^\mu 
		+ \partial_z \partial_\alpha h_{~ y}^\alpha
		- \Box h_{zy}
		- \partial_z \partial_y h
        \bigg) \\
    & = \frac{1}{2} \bigg(
        \partial_z \partial_y \eta^{zz} h_{zz}
        + \partial_z \partial_x \eta^{xx} h_{xy} 
        - \partial_z \partial_y (
        \eta^{tt} h_{tt} + \eta^{zz} h_{zz})
        \bigg) \\
    & = \frac{1}{2} \bigg(
        - \partial_z \partial_x h_{xy} 
        - \partial_z \partial_y h_{tt}
        \bigg) \\
    & = \frac{1}{2} \bigg(
        \partial_z \partial_x h_{{BIC}_{\times}}
        - \partial_z \partial_y h_{{BIC}_{\times}}
        \bigg) \\
    & = \frac{A k_z ~ sign(z)}{2} \bigg(
        k_x
        - k_y
        \bigg) 
        ~ (sin(k_x x) sin(k_y y) - cos(k_x x) cos(k_y y)) ~ cos( \omega ct/c) ~ 
        e^{- k_z |z|} \\
\end{split}        
\end{equation}

\begin{equation} \label{rzzx}
\begin{split}
    R_{{zz}_{\times}} 
    & = \frac{1}{2} \bigg(
        \partial_\mu \partial_z h_{~ z} ^\mu 
		+ \partial_z \partial_\alpha h_{~ z}^\alpha
		- \Box h_{zz}
		- \partial_z \partial_z h
        \bigg) \\
    & = \frac{1}{2} \bigg(
    \partial_z^2 \eta^{zz} h_{zz}
    + \partial_z^2 \eta^{zz} h_{zz}
    - \eta^{tt} \partial_t^2 h_{zz}
    - \eta^{xx} \partial_x^2 h_{zz}
    - \eta^{yy} \partial_y^2 h_{zz}
    - \eta^{zz} \partial_z^2 h_{zz}
    - \partial_z^2 (
    \eta^{tt} h_{tt} + \eta^{zz} h_{zz})
    \bigg) \\
    & = \frac{1}{2} \bigg(
    - \partial_z^2 h_{zz}
    - \partial_z^2 h_{zz}
    - \partial_t^2 h_{zz}
    + \partial_x^2 h_{zz}
    + \partial_y^2 h_{zz}
    + \partial_z^2 h_{zz}
    - \partial_z^2 h_{tt} 
    + \partial_z^2 h_{zz}
    \bigg) \\
    & = \frac{1}{2} \bigg(
    - \partial_t^2 h_{zz}
    + \partial_x^2 h_{zz}
    + \partial_y^2 h_{zz}
    - \partial_z^2 h_{tt}
    \bigg) \\
    & = \frac{1}{2} \bigg(
    \partial_t^2 h_{{BIC}_{\times}}
    - \partial_x^2 h_{{BIC}_{\times}}
    - \partial_y^2 h_{{BIC}_{\times}}
    - \partial_z^2 h_{{BIC}_{\times}} 
    \bigg) \\
    & = \frac{h_{{BIC}_{\times}}}{2} \bigg(
        - \omega^2/c^2
        + k_x^2
        + k_y^2
        - k_z( k_z ~ sign^2(z) - 2\delta (z)) 
        \bigg) 
    \\
    \therefore
    & = \frac{h_{{BIC}_{\times}}}{2} \bigg(
        - \omega^2/c^2
        + k_x^2 
        + k_y^2
        - k_z^2 
        \bigg) ~ for ~ z \neq 0 ~ ; ~ 
        \\
    & = \frac{(cos(k_x x)sin(k_y y) + sin(k_x x)cos(k_y y)) ~ cos(\omega ct/c)}{2} \bigg(
        - \omega^2/c^2
        + k_x^2 
        + k_y^2
        + 2 k_z \delta (0)
        \bigg) ~ for ~ z = 0
\end{split}        
\end{equation}

For this flat metric perturbation, only the $R_{{tz}_{\times}}$ and $R_{{zt}_{\times}}$ terms are zero. However, if we assume a square lattice for the perturbation ($k_x = k_y$), the following terms become null:

\begin{equation}
    R_{{tx}_{\times}} = R_{{xt}_{\times}}
    = R_{{ty}_{\times}} = R_{{yt}_{\times}}
    = R_{{xx}_{\times}}
    = R_{{xz}_{\times}} = R_{{zx}_{\times}}
    = R_{{yy}_{\times}}
    = R_{{yz}_{\times}} = R_{{zy}_{\times}}
    = 0
\end{equation}

The remaining non-zero terms are then:

\begin{equation}
    R_{{tt}_{\times}} 
        = \frac{h_{{BIC}_{\times}}}{2} \bigg(
        - k_x^2
        - k_y^2
        + k_z( k_z ~ sign^2(z) - 2\delta (z)) 
        + \omega^2/c^2
        \bigg) 
\end{equation}

\begin{equation}
\begin{split}
    R_{{xy}_{\times}}
    = R_{{yx}_{\times}}
    = & \frac{h_{{BIC}_{\times}}}{2} \bigg(
        - \omega^2/c^2
        - k_z( k_z ~ sign^2(z) - 2\delta (z)) 
        + k_y k_x 
        + k_y k_x
        \bigg) 
    \\
    = & \frac{h_{{BIC}_{\times}}}{2} \bigg(
        - \omega^2/c^2
        - k_z( k_z ~ sign^2(z) - 2\delta (z)) 
        + k_x^2
        + k_y^2
        \bigg) \\
    \end{split}
\end{equation}

\begin{equation}
     R_{{zz}_{\times}} 
    = \frac{h_{{BIC}_{\times}}}{2} \bigg(
        - \omega^2/c^2
        + k_x^2
        + k_y^2
        - k_z( k_z ~ sign^2(z) - 2\delta (z)) 
        \bigg) 
\end{equation}

implying that, for $z \neq 0$, all these terms are nullified if the following dispersion is obeyed by the ($\times$)-polarized BIC:

\begin{equation} \label{eq_dispersion_x_SM}
    \omega^2 = c^2(k_x^2 + k_y^2 - k_z^2)
\end{equation}

of similar form to that of Eq.~\ref{eq_dispersion_+}. At $z = 0$, the terms imply a distributional stress-energy tensor $^{(\Sigma)}T^{\mu \nu}_{_{\times}}$, similarly to the ($+$)-polarized BIC.

			
\subsection*{Ricci scalar}
Finally, we may calculate the Ricci scalar $R_{{}_{(+,\times)}}$ for the proposed bound perturbations.

\begin{equation} \label{lg_eq_RS_BIC}
\begin{split}
	R_{{}_{(+,\times)}} = 
	R_{{~ \nu}_{(+,\times)}}^\nu =
	\eta^{\nu \sigma} R_{{\sigma \nu}_{(+,\times)}} 
    = \eta^{tt} R_{tt_{(+,\times)}} + \eta^{xx} R_{xx_{(+,\times)}} + 
	\eta^{yy} R_{yy_{(+,\times)}} + \eta^{zz} R_{zz_{(+,\times)}}
\end{split}
\end{equation}

\subsubsection{$h_{xx}, h_{yy}$ BIC}
Using the results obtained in Eqs.~\ref{rtt+}, \ref{rxx+}, \ref{ryy+} and \ref{rzz+}, the Ricci scalar of the (+)-polarized BIC is:

\begin{equation} \label{ricci_scalar_+}
\begin{split}
    R_{{}_+} 
    & = R_{tt_+} - R_{xx_+} - R_{yy_+} - R_{zz_+} 
    \\
    \\
    & = \frac{h_{{BIC}_+}}{2} \bigg(
    - k_x^2
    + (\omega/c)^2 
    + k_z( k_z ~ sign^2(z) - 2\delta (z))
    \bigg) 
    - \frac{h_{{BIC}_+}}{2} \bigg(
    - (\omega/c)^2 - k_z( k_z ~ sign^2(z) - 2\delta (z))  
    + k_x^2
    \bigg) \\
    & + \frac{h_{{BIC}_+}}{2} \bigg(
    - (\omega/c)^2 + k_x^2
    - k_z( k_z ~ sign^2(z) - 2\delta (z))  
    \bigg)
    - \frac{h_{{BIC}_+}}{2} \bigg(
    - (\omega/c)^2 + k_x^2
    - k_z( k_z ~ sign^2(z) - 2\delta (z))  
    \bigg) 
    \\ 
    \\
    & = h_{{BIC}_+} \bigg(
    (\omega/c)^2 - k_x^2
    + k_z( k_z ~ sign^2(z) - 2\delta (z))  
    \bigg) 
    \\
    \therefore
    & = h_{{BIC}_+} \bigg(
    \omega^2/c^2 - k_x^2 + k_z^2 
    \bigg) ~ for ~ z \neq 0 ~ ; ~ 
    = cos(k_x x) ~ cos(\omega ct/c) \bigg(
    \omega^2/c^2 
    - k_x^2
    - 2 k_z \delta (0)
    \bigg) ~ for ~ z = 0
\end{split}
\end{equation}

which equals zero ($R_{{}_+} = 0$) for the same dispersion calculated above (Eq.~\ref{eq_dispersion_+}) outside the plane $z = 0$ while being singular at the localization plane. 

\subsubsection{$h_{xy}, h_{yx}$ BIC}

Using the results obtained in Eqs.~\ref{rttx}, \ref{rxxx}, \ref{ryyx} and \ref{rzzx}, the Ricci scalar of the ($\times$)-polarized BIC is:

\begin{equation} \label{ricci_scalar_x}
\begin{split}
    R_{{}_{\times}} 
    & = R_{tt_{\times}} - R_{xx_\times} - R_{yy_\times} - R_{zz_\times} 
    \\
    \\
       & = \frac{h_{{BIC}_{\times}}}{2} \bigg(
        - k_x^2
        - k_y^2
        + k_z( k_z ~ sign^2(z) - 2\delta (z)) 
        + \omega^2/c^2
        \bigg) 
    - h_{{BIC}_{\times}} k_x \bigg(
    - k_y
    + k_x
    \bigg) \\
    & ~ - h_{{BIC}_{\times}} k_y \bigg(
    - k_x
    + k_y
    \bigg)
    - \frac{h_{{BIC}_{\times}}}{2} \bigg(
        - \omega^2/c^2
        + k_x^2
        + k_y^2
        - k_z( k_z ~ sign^2(z) - 2\delta (z)) 
        \bigg) 
    \\ 
    \\
    & = h_{{BIC}_{\times}} \bigg(
        - k_x^2
        - k_y^2
        + k_z( k_z ~ sign^2(z) - 2\delta (z)) 
        + \omega^2/c^2
        \bigg) 
    - h_{{BIC}_{\times}} k_x \bigg(
    - k_y
    + k_x
    \bigg) \\
    & ~ - h_{{BIC}_{\times}} k_y \bigg(
    - k_x
    + k_y
    \bigg)
\end{split}
\end{equation}

As for the Ricci tensor components, two conditions need to be met for a null Ricci scalar outside the localization plane $z = 0$, in case, a square lattice perturbation ($k_x = k_y$) and the dispersion shown in Eq.~\ref{eq_dispersion_x_SM}. 
In this case, $R_{{}_\times} = 0$. At $z = 0$, $R_{\times}$ is singular.


\subsection{Non-trivial scalar invariants}
It is meaningful to calculate the quadratic non-trivial scalar invariants for the proposed solutions, such as the square of the manifold curvature, and to check if these curvature scalars are non-zero and remain localized in $z$, as required for a true physically bound perturbation. We may calculate the Kretschmann scalar $K$:
\begin{equation}
    K_{_{(+,\times)}} = R_{\rho\sigma\mu\nu}R^{\rho\sigma\mu\nu}
\end{equation}
requiring the fully covariant ($R_{\rho\sigma\mu\nu}$) and the fully contra-variant ($R^{\rho\sigma\mu\nu}$) formulation of the RCT (mixed forms shown in SM Eqs.~\ref{lg_eq_RCTc_BIC_hxxhyy}-\ref{lg_eq_RCTc_BIC}). These can be calculated from the mixed RCT as:
\begin{equation}
    R_{\rho\sigma\mu\nu} = \eta_{\rho\alpha}R^\alpha_{\sigma\mu\nu} \quad ; \quad
    R^{\rho\sigma\mu\nu} = \eta^{\nu\gamma}\eta^{\mu\beta}\eta^{\sigma\alpha}R^\rho_{\alpha\beta\gamma}
\end{equation}
in which $\alpha, \beta$ and $\gamma$ are dummy summation indices. Instead of being stated separately, the non-zero components of $R_{\rho\sigma\mu\nu}$ and of $R^{\rho\sigma\mu\nu}$ for each bound solution ($+, \times$) will be shown as part of the scalar invariant calculations below. 
\subsubsection{$h_{xx}, h_{yy}$ BIC}
The Kretschmann scalar of the ($+$)-polarized BIC is:
\begin{equation}
\begin{split}
    K_{_{+}} = ~ & R_{\rho\sigma\mu\nu}R^{\rho\sigma\mu\nu} \\
      = ~ & R_{txtx}R^{txtx} + R_{txtz}R^{txtz} + R_{txxt}R^{txxt} + R_{txxz}R^{txxz} + R_{txzt}R^{txzt} + R_{txzx}R^{txzx}
      \\
      & + R_{tyty}R^{tyty} + R_{tyxy}R^{tyxy} + R_{tyyt}R^{tyyt} + R_{tyyx}R^{tyyx} + R_{tyyz}R^{tyyz} + R_{tyzy}R^{tyzy} 
      \\ 
      & + R_{tztx}R^{tztx} + R_{tztz}R^{tztz} + R_{tzxt}R^{tzxt} + R_{tzxz}R^{tzxz} + R_{tzzt}R^{tzzt} + R_{tzzx}R^{tzzx}
      \\
      & + R_{xttx}R^{xttx} + R_{xttz}R^{xttz} + R_{xtxt}R^{xtxt} + R_{xtxz}R^{xtxz} + R_{xtzt}R^{xtzt} + R_{xtzx}R^{xtzx}
      \\
      & + R_{xyty}R^{xyty} + R_{xyxy}R^{xyxy} + R_{xyyt}R^{xyyt} + R_{xyyx}R^{xyyx} + R_{xyyz}R^{xyyz} + R_{xyzy}R^{xyzy}
      \\ 
      & + R_{xztx}R^{xztx} + R_{xztz}R^{xztz} + R_{xzxt}R^{xzxt} + R_{xzxz}R^{xzxz} + R_{xzzt}R^{xzzt} + R_{xzzx}R^{xzzx}
      \\
      & + R_{ytty}R^{ytty} + R_{ytxy}R^{ytxy} + R_{ytyt}R^{ytyt} + R_{ytyx}R^{ytyx} + R_{ytyz}R^{ytyz} + R_{ytzy}R^{ytzy}
      \\
      & + R_{yxty}R^{yxty} + R_{yxxy}R^{yxxy} + R_{yxyt}R^{yxyt} + R_{yxyx}R^{yxyx} + R_{yxyz}R^{yxyz} + R_{yxzy}R^{yxzy}
      \\ 
      & + R_{yzty}R^{yzty} + R_{yzxy}R^{yzxy} + R_{yzyt}R^{yzyt} + R_{yzyx}R^{yzyx} + R_{yzyz}R^{yzyz} + R_{yzzy}R^{yzzy}
      \\
      & + R_{zttx}R^{zttx} + R_{zttz}R^{zttz} + R_{ztxt}R^{ztxt} + R_{ztxz}R^{ztxz} + R_{ztzt}R^{ztzt} + R_{ztzx}R^{ztzx}
      \\
      & + R_{zxtx}R^{zxtx} + R_{zxtz}R^{zxtz} + R_{zxxt}R^{zxxt} + R_{zxxz}R^{zxxz} + R_{zxzt}R^{zxzt} + R_{zxzx}R^{zxzx}
      \\ 
      & + R_{zyty}R^{zyty} + R_{zyxy}R^{zyxy} + R_{zyyt}R^{zyyt} + R_{zyyx}R^{zyyx} + R_{zyyz}R^{zyyz} + R_{zyzy}R^{zyzy} 
\end{split}
\end{equation}
For compactness, terms will be shown squared when $R_{\rho\sigma\mu\nu} = R^{\rho\sigma\mu\nu}$. When different, their common terms will be shown only once (squared). Thus, in order:
\begin{equation}
\begin{split}
    & K_{_{+}} =
    \big(A(k_x^2 - \omega^2/c^2)/2 ~ cos(k_x x) ~ cos(\omega ct/c) ~ e^{- k_z |z|}\big)^2
    + 
    \big(- A k_x k_z ~ sign(z)/2 ~ sin(k_x x) ~ cos(\omega ct/c) ~ e^{- k_z |z|}\big)^2
    \\
    & + \big(A(- k_x^2 + \omega^2/c^2)/2 ~ cos(k_x x) ~ cos(\omega ct/c) ~ e^{- k_z |z|}\big)^2
    + 
    \big(A \omega ~ sign(z)/2c ~ cos(k_x x) ~ sin(\omega ct/c) ~ e^{- k_z |z|} \big)^2 (- k_z)\times(k_z)
    \\
    & + \big(A k_x k_z ~ sign(z)/2 ~ sin(k_x x) ~ cos(\omega ct/c) ~ e^{- k_z |z|}\big)^2
    +
    \big(A \omega ~ sign(z)/2c ~ cos(k_x x) ~ sin(\omega ct/c) ~ e^{- k_z |z|}\big)^2 (k_z)\times(- k_z)
    \\
    & + \big(A \omega^2/2c^2 ~ cos(k_x x) ~ cos(\omega ct/c) ~ e^{- k_z |z|}\big)^2
    +
    \big(A \omega/2c ~ sin(k_x x) ~ sin(\omega ct/c) ~ e^{- k_z |z|}\big)^2 (- k_x) \times (k_x)
    \\
    & + \big(- A \omega^2/2c^2 ~ cos(k_x x) ~ cos(\omega ct/c) ~ e^{- k_z |z|}\big)^2
    +
    \big(A \omega/2c ~ sin(k_x x) ~ sin(\omega ct/c) ~ e^{- k_z |z|}\big)^2 (k_x) \times (- k_x)
    \\
    & + \big(A \omega ~ sign(z)/2c ~ cos(k_x x) ~ sin(\omega ct/c) ~ e^{- k_z |z|}\big)^2 (k_z)\times(- k_z) 
    +
    \big(A \omega ~ sign(z)/2c ~ cos(k_x x) ~ sin(\omega ct/c) ~ e^{- k_z |z|}\big)^2 (- k_z)\times( k_z)
    \\
    & + \big(- A k_x k_z ~ sign(z)/2 ~ sin(k_x x) ~ cos(\omega ct/c) ~ e^{- k_z |z|}\big)^2
    \\ & +
    \big(A (- k_z(k_z sign^2(z) - 2\delta(z)) - \omega^2/c^2)/2 ~ cos(k_x x) ~ cos(\omega ct/c) ~ e^{- k_z |z|}\big)^2
    \\
    & + \big(A k_x k_z ~ sign(z)/2 ~ sin(k_x x) ~ cos(\omega ct/c) ~ e^{- k_z |z|}\big)^2
    +
    \big(A \omega/2c ~ sin(k_x x) ~ sin(\omega ct/c) ~ e^{- k_z |z|}\big)^2 (k_x) \times (- k_x)
    \\
    & + \big(A (k_z(k_z sign^2(z) - 2\delta(z)) + \omega^2/c^2)/2 ~ cos(k_x x) ~ cos(\omega ct/c) ~ e^{- k_z |z|}\big)^2
    +
    \big(A \omega/2c ~ sin(k_x x) ~ sin(\omega ct/c) ~ e^{- k_z |z|}\big)^2 (- k_x) \times (k_x)
    \\
    & + \big(A(- k_x^2 + \omega^2/c^2)/2 ~ cos(k_x x) ~ cos(\omega ct/c) ~ e^{- k_z |z|}\big)^2
    +
    \big(A k_x k_z ~ sign(z)/2 ~ sin(k_x x) ~ cos(\omega ct/c) ~ e^{- k_z |z|}\big)^2
    \\
    & + \big(A(k_x^2 - \omega^2/c^2)/2 ~ cos(k_x x) ~ cos(\omega ct/c) ~ e^{- k_z |z|}\big)^2
    +
    \big(A \omega ~ sign(z)/2c ~ cos(k_x x) ~ sin(\omega ct/c) ~ e^{- k_z |z|}\big)^2 ( k_z)\times(- k_z)
    \\
    & + \big(- A k_x k_z ~ sign(z)/2 ~ sin(k_x x) ~ cos(\omega ct/c) ~ e^{- k_z |z|}\big)^2
    +
    \big(A \omega ~ sign(z)/2c ~ cos(k_x x) ~ sin(\omega ct/c) ~ e^{- k_z |z|}\big)^2 (- k_z)\times(k_z) 
    \\
    & + \big(A \omega/2c ~ sin(k_x x) ~ sin(\omega ct/c) ~ e^{- k_z |z|}\big)^2 (- k_x) \times (k_x)
    +
    \big(A k_x^2/2 ~ cos(k_x x) ~ cos(\omega ct/c) ~ e^{- k_z |z|}\big)^2
    \\
    & + \big(A \omega/2c ~ sin(k_x x) ~ sin(\omega ct/c) ~ e^{- k_z |z|}\big)^2 (k_x) \times (- k_x)
    +
    \big(- A k_x^2/2 ~ cos(k_x x) ~ cos(\omega ct/c) ~ e^{- k_z |z|}\big)^2
    \\
    & + \big( A k_x k_z ~ sign(z)/2 ~ sin(k_x x) ~ cos(\omega ct/c) ~ e^{- k_z |z|}\big)^2
    +
    \big(- A k_x k_z ~ sign(z)/2 ~ sin(k_x x) ~ cos(\omega ct/c) ~ e^{- k_z |z|}\big)^2
    \\
    & + \big(A \omega ~ sign(z)/2c ~ cos(k_x x) ~ sin(\omega ct/c) ~ e^{- k_z |z|}\big)^2 (- k_z)\times( k_z) 
    +
    \big(A \omega/2c ~ sin(k_x x) ~ sin(\omega ct/c) ~ e^{- k_z |z|}\big)^2 (k_x) \times (- k_x)
    \\
    & + \big(A \omega ~ sign(z)/2c ~ cos(k_x x) ~ sin(\omega ct/c) ~ e^{- k_z |z|}\big)^2 (k_z)\times(- k_z)
    \\ & +
    \big(A(- k_x^2 + k_z(k_z sign^2(z) - 2\delta(z)))/2 ~ cos(k_x x) ~ cos(\omega ct/c) ~ e^{- k_z |z|}\big)^2
    \\
    & + \big(A \omega/2c ~ sin(k_x x) ~ sin(\omega ct/c) ~ e^{- k_z |z|}\big)^2 (- k_x) \times (k_x)
    +
    \big(A(k_x^2 - k_z(k_z sign^2(z) - 2\delta(z)))/2 ~ cos(k_x x) ~ cos(\omega ct/c) ~ e^{- k_z |z|}\big)^2
    \\
    & + (cont.)
\end{split}
\end{equation}
\newpage
\begin{align*}
\begin{split}
    & (...) 
    + \big(- A \omega^2/2c^2 ~ cos(k_x x) ~ cos(\omega ct/c) ~ e^{- k_z |z|}\big)^2 
    +
    \big(A \omega/2c ~ sin(k_x x) ~ sin(\omega ct/c) ~ e^{- k_z |z|}\big)^2 (k_x) \times (- k_x)
    \\
    & + \big(A \omega^2/2c^2 ~ cos(k_x x) ~ cos(\omega ct/c) ~ e^{- k_z |z|}\big)^2
    +
    \big(A \omega/2c ~ sin(k_x x) ~ sin(\omega ct/c) ~ e^{- k_z |z|}\big)^2 (- k_x) \times (k_x)
    \\
    & + \big(A\omega ~ sign(z)/2c ~ cos(k_x x) ~ sin(\omega ct/c) ~ e^{- k_z |z|}\big)^2 (- k_z)\times(k_z) 
    +
    \big(A\omega ~ sign(z)/2c ~ cos(k_x x) ~ sin(\omega ct/c) ~ e^{- k_z |z|}\big)^2 ( k_z)\times(- k_z) 
    \\
    & + \big(A \omega/2c ~ sin(k_x x) ~ sin(\omega ct/c) ~ e^{- k_z |z|}\big)^2 (k_x) \times (- k_x)
    +
    \big(- A k_x^2/2 ~ cos(k_x x) ~ cos(\omega ct/c) ~ e^{- k_z |z|}\big)^2
    \\
    & + \big(A \omega/2c ~ sin(k_x x) ~ sin(\omega ct/c) ~ e^{- k_z |z|}\big)^2 (- k_x) \times (k_x)
    +
    \big(A k_x^2/2 ~ cos(k_x x) ~ cos(\omega ct/c) ~ e^{- k_z |z|}\big)^2
    \\
    & + \big(- A k_x k_z ~ sign(z)/2 ~ sin(k_x x) ~ cos(\omega ct/c) ~ e^{- k_z |z|}\big)^2
    +
    \big(A k_x k_z ~ sign(z)/2 ~ sin(k_x x) ~ cos(\omega ct/c) ~ e^{- k_z |z|}\big)^2
    \\
    & + \big(A \omega ~ sign(z)/2c ~ cos(k_x x) ~ sin(\omega ct/c) ~ e^{- k_z |z|}\big)^2 (k_z)\times(- k_z)
    +
    \big(A k_x k_z ~ sign(z)/2 ~ sin(k_x x) ~ cos(\omega ct/c) ~ e^{- k_z |z|}\big)^2
    \\
    & + \big(A\omega ~ sign(z)/2c ~ cos(k_x x) ~ sin(\omega ct/c) ~ e^{- k_z |z|} \big)^2 (- k_z)\times(k_z)
    +
    \big(- A k_x k_z ~ sign(z)/2 ~ sin(k_x x) ~ cos(\omega ct/c) ~ e^{- k_z |z|}\big)^2
    \\
    & + \big(- A k_z(k_z sign^2(z) - 2\delta(z))/2 ~ cos(k_x x) ~ cos(\omega ct/c) ~ e^{- k_z |z|}\big)^2
    +
    \big(A k_z(k_z sign^2(z) - 2\delta(z))/2 ~ cos(k_x x) ~ cos(\omega ct/c) ~ e^{- k_z |z|}\big)^2
    \\
    & + \big(A k_x k_z ~ sign(z)/2 ~ sin(k_x x) ~ cos(\omega ct/c) ~ e^{- k_z |z|}\big)^2
    + 
    \big(A (k_z(k_z sign^2(z) - 2\delta(z)) + \omega^2/c^2)/2 ~ cos(k_x x) ~ cos(\omega ct/c) ~ e^{- k_z |z|}\big)^2
    \\
    & + \big(- A k_x k_z ~ sign(z)/2 ~ sin(k_x x) ~ cos(\omega ct/c) ~ e^{- k_z |z|}\big)^2
    +
    \big(A \omega/2c ~ sin(k_x x) ~ sin(\omega ct/c) ~ e^{- k_z |z|}\big)^2 (- k_x) \times (k_x)
    \\
    & + \big(A (- k_z(k_z sign^2(z) - 2\delta(z)) - \omega^2/c^2)/2 ~ cos(k_x x) ~ cos(\omega ct/c) ~ e^{- k_z |z|}\big)^2
    \\ & + 
    \big(A \omega/2c ~ sin(k_x x) ~ sin(\omega ct/c) ~ e^{- k_z |z|}\big)^2 (k_x) \times (- k_x)
    \\
    & + \big(A \omega ~ sign(z)/2c ~ cos(k_x x) ~ sin(\omega ct/c) ~ e^{- k_z |z|}\big)^2 (k_z)\times(- k_z)
    +
    \big(A \omega/2c ~ sin(k_x x) ~ sin(\omega ct/c) ~ e^{- k_z |z|}\big)^2 (- k_x) \times (k_x)
    \\
    & + \big(A\omega ~ sign(z)/2c ~ cos(k_x x) ~ sin(\omega ct/c) ~ e^{- k_z |z|}\big)^2 (- k_z)\times( k_z) 
    \\ & +
    \big(A(k_x^2 - k_z(k_z sign^2(z) - 2\delta(z)))/2 ~ cos(k_x x) ~ cos(\omega ct/c) ~ e^{- k_z |z|}\big)^2
    \\
    & + \big(A \omega/2c ~ sin(k_x x) ~ sin(\omega ct/c) ~ e^{- k_z |z|}\big)^2 (k_x) \times (- k_x)
    \\ & +
    \big(A(- k_x^2 + k_z(k_z sign^2(z) - 2\delta(z)))/2 ~ cos(k_x x) ~ cos(\omega ct/c) ~ e^{- k_z |z|}\big)^2
    \\
    & + \big(A\omega ~ sign(z)/2c ~ cos(k_x x) ~ sin(\omega ct/c) ~ e^{- k_z |z|}\big)^2 (- k_z)\times( k_z) 
    +
    \big(- A k_x k_z ~ sign(z)/2 ~ sin(k_x x) ~ cos(\omega ct/c) ~ e^{- k_z |z|}\big)^2
    \\
    & + \big(A \omega ~ sign(z)/2c ~ cos(k_x x) ~ sin(\omega ct/c) ~ e^{- k_z |z|}\big)^2 (k_z)\times(- k_z)
    +
    \big(A k_x k_z ~ sign(z)/2 ~ sin(k_x x) ~ cos(\omega ct/c) ~ e^{- k_z |z|}\big)^2
    \\
    & + \big(A k_z(k_z sign^2(z) - 2\delta(z))/2 ~ cos(k_x x) ~ cos(\omega ct/c) ~ e^{- k_z |z|}\big)^2
    \\ & +
    \big(- A k_z(k_z sign^2(z) - 2\delta(z))/2 ~ cos(k_x x) ~ cos(\omega ct/c) ~ e^{- k_z |z|}\big)^2
\end{split}
\end{align*}
After the multiplications, we may combine the identical terms:
\begin{equation}
\begin{split}
     K_{_{+}} = ~ & A^2\big[3k_x^4 - 2k_x^2 \omega^2/c^2 + 3\omega^4/c^4 + 3k_z^4 ~ sign^4(z) + 2 k_z^2 \omega^2 ~ sign^2(z)/c^2 - 2k_x^2 k_z^2 ~ sign^2(z) 
    \\ & + (- 12 k_z^2 ~ sign^2(z) - 4 k_z \omega^2/c^2  + 4 k_x^2 k_z) \delta(z) 
    + 12 k_z^2 \delta^2(z) \big] 
    cos^2(k_x x) ~ cos^2(\omega ct/c) ~ e^{- 2 k_z |z|}
    \\
    & 
    + 4 A^2 k_x^2 k_z^2 ~ sign^2(z) ~
    sin^2(k_x x) ~ cos^2(\omega ct/c) ~ e^{- 2 k_z |z|} 
    - 4 A^2 k_z^2 \omega^2 ~ sign^2(z)/c^2
    cos^2(k_x x) ~ sin^2(\omega ct/c) ~ e^{- 2 k_z |z|} 
    \\
    & - 4 A^2 k_x^2 \omega^2/c^2
    sin^2(k_x x) ~ sin^2(\omega ct/c) ~ e^{- 2 k_z |z|} \\
\end{split}
\end{equation}
$K_{_+}$ contains Dirac delta terms and also the square of distributions ($\delta^2(z)$), making it ill-defined at the hypersurface $\Sigma ~ (z = 0)$. However,
for the subsequent analysis restricted to where the EFEs yield a vacuum solution, i.e. outside the localization plane, we may split $K_{_+}$ into a well-behaved, smooth ($C^\infty$), part ($K_{S_+}$) outside $\Sigma$ and one with the ill-defined terms ($K_{\Sigma_+}$) containing the distributions and their product:
\begin{equation}
    K_{_+} = K_{S_+} + K_{\Sigma_+}
\end{equation}
where
\begin{equation}
\begin{split}
    K_{\Sigma_+} = & A^2\big[(- 12 k_z^2 ~ sign^2(z) - 4 k_z \omega^2/c^2  + 4 k_x^2 k_z) \delta(z) 
    + 12 k_z^2 \delta^2(z)\big]
    ~ cos(k_x x)^2 ~ cos^2(\omega ct/c) ~ e^{- 2 k_z |z|}
\end{split}
\end{equation}
which becomes zero outside $\Sigma$. Also, noting that for $z \neq 0$, $sign^2(z) = sign^4(z) = 1$, it results for the well-behaved scalar invariant:
\begin{equation}
\begin{split}
     K_{S_{+}} = ~ & A^2\big( 3k_x^4 - 2k_x^2 \omega^2/c^2 + 3\omega^4/c^4 + 3k_z^4 + 2 k_z^2 \omega^2/c^2 - 2k_x^2 k_z^2 \big) 
    ~ cos^2(k_x x) ~ cos^2(\omega ct/c) ~ e^{- 2 k_z |z|}
    \\
    & 
    + 4 A^2 k_x^2 k_z^2 ~
    sin^2(k_x x) ~ cos^2(\omega ct/c) ~ e^{- 2 k_z |z|} 
    - 4 A^2 k_z^2 \omega^2/c^2
    cos^2(k_x x) ~ sin^2(\omega ct/c) ~ e^{- 2 k_z |z|} 
    \\
    & - 4 A^2 k_x^2 \omega^2/c^2
    sin^2(k_x x) ~ sin^2(\omega ct/c) ~ e^{- 2 k_z |z|} \\
\end{split}
\end{equation}
To generate a perfect square of the dispersion (Eq.~\ref{eq_dispersion_+}, thus canceling it), we may add and subtract $4 k_x^2 k_z^2$, $4 k_z^2 \omega^2/c^2$ and $4 k_x^2 \omega^2/c^2$ to and from the first term:
\begin{equation}
\begin{split}
     K_{S_{+}} = ~ & A^2\big(3k_x^4 - 6k_x^2 \omega^2/c^2 + 3\omega^4/c^4 + 3k_z^4 + 6 k_z^2 \omega^2/c^2 - 6k_x^2 k_z^2 \big) 
    cos^2(k_x x) ~ cos^2(\omega ct/c) ~ e^{- 2 k_z |z|}
    \\
    &
    + A^2\big(4k_x^2 \omega^2/c^2 - 4 k_z^2 \omega^2/c^2 + 4k_x^2 k_z^2 \big) 
    cos^2(k_x x) ~ cos^2(\omega ct/c) ~ e^{- 2 k_z |z|}
    \\
    & 
    + 4 A^2 k_x^2 k_z^2
    sin^2(k_x x) ~ cos^2(\omega ct/c) ~ e^{- 2 k_z |z|} 
    - 4 A^2 k_z^2 \omega^2/c^2
    cos^2(k_x x) ~ sin^2(\omega ct/c) ~ e^{- 2 k_z |z|} 
    \\
    & - 4 A^2 k_x^2 \omega^2/c^2
    sin^2(k_x x) ~ sin^2(\omega ct/c) ~ e^{- 2 k_z |z|} \\
\end{split}
\end{equation}
While now there is a perfect square of the dispersion, the extra terms can be combined by their common factors with the last three terms:
\begin{equation}
\begin{split}
     K_{S_{+}} = ~ & 3 A^2 \big(k_x^2 - k_z^2 - \omega^2/c^2 \big) 
    cos^2(k_x x) ~ cos^2(\omega ct/c) ~ e^{- 2 k_z |z|}
    \\
    &
    + 4 A^2 k_x^2 \omega^2/c^2 ~ e^{- 2 k_z |z|} 
    \big(cos^2(k_x x) ~ cos^2(\omega ct/c) -  sin^2(k_x x) ~ sin^2(\omega ct/c) \big) 
    \\
    &
    - 4 A^2 k_z^2 \omega^2/c^2 ~ e^{- 2 k_z |z|} ~ cos^2(k_x x)
    \big(cos^2(\omega ct/c) + sin^2(\omega ct/c)\big)
    \\
    & 
    + 4 A^2 k_x^2 k_z^2 ~ e^{- 2 k_z |z|} ~ cos^2(\omega ct/c)
    \big(cos^2(k_x x) + sin^2(k_x x)\big)
\end{split}
\end{equation}
Thus,
\begin{equation}
\begin{split}
     K_{S_{+}} = ~ & 4 A^2 k_x^2 \omega^2/c^2 ~ e^{- 2 k_z |z|} 
    \big(cos^2(k_x x) ~ cos^2(\omega ct/c) -  sin^2(k_x x) ~ sin^2(\omega ct/c) \big) 
    \\
    &
    - 4 A^2 k_z^2 \omega^2/c^2 ~ e^{- 2 k_z |z|} ~ cos^2(k_x x)
    \\
    & 
    + 4 A^2 k_x^2 k_z^2 ~ e^{- 2 k_z |z|} ~ cos^2(\omega ct/c)
\end{split}
\end{equation}
Finally, by adding and subtracting $4 A^2 k_x^2 \omega^2/c^2 e^{- 2 k_z |z|} cos^2(\omega ct/c)$ and $4 A^2 k_x^2 \omega^2/c^2 e^{- 2 k_z |z|} cos^2(k_x x)$, it may be further simplified. Rearranging:
\begin{equation}
\begin{split}
     K_{S_{+}} = ~ & 4 A^2 k_x^2 \omega^2/c^2 ~ e^{- 2 k_z |z|} 
     \big(- cos^2 (k_x x) -  sin^2(k_x x) ~ sin^2(\omega ct/c) \big) 
    \\
    &
    - 4 A^2 \omega^2/c^2 ~ e^{- 2 k_z |z|} ~ \big(k_z^2 cos^2(k_x x) - k_x^2 cos^2(k_x x)\big)
    \\
    & 
    + 4 A^2 k_x^2 e^{- 2 k_z |z|}  ~ cos^2(\omega ct/c) ~ (k_z^2 + \omega^2/c^2)
    \\
    & 
    + 4 A^2 k_x^2 e^{- 2 k_z |z|}  ~ cos^2(\omega ct/c) ~ \big(-\omega^2/c^2 + \omega^2/c^2 ~ cos^2(k_xx)\big)
\end{split}
\end{equation}
As $-\omega^2/c^2 + \omega^2/c^2 cos^2(k_xx) = -\omega^2/c^2 sin^2(k_xx)$ and by invoking the dispersion (Eq.~\ref{eq_dispersion_+}):
\begin{equation}
\begin{split}
     K_{S_{+}} = ~ & 4 A^2 k_x^2 \omega^2/c^2 ~ e^{- 2 k_z |z|} 
     \big(- cos^2 (k_x x) -  sin^2(k_x x) ~ sin^2(\omega ct/c) \big) 
    \\
    &
    - 4 A^2 \omega^2/c^2 ~ e^{- 2 k_z |z|} ~ \big(-\omega^2/c^2 ~ cos^2(k_x x)\big)
    \\
    & 
    + 4 A^2 k_x^2 e^{- 2 k_z |z|}  ~ cos^2(\omega ct/c) ~ (k_x^2)
    \\
    & 
    + 4 A^2 k_x^2 e^{- 2 k_z |z|}  ~ cos^2(\omega ct/c) ~ \big(-\omega^2/c^2 ~ sin^2(k_xx)\big)
\end{split}
\end{equation}
Yielding:
\begin{equation}
\begin{split}
     K_{S_{+}} = ~ & 4 A^2 k_x^2 \omega^2/c^2 ~ e^{- 2 k_z |z|} 
     \big[
     - cos^2 (k_x x) - sin^2(k_x x) \big( sin^2(\omega ct/c) + cos^2(\omega ct/c) \big)
     \big]
    \\
    &
    + 4 A^2 \omega^4/c^4 ~ e^{- 2 k_z |z|} ~ cos^2(k_x x)
    + 4 A^2 k_x^4 e^{- 2 k_z |z|}  ~ cos^2(\omega ct/c)
\end{split}
\end{equation}
At last,
\begin{equation} \label{kretschmann_+}
\begin{split}
     K_{S_{+}} = ~ & 4 A^2 
     \big(k_x^4 cos^2(\omega ct/c)
     + \omega^4/c^4 ~ cos^2(k_x x)
     - k_x^2 \omega^2/c^2 \big)
     e^{- 2 k_z |z|}
\end{split}
\end{equation}
resulting in a scalar $K_{S_{+}}$ that decays exponentially from $z = 0$ and that is without any divergences/singularities for all spacetime coordinates outside the $\Sigma$ hypersurface.
\newpage
\subsubsection{$h_{xy}, h_{yx}$ BIC}
The Kretschmann scalar of the ($\times$)-polarized BIC is:
\begin{equation}
\begin{split}
    K_{_{\times}} = ~ & R_{\rho\sigma\mu\nu}R^{\rho\sigma\mu\nu} \\ 
      = ~ & R_{txtx}R^{txtx} + R_{txty}R^{txty} + R_{txtz}R^{txtz} + R_{txxt}R^{txxt} + R_{txxy}R^{txxy} \\ & + R_{txyt}R^{txyt} + R_{txyx}R^{txyx} + R_{txyz}R^{txyz} + R_{txzt}R^{txzt} + R_{txzy}R^{txzy}
      \\ 
      + & R_{tytx}R^{tytx} + R_{tyty}R^{tyty} + R_{tytz}R^{tytz} + R_{tyxt}R^{tyxt} + R_{tyxy}R^{tyxy} \\ & + R_{tyxz}R^{tyxz} + R_{tyyt}R^{tyyt} + R_{tyyx}R^{tyyx} + R_{tyzt}R^{tyzt} + R_{tyzx}R^{tyzx} 
      \\ 
      + & R_{tztx}R^{tztx} + R_{tzty}R^{tzty} + R_{tztz}R^{tztz} + R_{tzxt}R^{tzxt} + R_{tzxz}R^{tzxz} \\ & + R_{tzyt}R^{tzyt} + R_{tzyz}R^{tzyz} + R_{tzzt}R^{tzzt} + R_{tzzx}R^{tzzx} + R_{tzzy}R^{tzzy}
      \\ 
      + & R_{xttx}R^{xttx} + R_{xtty}R^{xtty} + R_{xttz}R^{xttz} + R_{xtxt}R^{xtxt} + R_{xtxy}R^{xtxy} \\ & + R_{xtyt}R^{xtyt} + R_{xtyx}R^{xtyx} + R_{xtyz}R^{xtyz} + R_{xtzt}R^{xtzt} + R_{xtzy}R^{xtzy}
      \\ 
      + & R_{xytx}R^{xytx} + R_{xyty}R^{xyty} + R_{xyxt}R^{xyxt} + R_{xyxy}R^{xyxy} + R_{xyxz}R^{xyxz} \\ & + R_{xyyt}R^{xyyt} + R_{xyyx}R^{xyyx} + R_{xyyz}R^{xyyz} + R_{xyzx}R^{xyzx} + R_{xyzy}R^{xyzy}
      \\ 
      + & R_{xzty}R^{xzty} + R_{xztz}R^{xztz} + R_{xzxy}R^{xzxy} + R_{xzxz}R^{xzxz} + R_{xzyt}R^{xzyt} \\ & + R_{xzyx}R^{xzyx} + R_{xzyz}R^{xzyz} + R_{xzzt}R^{xzzt} + R_{xzzx}R^{xzzx} + R_{xzzy}R^{xzzy}
      \\ 
      + & R_{yttx}R^{yttx} + R_{ytty}R^{ytty} + R_{yttz}R^{yttz} + R_{ytxt}R^{ytxt} + R_{ytxy}R^{ytxy} \\ & + R_{ytxz}R^{ytxz} + R_{ytyt}R^{ytyt} + R_{ytyx}R^{ytyx} + R_{ytzt}R^{ytzt} + R_{ytzx}R^{ytzx}
      \\ 
      + & R_{yxtx}R^{yxtx} + R_{yxty}R^{yxty} + R_{yxxt}R^{yxxt} + R_{yxxy}R^{yxxy} + R_{yxxz}R^{yxxz} \\ & + R_{yxyt}R^{yxyt} + R_{yxyx}R^{yxyx} + R_{yxyz}R^{yxyz} + R_{yxzx}R^{yxzx} + R_{yxzy}R^{yxzy}
      \\ 
      + & R_{yztx}R^{yztx} + R_{yztz}R^{yztz} + R_{yzxt}R^{yzxt} + R_{yzxy}R^{yzxy} + R_{yzxz}R^{yzxz} \\ & + R_{yzyx}R^{yzyx} + R_{yzyz}R^{yzyz} + R_{yzzt}R^{yzzt} + R_{yzzx}R^{yzzx} + R_{yzzy}R^{yzzy}
      \\ 
      + & R_{zttx}R^{zttx} + R_{ztty}R^{ztty} + R_{zttz}R^{zttz} + R_{ztxt}R^{ztxt} + R_{ztxz}R^{ztxz} \\ & + R_{ztyt}R^{ztyt} + R_{ztyz}R^{ztyz} + R_{ztzt}R^{ztzt} + R_{ztzx}R^{ztzx} + R_{ztzy}R^{ztzy}
      \\ 
      + & R_{zxty}R^{zxty} + R_{zxtz}R^{zxtz} + R_{zxxy}R^{zxxy} + R_{zxxz}R^{zxxz} + R_{zxyt}R^{zxyt} \\ & + R_{zxyx}R^{zxyx} + R_{zxyz}R^{zxyz} + R_{zxzt}R^{zxzt} + R_{zxzx}R^{zxzx} + R_{zxzy}R^{zxzy}
      \\ 
      + & R_{zytx}R^{zytx} + R_{zytz}R^{zytz} + R_{zyxt}R^{zyxt} + R_{zyxy}R^{zyxy} + R_{zyxz}R^{zyxz} \\ & + R_{zyyx}R^{zyyx} + R_{zyyz}R^{zyyz} + R_{zyzt}R^{zyzt} + R_{zyzx}R^{zyzx} + R_{zyzy}R^{zyzy}  
\end{split}
\end{equation}
\begin{equation}
\begin{split}
    K_{_{\times}} =
    \big(A k_x^2/2 ~ \big(cos(k_x x)sin(k_yy) + sin(k_xx)cos(k_yy)\big) ~ cos(\omega ct/c) ~ e^{- k_z |z|}\big)^2
    \\ + 
    \big(A (k_x k_y - \omega^2/c^2)/2 ~ \big(cos(k_x x)sin(k_yy) + sin(k_xx)cos(k_yy)\big) ~ cos(\omega ct/c) ~ e^{- k_z |z|}\big)^2
    \\ + 
    \big(-A k_x k_z ~ sign(z)/2 ~ \big(sin(k_x x)sin(k_yy) - cos(k_xx)cos(k_yy)\big) ~ cos(\omega ct/c) ~ e^{- k_z |z|}\big)^2
    \\ + 
    \big(- A k_x^2/2 ~ \big(cos(k_x x)sin(k_yy) + sin(k_xx)cos(k_yy)\big) ~ cos(\omega ct/c) ~ e^{- k_z |z|}\big)^2
    \\ +
    \big(A \omega/2c ~ \big(sin(k_x x)sin(k_yy) - cos(k_xx)cos(k_yy)\big) ~ sin(\omega ct/c) ~ e^{- k_z |z|}\big)^2 (k_x) \times (- k_x)
    \\ +
    \big(A (- k_x k_y + \omega^2/c^2)/2 ~ \big(cos(k_x x)sin(k_yy) + sin(k_xx)cos(k_yy)\big) ~ cos(\omega ct/c) ~ e^{- k_z |z|}\big)^2
    \\ +
    \big(A \omega/2c ~ \big(sin(k_x x)sin(k_yy) - cos(k_xx)cos(k_yy)\big) ~ sin(\omega ct/c) ~ e^{- k_z |z|}\big)^2 (- k_x) \times (k_x)
    \\ +
    \big(A \omega ~ sign(z)/2c ~ \big(cos(k_x x)sin(k_yy) + sin(k_xx)cos(k_yy)\big) ~ sin(\omega ct/c) ~ e^{- k_z |z|}\big)^2 (- k_z) \times (k_z)
    \\ + 
    \big(A k_x k_z ~ sign(z)/2 ~ \big(sin(k_x x)sin(k_yy) - cos(k_xx)cos(k_yy)\big) ~ cos(\omega ct/c) ~ e^{- k_z |z|}\big)^2
    \\ +
    \big(A \omega ~ sign(z)/2c ~ \big(cos(k_x x)sin(k_yy) + sin(k_xx)cos(k_yy)\big) ~ sin(\omega ct/c) ~ e^{- k_z |z|}\big)^2 (k_z) \times (- k_z)
    \\
    + 
    \big(A (k_x k_y - \omega^2/c^2)/2 ~ \big(cos(k_x x)sin(k_yy) + sin(k_xx)cos(k_yy)\big) ~ cos(\omega ct/c) ~ e^{- k_z |z|}\big)^2
    \\ + ~ (cont.)
\end{split}
\end{equation}
\begin{align*}
\begin{split}
    (...) +
    \big(A k_y^2/2 ~ \big(cos(k_x x)sin(k_yy) + sin(k_xx)cos(k_yy)\big) ~ cos(\omega ct/c) ~ e^{- k_z |z|}\big)^2
    \\ + 
    \big(- A k_y k_z ~ sign(z)/2 ~ \big(sin(k_x x)sin(k_yy) - cos(k_xx)cos(k_yy)\big) ~ cos(\omega ct/c) ~ e^{- k_z |z|}\big)^2
    \\ + 
    \big(A (- k_x k_y + \omega^2/c^2)/2 ~ \big(cos(k_x x)sin(k_yy) + sin(k_xx)cos(k_yy)\big) ~ cos(\omega ct/c) ~ e^{- k_z |z|}\big)^2
    \\ +
    \big(A \omega/2c ~ \big(sin(k_x x)sin(k_yy) - cos(k_xx)cos(k_yy)\big) ~ sin(\omega ct/c) ~ e^{- k_z |z|}\big)^2 (- k_y) \times (k_y)
    \\ +
    \big(A \omega ~ sign(z)/2c ~ \big(cos(k_x x)sin(k_yy) + sin(k_xx)cos(k_yy)\big) ~ sin(\omega ct/c) ~ e^{- k_z |z|}\big)^2 (- k_z) \times (k_z)
    \\ +
    \big(- A k_y^2/2 ~ \big(cos(k_x x)sin(k_yy) + sin(k_xx)cos(k_yy)\big) ~ cos(\omega ct/c) ~ e^{- k_z |z|}\big)^2
    \\ +
    \big(A \omega/2c ~ \big(sin(k_x x)sin(k_yy) - cos(k_xx)cos(k_yy)\big) ~ sin(\omega ct/c) ~ e^{- k_z |z|}\big)^2 (k_y) \times (- k_y)
    \\ + 
    \big(A k_y k_z ~ sign(z)/2 ~ \big(sin(k_x x)sin(k_yy) - cos(k_xx)cos(k_yy)\big) ~ cos(\omega ct/c) ~ e^{- k_z |z|}\big)^2
    \\ +
    \big(A \omega ~ sign(z)/2c ~ \big(cos(k_x x)sin(k_yy) + sin(k_xx)cos(k_yy)\big) ~ sin(\omega ct/c) ~ e^{- k_z |z|}\big)^2 (k_z) \times (-k_z)
    \\
    + 
    \big(- A k_x k_z ~ sign(z)/2 ~ \big(sin(k_x x)sin(k_yy) - cos(k_xx)cos(k_yy)\big) ~ cos(\omega ct/c) ~ e^{- k_z |z|}\big)^2
    \\ + 
    \big(- A k_y k_z ~ sign(z)/2 ~ \big(sin(k_x x)sin(k_yy) - cos(k_xx)cos(k_yy)\big) ~ cos(\omega ct/c) ~ e^{- k_z |z|}\big)^2
    \\ + 
    \big(A (- k_z(k_z ~sign^2(z) - 2\delta(z)) - \omega^2/c^2)/2 ~ \big(cos(k_x x)sin(k_yy) + sin(k_xx)cos(k_yy)\big) ~ cos(\omega ct/c) ~ e^{- k_z |z|}\big)^2
    \\ + 
    \big(A k_x k_z ~ sign(z)/2 ~ \big(sin(k_x x)sin(k_yy) - cos(k_xx)cos(k_yy)\big) ~ cos(\omega ct/c) ~ e^{- k_z |z|}\big)^2
    \\ +
    \big(A \omega/2c ~ \big(sin(k_x x)sin(k_yy) - cos(k_xx)cos(k_yy)\big) ~ sin(\omega ct/c) ~ e^{- k_z |z|}\big)^2 (k_x) \times (- k_x)
    \\ + 
    \big(A k_y k_z ~ sign(z)/2 ~ \big(sin(k_x x)sin(k_yy) - cos(k_xx)cos(k_yy)\big) ~ cos(\omega ct/c) ~ e^{- k_z |z|}\big)^2
    \\ +
    \big(A \omega/2c ~ \big(sin(k_x x)sin(k_yy) - cos(k_xx)cos(k_yy)\big) ~ sin(\omega ct/c) ~ e^{- k_z |z|}\big)^2 (k_y) \times (- k_y)
    \\ + 
    \big(A (k_z(k_z ~sign^2(z) - 2\delta(z)) + \omega^2/c^2)/2 ~ \big(cos(k_x x)sin(k_yy) + sin(k_xx)cos(k_yy)\big) ~ cos(\omega ct/c) ~ e^{- k_z |z|}\big)^2
    \\ +
    \big(A \omega/2c ~ \big(sin(k_x x)sin(k_yy) - cos(k_xx)cos(k_yy)\big) ~ sin(\omega ct/c) ~ e^{- k_z |z|}\big)^2 (- k_x) \times (k_x)
    \\ +
    \big(A \omega/2c ~ \big(sin(k_x x)sin(k_yy) - cos(k_xx)cos(k_yy)\big) ~ sin(\omega ct/c) ~ e^{- k_z |z|}\big)^2 (- k_y) \times (k_y)
    \\
    +
    \big(- A k_x^2/2 ~ \big(cos(k_x x)sin(k_yy) + sin(k_xx)cos(k_yy)\big) ~ cos(\omega ct/c) ~ e^{- k_z |z|}\big)^2
    \\ + 
    \big(A (- k_x k_y + \omega^2/c^2)/2 ~ \big(cos(k_x x)sin(k_yy) + sin(k_xx)cos(k_yy)\big) ~ cos(\omega ct/c) ~ e^{- k_z |z|}\big)^2
    \\ + 
    \big(A k_x k_z ~ sign(z)/2 ~ \big(sin(k_x x)sin(k_yy) - cos(k_xx)cos(k_yy)\big) ~ cos(\omega ct/c) ~ e^{- k_z |z|}\big)^2
    \\ +
    \big(A k_x^2/2 ~ \big(cos(k_x x)sin(k_yy) + sin(k_xx)cos(k_yy)\big) ~ cos(\omega ct/c) ~ e^{- k_z |z|}\big)^2
    \\ +
    \big(A \omega/2c ~ \big(sin(k_x x)sin(k_yy) - cos(k_xx)cos(k_yy)\big) ~ sin(\omega ct/c) ~ e^{- k_z |z|}\big)^2 (- k_x) \times (k_x)
    \\ +
    \big(A (k_x k_y - \omega^2/c^2)/2 ~ \big(cos(k_x x)sin(k_yy) + sin(k_xx)cos(k_yy)\big) ~ cos(\omega ct/c) ~ e^{- k_z |z|}\big)^2
    \\ +
    \big(A \omega/2c ~ \big(sin(k_x x)sin(k_yy) - cos(k_xx)cos(k_yy)\big) ~ sin(\omega ct/c) ~ e^{- k_z |z|}\big)^2 (k_x) \times (- k_x)
    \\ +
    \big(A \omega ~sign(z)/2c ~ \big(cos(k_x x)sin(k_yy) + sin(k_xx)cos(k_yy)\big) ~ sin(\omega ct/c) ~ e^{- k_z |z|}\big)^2 (k_z) \times (- k_z)
    \\ + 
    \big(- A k_x k_z ~ sign(z)/2 ~ \big(sin(k_x x)sin(k_yy) - cos(k_xx)cos(k_yy)\big) ~ cos(\omega ct/c) ~ e^{- k_z |z|}\big)^2
     \\ +
    \big(A \omega ~ sign(z)/2c ~ \big(cos(k_x x)sin(k_yy) + sin(k_xx)cos(k_yy)\big) ~ sin(\omega ct/c) ~ e^{- k_z |z|}\big)^2 (- k_z) \times (k_z)
    \\
    +
    \big(A \omega/2c ~ \big(sin(k_x x)sin(k_yy) - cos(k_xx)cos(k_yy)\big) ~ sin(\omega ct/c) ~ e^{- k_z |z|}\big)^2 (k_x) \times (- k_x)
    \\ +
    \big(A \omega/2c ~ \big(sin(k_x x)sin(k_yy) - cos(k_xx)cos(k_yy)\big) ~ sin(\omega ct/c) ~ e^{- k_z |z|}\big)^2 (- k_y) \times (k_y)
    \\ +
    \big(A \omega/2c ~ \big(sin(k_x x)sin(k_yy) - cos(k_xx)cos(k_yy)\big) ~ sin(\omega ct/c) ~ e^{- k_z |z|}\big)^2 (- k_x) \times (k_x)
    \\ +
    \big(A k_x k_y ~ \big(cos(k_x x)sin(k_yy) - sin(k_xx)cos(k_yy)\big) ~ cos(\omega ct/c) ~ e^{- k_z |z|} \big)^2
    \\ + 
    \big(- A k_x k_z ~sign(z)/2 ~ \big(sin(k_x x)sin(k_yy) - cos(k_xx)cos(k_yy)\big) ~ cos(\omega ct/c) ~ e^{- k_z |z|}\big)^2
    \\ +
    \big(A \omega/2c ~ \big(sin(k_x x)sin(k_yy) - cos(k_xx)cos(k_yy)\big) ~ sin(\omega ct/c) ~ e^{- k_z |z|}\big)^2 (k_y) \times (- k_y)
    \\ +
    \big(- A k_x k_y ~ \big(cos(k_x x)sin(k_yy) - sin(k_xx)cos(k_yy)\big) ~ cos(\omega ct/c) ~ e^{- k_z |z|} \big)^2
    \\ + 
    \big(A k_y k_z ~ sign(z)/2 ~ \big(sin(k_x x)sin(k_yy) - cos(k_xx)cos(k_yy)\big) ~ cos(\omega ct/c) ~ e^{- k_z |z|}\big)^2
    \\ + 
    \big(A k_x k_z ~sign(z)/2 ~ \big(sin(k_x x)sin(k_yy) - cos(k_xx)cos(k_yy)\big) ~ cos(\omega ct/c) ~ e^{- k_z |z|}\big)^2
    \\ + 
    \big(- A k_y k_z ~ sign(z)/2 ~ \big(sin(k_x x)sin(k_yy) - cos(k_xx)cos(k_yy)\big) ~ cos(\omega ct/c) ~ e^{- k_z |z|}\big)^2
    \\ + ~ (cont.)
\end{split}
\end{align*}
\newpage
\begin{align*}
\begin{split}
    (...) +
    \big(A \omega ~ sign(z)/2c ~ \big(cos(k_x x)sin(k_yy) + sin(k_xx)cos(k_yy)\big) ~ sin(\omega ct/c) ~ e^{- k_z |z|}\big)^2 (- k_z) \times (k_z)
    \\ +
    \big(A \omega/2c ~ \big(sin(k_x x)sin(k_yy) - cos(k_xx)cos(k_yy)\big) ~ sin(\omega ct/c) ~ e^{- k_z |z|}\big)^2 (k_x) \times (- k_x)
    \\ + 
    \big(- A k_x k_z ~sign(z)/2 ~ \big(sin(k_x x)sin(k_yy) - cos(k_xx)cos(k_yy)\big) ~ cos(\omega ct/c) ~ e^{- k_z |z|}\big)^2
    \\ + 
    \big(- A k_x^2/2 ~ \big(cos(k_x x)sin(k_yy) + sin(k_xx)cos(k_yy)\big) ~ cos(\omega ct/c) ~ e^{- k_z |z|}\big)^2
    \\ +
    \big(A \omega ~sign(z)/2c ~ \big(cos(k_x x)sin(k_yy) + sin(k_xx)cos(k_yy)\big) ~ sin(\omega ct/c) ~ e^{- k_z |z|}\big)^2 (k_z) \times (- k_z)
    \\ + 
    \big(A k_x k_z ~sign(z)/2 ~ \big(sin(k_x x)sin(k_yy) - cos(k_xx)cos(k_yy)\big) ~ cos(\omega ct/c) ~ e^{- k_z |z|}\big)^2
    \\ + 
    \big(A (- k_x k_y + k_z(k_z sign(z) - 2\delta(z)))/2 ~ \big(cos(k_x x)sin(k_yy) + sin(k_xx)cos(k_yy)\big) ~ cos(\omega ct/c) ~ e^{- k_z |z|}\big)^2
    \\ +
    \big(A \omega/2c ~ \big(sin(k_x x)sin(k_yy) - cos(k_xx)cos(k_yy)\big) ~ sin(\omega ct/c) ~ e^{- k_z |z|}\big)^2 (- k_x) \times (k_x)
    \\ + 
    \big(A k_x^2/2 ~ \big(cos(k_x x)sin(k_yy) + sin(k_xx)cos(k_yy)\big) ~ cos(\omega ct/c) ~ e^{- k_z |z|}\big)^2
    \\ + 
    \big(A (k_x k_y - k_z(k_z sign(z) - 2\delta(z)))/2 ~ \big(cos(k_x x)sin(k_yy) + sin(k_xx)cos(k_yy)\big) ~ cos(\omega ct/c) ~ e^{- k_z |z|}\big)^2
    \\
    + 
    \big(A (- k_x k_y + \omega^2/c^2)/2 ~ \big(cos(k_x x)sin(k_yy) + sin(k_xx)cos(k_yy)\big) ~ cos(\omega ct/c) ~ e^{- k_z |z|}\big)^2
    \\ + 
    \big(- A k_y^2/2 ~ \big(cos(k_x x)sin(k_yy) + sin(k_xx)cos(k_yy)\big) ~ cos(\omega ct/c) ~ e^{- k_z |z|}\big)^2
    \\ + 
    \big(A k_y k_z ~ sign(z)/2 ~ \big(sin(k_x x)sin(k_yy) - cos(k_xx)cos(k_yy)\big) ~ cos(\omega ct/c) ~ e^{- k_z |z|}\big)^2
    \\ + 
    \big(A (k_x k_y - \omega^2/c^2)/2 ~ \big(cos(k_x x)sin(k_yy) + sin(k_xx)cos(k_yy)\big) ~ cos(\omega ct/c) ~ e^{- k_z |z|}\big)^2
    \\ +
    \big(A \omega/2c ~ \big(sin(k_x x)sin(k_yy) - cos(k_xx)cos(k_yy)\big) ~ sin(\omega ct/c) ~ e^{- k_z |z|}\big)^2 (k_y) \times (- k_y)
    \\ +
    \big(A \omega ~sign(z)/2c ~ \big(cos(k_x x)sin(k_yy) + sin(k_xx)cos(k_yy)\big) ~ sin(\omega ct/c) ~ e^{- k_z |z|}\big)^2 (k_z) \times (- k_z)
    \\ + 
    \big(A k_y^2/2 ~ \big(cos(k_x x)sin(k_yy) + sin(k_xx)cos(k_yy)\big) ~ cos(\omega ct/c) ~ e^{- k_z |z|}\big)^2
    \\ +
    \big(A \omega/2c ~ \big(sin(k_x x)sin(k_yy) - cos(k_xx)cos(k_yy)\big) ~ sin(\omega ct/c) ~ e^{- k_z |z|}\big)^2 (- k_y) \times (k_y)
    \\ + 
    \big(- A k_y k_z ~ sign(z)/2 ~ \big(sin(k_x x)sin(k_yy) - cos(k_xx)cos(k_yy)\big) ~ cos(\omega ct/c) ~ e^{- k_z |z|}\big)^2
    \\ +
    \big(A \omega ~sign(z)/2c ~ \big(cos(k_x x)sin(k_yy) + sin(k_xx)cos(k_yy)\big) ~ sin(\omega ct/c) ~ e^{- k_z |z|}\big)^2 (- k_z) \times (k_z)
    \\
    + 
    \big(A \omega/2c ~ \big(sin(k_x x)sin(k_yy) - cos(k_xx)cos(k_yy)\big) ~ sin(\omega ct/c) ~ e^{- k_z |z|}\big)^2 (- k_x) \times (k_x)
    \\ +
    \big(A \omega/2c ~ \big(sin(k_x x)sin(k_yy) - cos(k_xx)cos(k_yy)\big) ~ sin(\omega ct/c) ~ e^{- k_z |z|}\big)^2 (k_y) \times (- k_y)
    \\ +
    \big(A \omega/2c ~ \big(sin(k_x x)sin(k_yy) - cos(k_xx)cos(k_yy)\big) ~ sin(\omega ct/c) ~ e^{- k_z |z|}\big)^2 (k_x) \times (- k_x)
    \\ +
    \big(- A k_x k_y ~ \big(cos(k_x x)sin(k_yy) - sin(k_xx)cos(k_yy)\big) ~ cos(\omega ct/c) ~ e^{- k_z |z|} \big)^2
    \\ + 
    \big(A k_x k_z ~sign(z)/2 ~ \big(sin(k_x x)sin(k_yy) - cos(k_xx)cos(k_yy)\big) ~ cos(\omega ct/c) ~ e^{- k_z |z|}\big)^2
    \\ +
    \big(A \omega/2c ~ \big(sin(k_x x)sin(k_yy) - cos(k_xx)cos(k_yy)\big) ~ sin(\omega ct/c) ~ e^{- k_z |z|}\big)^2 (- k_y) \times (k_y)
    \\ +
    \big(A k_x k_y ~ \big(cos(k_x x)sin(k_yy) - sin(k_xx)cos(k_yy)\big) ~ cos(\omega ct/c) ~ e^{- k_z |z|} \big)^2
    \\ + 
    \big(- A k_y k_z ~ sign(z)/2 ~ \big(sin(k_x x)sin(k_yy) - cos(k_xx)cos(k_yy)\big) ~ cos(\omega ct/c) ~ e^{- k_z |z|}\big)^2
    \\ + 
    \big(- A k_x k_z ~sign(z)/2 ~ \big(sin(k_x x)sin(k_yy) - cos(k_xx)cos(k_yy)\big) ~ cos(\omega ct/c) ~ e^{- k_z |z|}\big)^2
    \\ + 
    \big(A k_y k_z ~ sign(z)/2 ~ \big(sin(k_x x)sin(k_yy) - cos(k_xx)cos(k_yy)\big) ~ cos(\omega ct/c) ~ e^{- k_z |z|}\big)^2
    \\
    + 
    \big(A \omega ~ sign(z)/2c ~ \big(cos(k_x x)sin(k_yy) + sin(k_xx)cos(k_yy)\big) ~ sin(\omega ct/c) ~ e^{- k_z |z|}\big)^2 (- k_z) \times (k_z)
    \\ +
    \big(A \omega/2c ~ \big(sin(k_x x)sin(k_yy) - cos(k_xx)cos(k_yy)\big) ~ sin(\omega ct/c) ~ e^{- k_z |z|}\big)^2 (k_y) \times (- k_y)
    \\ +
    \big(A \omega ~sign(z)/2c ~ \big(cos(k_x x)sin(k_yy) + sin(k_xx)cos(k_yy)\big) ~ sin(\omega ct/c) ~ e^{- k_z |z|}\big)^2 (k_z) \times (- k_z)
    \\ + 
    \big(A k_y k_z ~ sign(z)/2 ~ \big(sin(k_x x)sin(k_yy) - cos(k_xx)cos(k_yy)\big) ~ cos(\omega ct/c) ~ e^{- k_z |z|}\big)^2
    \\ + 
    \big(A (- k_x k_y + k_z(k_z sign(z) - 2\delta(z)))/2 ~ \big(cos(k_x x)sin(k_yy) + sin(k_xx)cos(k_yy)\big) ~ cos(\omega ct/c) ~ e^{- k_z |z|}\big)^2
    \\ + 
    \big(- A k_y k_z ~ sign(z)/2 ~ \big(sin(k_x x)sin(k_yy) - cos(k_xx)cos(k_yy)\big) ~ cos(\omega ct/c) ~ e^{- k_z |z|}\big)^2
    \\ + 
    \big(- A k_y^2/2 ~ \big(cos(k_x x)sin(k_yy) + sin(k_xx)cos(k_yy)\big) ~ cos(\omega ct/c) ~ e^{- k_z |z|}\big)^2
    \\ +
    \big(A \omega/2c ~ \big(sin(k_x x)sin(k_yy) - cos(k_xx)cos(k_yy)\big) ~ sin(\omega ct/c) ~ e^{- k_z |z|}\big)^2 (- k_y) \times (k_y)
    \\ + ~ (cont.)
\end{split}
\end{align*}
\newpage
\begin{align*}
\begin{split}
    (...) +
    \big(A (k_x k_y - k_z(k_z sign(z) - 2\delta(z)))/2 ~ \big(cos(k_x x)sin(k_yy) + sin(k_xx)cos(k_yy)\big) ~ cos(\omega ct/c) ~ e^{- k_z |z|}\big)^2
    \\ + 
    \big(A k_y^2/2 ~ \big(cos(k_x x)sin(k_yy) + sin(k_xx)cos(k_yy)\big) ~ cos(\omega ct/c) ~ e^{- k_z |z|}\big)^2
    \\
    + 
    \big(A k_x k_z ~sign(z)/2 ~ \big(sin(k_x x)sin(k_yy) - cos(k_xx)cos(k_yy)\big) ~ cos(\omega ct/c) ~ e^{- k_z |z|}\big)^2
    \\ + 
    \big(A k_y k_z ~ sign(z)/2 ~ \big(sin(k_x x)sin(k_yy) - cos(k_xx)cos(k_yy)\big) ~ cos(\omega ct/c) ~ e^{- k_z |z|}\big)^2
    \\ + 
    \big(A (k_z(k_z sign(z) - 2\delta(z)) + \omega^2/c^2)/2 ~ \big(cos(k_x x)sin(k_yy) + sin(k_xx)cos(k_yy)\big) ~ cos(\omega ct/c) ~ e^{- k_z |z|}\big)^2
    \\ + 
    \big(- A k_x k_z ~sign(z)/2 ~ \big(sin(k_x x)sin(k_yy) - cos(k_xx)cos(k_yy)\big) ~ cos(\omega ct/c) ~ e^{- k_z |z|}\big)^2
    \\ +
    \big(A \omega/2c ~ \big(sin(k_x x)sin(k_yy) - cos(k_xx)cos(k_yy)\big) ~ sin(\omega ct/c) ~ e^{- k_z |z|}\big)^2 (- k_x) \times (k_x)
    \\ + 
    \big(- A k_y k_z ~ sign(z)/2 ~ \big(sin(k_x x)sin(k_yy) - cos(k_xx)cos(k_yy)\big) ~ cos(\omega ct/c) ~ e^{- k_z |z|}\big)^2
    \\ +
    \big(A \omega/2c ~ \big(sin(k_x x)sin(k_yy) - cos(k_xx)cos(k_yy)\big) ~ sin(\omega ct/c) ~ e^{- k_z |z|}\big)^2 (- k_y) \times (k_y)
    \\ + 
    \big(A (- k_z(k_z sign(z) - 2\delta(z)) - \omega^2/c^2)/2 ~ \big(cos(k_x x)sin(k_yy) + sin(k_xx)cos(k_yy)\big) ~ cos(\omega ct/c) ~ e^{- k_z |z|}\big)^2
    \\ +
    \big(A \omega/2c ~ \big(sin(k_x x)sin(k_yy) - cos(k_xx)cos(k_yy)\big) ~ sin(\omega ct/c) ~ e^{- k_z |z|}\big)^2 (k_x) \times (- k_x)
    \\ +
    \big(A \omega/2c ~ \big(sin(k_x x)sin(k_yy) - cos(k_xx)cos(k_yy)\big) ~ sin(\omega ct/c) ~ e^{- k_z |z|}\big)^2 (k_y) \times (- k_y)
    \\
    + 
    \big(A \omega ~sign(z)/2c ~ \big(cos(k_x x)sin(k_yy) + sin(k_xx)cos(k_yy)\big) ~ sin(\omega ct/c) ~ e^{- k_z |z|}\big)^2 (k_z) \times (- k_z)
    \\ +
    \big(A \omega/2c ~ \big(sin(k_x x)sin(k_yy) - cos(k_xx)cos(k_yy)\big) ~ sin(\omega ct/c) ~ e^{- k_z |z|}\big)^2 (- k_x) \times (k_x)
    \\ + 
    \big(A k_x k_z ~sign(z)/2 ~ \big(sin(k_x x)sin(k_yy) - cos(k_xx)cos(k_yy)\big) ~ cos(\omega ct/c) ~ e^{- k_z |z|}\big)^2
    \\ + 
    \big(A k_x^2/2 ~ \big(cos(k_x x)sin(k_yy) + sin(k_xx)cos(k_yy)\big) ~ cos(\omega ct/c) ~ e^{- k_z |z|}\big)^2
    \\ + 
    \big(A \omega ~sign(z)/2c ~ \big(cos(k_x x)sin(k_yy) + sin(k_xx)cos(k_yy)\big) ~ sin(\omega ct/c) ~ e^{- k_z |z|}\big)^2 (- k_z) \times (k_z)
    \\ + 
    \big(- A k_x k_z ~sign(z)/2 ~ \big(sin(k_x x)sin(k_yy) - cos(k_xx)cos(k_yy)\big) ~ cos(\omega ct/c) ~ e^{- k_z |z|}\big)^2
    \\ +
    \big(A (k_x k_y - k_z(k_z sign(z) - 2\delta(z)))/2 ~ \big(cos(k_x x)sin(k_yy) + sin(k_xx)cos(k_yy)\big) ~ cos(\omega ct/c) ~ e^{- k_z |z|}\big)^2
    \\ +
    \big(A \omega/2c ~ \big(sin(k_x x)sin(k_yy) - cos(k_xx)cos(k_yy)\big) ~ sin(\omega ct/c) ~ e^{- k_z |z|}\big)^2 (k_x) \times (- k_x)
    \\ + 
    \big(- A k_x^2/2 ~ \big(cos(k_x x)sin(k_yy) + sin(k_xx)cos(k_yy)\big) ~ cos(\omega ct/c) ~ e^{- k_z |z|}\big)^2
    \\ +
    \big(A (- k_x k_y + k_z(k_z sign(z) - 2\delta(z)))/2 ~ \big(cos(k_x x)sin(k_yy) + sin(k_xx)cos(k_yy)\big) ~ cos(\omega ct/c) ~ e^{- k_z |z|}\big)^2
    \\
    +
    \big(A \omega ~sign(z)/2c ~ \big(cos(k_x x)sin(k_yy) + sin(k_xx)cos(k_yy)\big) ~ sin(\omega ct/c) ~ e^{- k_z |z|}\big)^2 (k_z) \times (- k_z)
    \\ +
    \big(A \omega/2c ~ \big(sin(k_x x)sin(k_yy) - cos(k_xx)cos(k_yy)\big) ~ sin(\omega ct/c) ~ e^{- k_z |z|}\big)^2 (- k_y) \times (k_y)
    \\ + 
    \big(A \omega ~sign(z)/2c ~ \big(cos(k_x x)sin(k_yy) + sin(k_xx)cos(k_yy)\big) ~ sin(\omega ct/c) ~ e^{- k_z |z|}\big)^2 (- k_z) \times (k_z)
    \\ + 
    \big(- A k_y k_z ~ sign(z)/2 ~ \big(sin(k_x x)sin(k_yy) - cos(k_xx)cos(k_yy)\big) ~ cos(\omega ct/c) ~ e^{- k_z |z|}\big)^2
    \\ +
    \big(A (k_x k_y - k_z(k_z sign(z) - 2\delta(z)))/2 ~ \big(cos(k_x x)sin(k_yy) + sin(k_xx)cos(k_yy)\big) ~ cos(\omega ct/c) ~ e^{- k_z |z|}\big)^2
    \\ + 
    \big(A k_y k_z ~ sign(z)/2 ~ \big(sin(k_x x)sin(k_yy) - cos(k_xx)cos(k_yy)\big) ~ cos(\omega ct/c) ~ e^{- k_z |z|}\big)^2
    \\ + 
    \big(A k_y^2/2 ~ \big(cos(k_x x)sin(k_yy) + sin(k_xx)cos(k_yy)\big) ~ cos(\omega ct/c) ~ e^{- k_z |z|}\big)^2
    \\ +
    \big(A \omega/2c ~ \big(sin(k_x x)sin(k_yy) - cos(k_xx)cos(k_yy)\big) ~ sin(\omega ct/c) ~ e^{- k_z |z|}\big)^2 (k_y) \times (- k_y)
    \\ +
    \big(A (- k_x k_y + k_z(k_z sign(z) - 2\delta(z)))/2 ~ \big(cos(k_x x)sin(k_yy) + sin(k_xx)cos(k_yy)\big) ~ cos(\omega ct/c) ~ e^{- k_z |z|}\big)^2
    \\ + 
    \big(- A k_y^2/2 ~ \big(cos(k_x x)sin(k_yy) + sin(k_xx)cos(k_yy)\big) ~ cos(\omega ct/c) ~ e^{- k_z |z|}\big)^2
\end{split}
\end{align*}
Squaring, multiplying and combining similar terms leads to:
\begin{equation}
\begin{split}
    K_{_{\times}} = & ~ A^2\big[2k_x^4 + 8k_x^2 k_y^2 - 4 k_x k_y \omega^2/c^2 + 3\omega^4/c^4 + 2k_y^4 + 3k_z^4 ~ sign^4(z) + 2k_z^2 \omega^2 ~ sign^2(z)/c^2 - 4k_x k_y k_z^2 ~ sign^2(z) 
    \\ & + (- 12 k_z^2 ~ sign^2(z) - 4 k_z \omega^2/c^2  + 8 k_x k_y k_z) \delta(z) 
    + 12 k_z^2 \delta^2(z) \big] 
    \\ & \times
    \big(cos(k_x x)sin(k_yy) + sin(k_xx)cos(k_yy)\big)^2 ~ cos^2(\omega ct/c) ~ e^{- 2 k_z |z|}
    \\ &
    + 4 A^2(k_x^2 + k_y^2) k_z^2 ~ sign^2(z) \big(sin(k_x x)sin(k_yy) - cos(k_xx)cos(k_yy)\big)^2 cos^2(\omega ct/c) ~ e^{- 2 k_z |z|}
    \\ & 
    - 4 A^2(k_x^2 + k_y^2) \omega^2/c^2 \big(sin(k_x x)sin(k_yy) - cos(k_xx)cos(k_yy)\big)^2 ~ sin^2(\omega ct/c) ~ e^{- 2 k_z |z|}
    \\ & 
    - 4k_z^2 \omega^2 ~ sign^2(z)/c^2 \big(cos(k_x x)sin(k_yy) + sin(k_xx)cos(k_yy)\big)^2 ~ sin^2(\omega ct/c) ~ e^{- 2 k_z |z|} 
\end{split}
\end{equation}
Similarly to $K_{_+}$, $K_{_\times}$ contains Dirac delta terms and also the square of distributions ($\delta^2(z)$), making it ill-defined at the hypersurface $\Sigma ~ (z = 0)$. Again, it will be split into a well-behaved, smooth, part ($K_{S_\times}$) outside $\Sigma$ and one with the ill-defined terms ($K_{\Sigma_\times}$) containing the distributions and their product:
\begin{equation}
    K_{_\times} = K_{S_\times} + K_{\Sigma_\times}
\end{equation}
where
\begin{equation}
\begin{split}
    K_{\Sigma_\times} = & A^2(- 12 k_z^2 ~ sign^2(z) - 4 k_z \omega^2/c^2  + 8 k_x k_y k_z) \delta(z) 
    + 12 k_z^2 \delta^2(z))
    \\ & \times
    \big(cos(k_x x)sin(k_yy) + sin(k_xx)cos(k_yy)\big)^2 ~ cos^2(\omega ct/c) ~ e^{- 2 k_z |z|}
\end{split}
\end{equation}
Again, noting that for $z \neq 0$, $sign^2(z) = sign^4(z) = 1$, it results for the smooth $K_{S_\times}$:
\begin{equation}
\begin{split}
    K_{S_{\times}} = & ~ A^2(2k_x^4 + 8k_x^2 k_y^2 - 4 k_x k_y \omega^2/c^2 + 3\omega^4/c^4 + 2k_y^4 + 3k_z^4 + 2k_z^2\omega^2/c^2  - 4k_x k_y k_z^2)
    \\ & \times
    \big(cos(k_x x)sin(k_yy) + sin(k_xx)cos(k_yy)\big)^2 ~ cos^2(\omega ct/c) ~ e^{- 2 k_z |z|}
    \\ &
    + 4 A^2(k_x^2 + k_y^2) k_z^2 \big(sin(k_x x)sin(k_yy) - cos(k_xx)cos(k_yy)\big)^2 cos^2(\omega ct/c) ~ e^{- 2 k_z |z|}
    \\ & 
    - 4 A^2(k_x^2 + k_y^2) \omega^2/c^2 \big(sin(k_x x)sin(k_yy) - cos(k_xx)cos(k_yy)\big)^2 ~ sin^2(\omega ct/c) ~ e^{- 2 k_z |z|}
    \\ & 
    - 4k_z^2 \omega^2/c^2 \big(cos(k_x x)sin(k_yy) + sin(k_xx)cos(k_yy)\big)^2 ~ sin^2(\omega ct/c) ~ e^{- 2 k_z |z|}
\end{split}
\end{equation}
To generate perfect squares of the dispersion (Eq.~\ref{eq_dispersion_x_SM}) we may add/subtract the following terms within the first parenthesis: $4k_x^2 k_z^2$, $4k_x^2 \omega^2/c^2$, $4k_y^2 k_z^2$, $4k_y^2 \omega^2/c^2$ and $4k_z^2 \omega^2/c^2$. Therefore, $K_{S_{\times}}$ may be rewritten as:
\begin{equation}
\begin{split}
    K_{S_{\times}} = & ~ A^2\big[(2k_x^4 + 4k_x^2 k_y^2 - 4k_x^2 k_z^2 - 4k_x^2 \omega^2/c^2 + 2k_y^4 - 4k_y^2 k_z^2 - 4k_y^2 \omega^2/c^2 + 2k_z^4 + 4k_z^2\omega^2/c^2 + 2\omega^4/c^4)
    \\ & + 
    (4 k_x^2 k_y^2 - 4k_x k_y k_z^2 - 4k_x k_y \omega^2/c^2 + k_z^4 + 2k_z^2\omega^2/c^2 + \omega^4/c^4)
    \\ &
    + (4k_x^2 k_z^2 + 4k_x^2 \omega^2/c^2 + 4k_y^2 k_z^2 + 4k_y^2 \omega^2/c^2 - 4k_z^2\omega^2/c^2) \big]
    \\ & \times
    \big(cos(k_x x)sin(k_yy) 
    + sin(k_xx)cos(k_yy)\big)^2 ~ cos^2(\omega ct/c) ~ e^{- 2 k_z |z|}
    \\ & + 4 A^2(k_x^2 + k_y^2) k_z^2 \big(sin(k_x x)sin(k_yy) - cos(k_xx)cos(k_yy)\big)^2 cos^2(\omega ct/c) ~ e^{- 2 k_z |z|}
    \\ & 
    - 4 A^2(k_x^2 + k_y^2) \omega^2/c^2 \big(sin(k_x x)sin(k_yy) - cos(k_xx)cos(k_yy)\big)^2 ~ sin^2(\omega ct/c) ~ e^{- 2 k_z |z|}
    \\ & 
    - 4k_z^2 \omega^2/c^2 \big(cos(k_x x)sin(k_yy) + sin(k_xx)cos(k_yy)\big)^2 ~ sin^2(\omega ct/c) ~ e^{- 2 k_z |z|}
\end{split}
\end{equation}
Considering that:
\begin{equation}
    (2k_x k_y - k_z^2 - \omega^2/c^2)^2 = 4 k_x^2 k_y^2 - 4k_x k_y k_z^2 - 4k_x k_y \omega^2/c^2 + k_z^4 + 2k_z^2\omega^2/c^2 + \omega^4/c^4
\end{equation}
Then,
\begin{equation}
\begin{split}
    K_{S_{\times}} = & ~ A^2\big[2(k_x^2 + k_y^2 - k_z^2 - \omega^2/c^2)^2
    + 
    (2k_x k_y - k_z^2 - \omega^2/c^2)^2
    + (4k_x^2 k_z^2 + 4k_x^2 \omega^2/c^2 + 4k_y^2 k_z^2 + 4k_y^2 \omega^2/c^2 - 4k_z^2\omega^2/c^2) \big]
    \\ &  \times 
    \big(cos(k_x x)sin(k_yy) 
    + sin(k_xx)cos(k_yy)\big)^2 ~ cos^2(\omega ct/c) ~ e^{- 2 k_z |z|}
    \\ & + 4 A^2(k_x^2 + k_y^2) k_z^2 \big(sin(k_x x)sin(k_yy) - cos(k_xx)cos(k_yy)\big)^2 cos^2(\omega ct/c) ~ e^{- 2 k_z |z|}
    \\ & 
    - 4 A^2(k_x^2 + k_y^2) \omega^2/c^2 \big(sin(k_x x)sin(k_yy) - cos(k_xx)cos(k_yy)\big)^2 ~ sin^2(\omega ct/c) ~ e^{- 2 k_z |z|}
    \\ & 
    - 4k_z^2 \omega^2/c^2 \big(cos(k_x x)sin(k_yy) + sin(k_xx)cos(k_yy)\big)^2 ~ sin^2(\omega ct/c) ~ e^{- 2 k_z |z|}
\end{split}
\end{equation}
By the dispersion in Eq.~\ref{eq_dispersion_x_SM} and imposing a square unit cell ($k_x = k_y$), the first two terms become null. Combining the remaining similar terms:
\begin{equation}
\begin{split}
    K_{S_{\times}} = & ~ 4A^2 (k_x^2 + k_y^2) k_z^2 e^{- 2 k_z |z|} cos^2(\omega ct/c) \big[ \big(cos(k_x x)sin(k_yy) 
    + sin(k_xx)cos(k_yy)\big)^2 + \big(sin(k_x x)sin(k_yy) - cos(k_xx)cos(k_yy)\big)^2 \big]
    \\ &
    + 4 A^2(k_x^2 + k_y^2) \omega^2/c^2 e^{- 2 k_z |z|} 
    \big[ \big(cos(k_x x)sin(k_yy) + sin(k_xx)cos(k_yy)\big)^2 cos^2(\omega ct/c) 
    \\ & - \big(sin(k_x x)sin(k_yy) - cos(k_xx)cos(k_yy)\big)^2 
    sin^2(\omega ct/c) \big]
    \\ & 
    - 4k_z^2 \omega^2/c^2 e^{- 2 k_z |z|} 
    \big(cos(k_x x)sin(k_yy) + sin(k_xx)cos(k_yy)\big)^2 
    \big[ cos^2(\omega ct/c) + sin^2(\omega ct/c) \big]
\end{split}
\end{equation}
By recognizing that:
\begin{equation} \label{sincossq}
\begin{split}
    \big(cos(k_x x)sin(k_yy) 
    + sin(k_xx)cos(k_yy)\big)^2 + \big(sin(k_x x)sin(k_yy) - cos(k_xx)cos(k_yy)\big)^2 
    \\ = sin^2(k_xx + k_yy) ~ + cos^2(k_xx + k_yy) = 1 
\end{split}
\end{equation}
and by adding/subtracting $4 A^2(k_x^2 + k_y^2) \omega^2/c^2 ~ e^{- 2 k_z |z|} cos^2(\omega ct/c) \big(sin(k_x x)sin(k_yy) - cos(k_xx)cos(k_yy)\big)^2$:
\begin{equation}
\begin{split}
    K_{S_{\times}} = & ~ 4A^2 (k_x^2 + k_y^2) k_z^2 e^{- 2 k_z |z|} cos^2(\omega ct/c)
    + 4 A^2(k_x^2 + k_y^2) \omega^2/c^2 e^{- 2 k_z |z|} cos^2(\omega ct/c)
    \\ & \times \big[ 
    \big(cos(k_x x)sin(k_yy) + sin(k_xx)cos(k_yy)\big)^2 
    + \big(sin(k_x x)sin(k_yy) - cos(k_xx)cos(k_yy)\big)^2 
    \big]
    \\ & 
    - 4 A^2(k_x^2 + k_y^2) \omega^2/c^2 e^{- 2 k_z |z|} \big(sin(k_x x)sin(k_yy) - cos(k_xx)cos(k_yy)\big)^2 
    \big[
    sin^2(\omega ct/c) + cos^2(\omega ct/c) 
    \big]
    \\ & 
    - 4k_z^2 \omega^2/c^2 e^{- 2 k_z |z|} 
    \big(cos(k_x x)sin(k_yy) + sin(k_xx)cos(k_yy)\big)^2 
\end{split}
\end{equation}
Again, by Eq.~\ref{sincossq}:
\begin{equation}
\begin{split}
    K_{S_{\times}} = & ~ 4A^2 (k_x^2 + k_y^2) k_z^2 e^{- 2 k_z |z|} cos^2(\omega ct/c)
    + 4 A^2(k_x^2 + k_y^2) \omega^2/c^2 e^{- 2 k_z |z|} cos^2(\omega ct/c)
    \\ & 
    - 4 A^2(k_x^2 + k_y^2) \omega^2/c^2 e^{- 2 k_z |z|} \big(sin(k_x x)sin(k_yy) - cos(k_xx)cos(k_yy)\big)^2 
    \\ & 
    - 4k_z^2 \omega^2/c^2 e^{- 2 k_z |z|} 
    \big(cos(k_x x)sin(k_yy) + sin(k_xx)cos(k_yy)\big)^2 
\end{split}
\end{equation}
%
%
Finally, by adding/subtracting $4 A^2(k_x^2 + k_y^2) \omega^2/c^2 ~ e^{- 2 k_z |z|} \big(sin(k_x x)cos(k_yy) + sin(k_xx)cos(k_yy)\big)^2$ and by invoking the dispersion:
\begin{equation}
\begin{split}
    K_{S_{\times}} = & ~ 4A^2 (k_x^2 + k_y^2)^2 e^{- 2 k_z |z|} cos^2(\omega ct/c)
    \\ & 
    - 4 A^2(k_x^2 + k_y^2) \omega^2/c^2 e^{- 2 k_z |z|} 
    \big[
    \big(sin(k_x x)sin(k_yy) - cos(k_xx)cos(k_yy)\big)^2 
    + \big(sin(k_x x)cos(k_yy) + sin(k_xx)cos(k_yy)\big)^2
    \big]
    \\ & 
    + 4 \omega^2/c^2 e^{- 2 k_z |z|} 
    \big(cos(k_x x)sin(k_yy) + sin(k_xx)cos(k_yy)\big)^2
    \big[(k_x^2 + k_y^2) - k_z^2 \big]
\end{split}
\end{equation}
Being $(k_x^2 + k_y^2) - k_z^2 = \omega^2/c^2$ and by Eq.~\ref{sincossq}, it results:
\begin{equation} \label{kretschmann_x}
\begin{split}
    K_{S_{\times}} = 4A^2 
    \big[ 
    (k_x^2 + k_y^2)^2 cos^2(\omega ct/c)
    + \omega^4/c^4
    \big(cos(k_x x)sin(k_yy) + sin(k_xx)cos(k_yy)\big)^2
    - (k_x^2 + k_y^2) \omega^2/c^2
    \big] e^{- 2 k_z |z|}
\end{split}
\end{equation}
being of an identical form to the Kretschmann scalar $K_{S_{+}}$ of the ($+$)-polarized BIC (Eq.~\ref{kretschmann_+}), i.e., decaying exponentially from $z = 0$ and 
without divergences/singularities. However, both $K_{S_{+}}$ and $K_{S_{\times}}$ can assume negative values at certain spacetime coordinates ($- 4 A^2 k_x^2\omega^2/c^2$ and $- 4 A^2 (k_x^2 + k_y^2) \omega^2/c^2$ at their most negative, respectively). These finite second order scalars imply that the proposed bound metrics correspond to regular spacetimes outside the $\Sigma$ hypersurface. 
Note that as we are dealing with null Ricci tensors ($R_{{\sigma \nu}_{(+,\times)}} = 0
$) and Ricci scalars ($R_{{}_{(+, \times)}} = 0$) for $z \neq 0$, the calculated Kretschmann invariants are equivalent to the second order Weyl scalars ($C_{\rho \sigma \mu \nu} C^{\rho \sigma \mu \nu}$). For a $d = 4$ dimensions spacetime \cite{Cherubini2003}:
\begin{equation}
    R_{\rho \sigma \mu \nu} R^{\rho \sigma \mu \nu} = 
    C_{\rho \sigma \mu \nu} C^{\rho \sigma \mu \nu}
    + 2 
    R_{{\rho \sigma}_{(+,\times)}} 
    R^{\rho \sigma}_{_{(+,\times)}}
    - \frac{1}{3} R^2_{{}_{(+, \times)}}
    = C_{\rho \sigma \mu \nu} C^{\rho \sigma \mu \nu}
\end{equation}
\\
\newpage

\subsection*{Lorenz gauge}
So far we have demonstrated that the proposed bound metrics imply vacuum solutions of the linearized EFEs outside the localization plane $z = 0$ by explicitly calculating the Ricci tensor components and Ricci scalar for each metric perturbation tensor. This involves finding solutions of coupled differential equations containing different metric perturbation terms ($h_{{\mu \nu}_{(+, \times)}}$). A simpler, and in this case, equivalent approach can be used by using the Lorenz gauge.
In this section it will be shown that the trace-reversed perturbation tensors ($\overline{h}_{{\mu \nu}_{(+, \times)}}$) of the proposed localized solutions fulfill the condition of the Lorenz gauge:
\begin{equation} \label{lor_gauge}
    \partial^{\mu} \overline{h}_{{\mu \nu}_{(+, \times)}} = 0
\end{equation}
in which the trace-reversed perturbation tensor is:
\begin{equation}
    \overline{h}_{{\mu \nu}_{(+, \times)}} = h_{{\mu \nu}_{(+, \times)}} - \frac{1}{2} \eta_{\mu \nu} h_{{}_{(+, \times)}}
\end{equation}
where $h$ is the trace of the perturbation tensor ($h = h^\mu_\mu = \eta^{\mu \alpha}h_{\alpha \mu}$). The Lorenz gauge simplifies the linearized EFEs, equating the d'Alambertian of the trace-reversed perturbation to the respective components of the stress-energy tensor whereas 
in vacuum, they are further reduced to a set of homogeneous wave equations ($\Box \overline{h}_{\mu \nu} = 0$, $\Box = 
\partial_t\partial_t - \bigtriangleup$). 
\subsubsection{Localized perturbations are not a pure gauge}
Before proceeding with the Lorenz gauge evaluation, it will be demonstrated that the proposed bound metrics are not a pure gauge, i.e., that they do not correspond to a coordinate transformation of the Minkowsky flat metric. For that we may assume that there is vector field $\xi_\mu$ that transforms all the perturbation components into zero, thus leading to a flat metric, via a gauge transformation. This is equivalent to showing that a perturbation $h_{\mu \nu}$ can be generated by the vector field $\xi_\mu$. However, this will lead to contradictions among the required transformations regarding their dependences on spacetime variables. Note that, if we were dealing with pure gauge configurations, all the components of the Riemann curvature tensor (Eq.~\ref{lg_eq_RCTc_BIC_hxxhyy} - \ref{lg_eq_RCTc_BIC}) would be zero or nullified by a given dispersion, which is not the case. In linearized gravity we have the following gauge symmetry:
\begin{equation} \label{gaugesym}
    h_{\mu \nu} \rightarrow h_{\mu \nu} + \partial_\mu \xi_\nu + \partial_\nu \xi_\mu 
\end{equation}
that results from a coordinate transformation $\tilde{x}^\alpha = x^\alpha + \xi^\alpha$, $||\xi^\alpha|| \ll 1$ and $|| \partial_\beta \xi^\alpha|| \ll 1$ (i.e. small displacement and small changes in the displacement as a function of the spacetime coordinates). In other words, these imply that:
\begin{equation}
    \tilde{h}_{\mu \nu} = h_{\mu \nu} + \partial_\mu \xi_\nu + \partial_\nu \xi_\mu 
\end{equation}
As we want the new perturbation $\tilde{h}_{\mu \nu}$ to be zero for all spacetime indices, corresponding to a flat metric:
\begin{equation} \label{flatmetricgauge}
    0 = h_{\mu \nu} + \partial_\mu \xi_\nu + \partial_\nu \xi_\mu 
\end{equation}
\subsubsection{$h_{xx}, h_{yy}$ BIC - Pure gauge hypothesis}
The non-zero perturbation components of the ($+$)-polarized BIC (Eq.~\ref{eq_BIC_+} and \ref{eq_hbic_plus}) and the gauge transformation to a flat metric (Eq.~\ref{flatmetricgauge}) result in:
\begin{equation}
\begin{split}
    h_{tt} + 2 \partial_{t} \xi_{t} = 
    h_{BIC_+} + 2 \partial_{t} \xi_{t} =
    0 
    \\
    \partial_t \xi_x + \partial_x \xi_t = 0
    \\
    \partial_t \xi_y + \partial_y \xi_t = 0
    \\
    \partial_t \xi_z + \partial_z \xi_t = 0
    \\
    h_{xx} + 2 \partial_{x} \xi_{x} = 
    - h_{BIC_+} + 2 \partial_{x} \xi_{x} =
    0 
    \\
    \partial_x \xi_y + \partial_y \xi_x = 0
    \\
    \partial_x \xi_z + \partial_x \xi_z = 0
    \\
    h_{yy} + 2 \partial_{y} \xi_{y} = 
    h_{BIC_+} + 2 \partial_{y} \xi_{y} =
    0 
    \\
    \partial_y \xi_z + \partial_z \xi_y = 0
    \\
    h_{zz} + 2 \partial_{z} \xi_{z} = 
    - h_{BIC_+} + 2 \partial_{z} \xi_{z} =
    0 
    \\
\end{split}
\end{equation}
Integrating the first equation $h_{BIC_+} = - 2 \partial_{t} \xi_{t}$ in $t$ results in the following solution for $\xi_t$:
\begin{equation}
    \xi_t = - \frac{A c}{2 \omega} ~ cos(k_x x) ~ sin(\omega ct/c) ~ e^{- k_z |z|} + f(x,y,z)
\end{equation}
where the function $f(x,y,z)$ generalizes the possible integration constants (relative to $t$). Similarly, integrating $h_{BIC_+} = 2 \partial_{x} \xi_{x}$, in $x$ gives the solution:
\begin{equation}
    \xi_x = \frac{A}{2 k_x} ~ sin(k_x x) ~ cos(\omega ct/c) ~ e^{- k_z |z|} + g(t,y,z)
\end{equation}
The last two equations should also fulfill the condition $\partial_t \xi_x + \partial_x \xi_t = 0$, thus:
\begin{equation}
\begin{split}
    \partial_t \big( \frac{A}{2 k_x} ~ sin(k_x x) ~ cos(\omega ct/c) ~ e^{- k_z |z|} + g(t,y,z) \big) 
    + \partial_x \big( - \frac{A c}{2 \omega} ~ cos(k_x x) ~ sin(\omega ct/c) ~ e^{- k_z |z|} + f(x,y,z) \big) 
    = 0
    \\
    \therefore
     \frac{A \omega}{2 k_x c} ~ sin(k_x x) ~ sin(\omega ct/c) ~ e^{- k_z |z|} + \partial_t \big( g(t,y,z) \big) 
    - \frac{A k_x c}{2 \omega} ~ sin(k_x x) ~ sin(\omega ct/c) ~ e^{- k_z |z|} + \partial_x \big( f(x,y,z) \big) 
    = 0
\end{split}
\end{equation}
Now, by deriving the expression with respect to $x$ and noting that the order of partial derivatives does not matter:
\begin{equation}
\begin{split}
     \frac{A \omega}{2c} ~ cos(k_x x) ~ sin(\omega ct/c) ~ e^{- k_z |z|} + \partial_t \big( \partial_x (g(t,y,z)) \big) 
    - \frac{A k_x^2 c}{2 \omega} ~ cos(k_x x) ~ sin(\omega ct/c) ~ e^{- k_z |z|} + \partial^2_x \big( f(x,y,z) \big) 
    = 0
\end{split}
\end{equation}
Implying
\begin{equation}
\begin{split}
    \partial^2_x \big( f(x,y,z) \big) =
     \frac{A}{2} \big[k_x^2 c / \omega - \omega /c \big]~ cos(k_x x) ~ sin(\omega ct/c) ~ e^{- k_z |z|}
     =
      \frac{A}{2 \omega c} \big[k_x^2 c^2 - \omega^2 \big]~ cos(k_x x) ~ sin(\omega ct/c) ~ e^{- k_z |z|}
\end{split}
\end{equation}
an equality that should hold for all values of $ct$. This is clearly not true as the function on the left-hand side is clearly independent of $ct$ and $k_x^2 c^2 \neq \omega^2$. This implies that the functional equality previous to the derivation is also spurious. It can be concluded that there is no vector field $\xi_\mu$ that fulfills the conditions $h_{BIC_+} = - 2 \partial_{t} \xi_{t}$, $h_{BIC_+} = 2 \partial_{x} \xi_{x}$ and $\partial_t \xi_x + \partial_x \xi_t = 0$ simultaneously, and thus that the perturbation $h_{\mu \nu_{+}}$ is not a pure gauge configuration. A similar reasoning can be applied to any other conditions above involving any pairs of $\xi_\mu$ components.
\subsubsection{$h_{xy}, h_{yx}$ BIC - Pure gauge hypothesis}
A similar reasoning can be applied to the ($\times$)-polarized BIC. The attempt to gauge-away the perturbation $h_{\mu \nu_{\times}}$ (Eq.~\ref{eq_BIC_x} and \ref{eq_hbic_x}) leads to:
\begin{equation}
\begin{split}
    h_{tt} + 2 \partial_{t} \xi_{t} = 
    h_{BIC_\times} + 2 \partial_{t} \xi_{t} =
    0 
    \\
    \partial_t \xi_x + \partial_x \xi_t = 0
    \\
    \partial_t \xi_y + \partial_y \xi_t = 0
    \\
    \partial_t \xi_z + \partial_z \xi_t = 0
    \\
    2 \partial_{x} \xi_{x} =
    0 
    \\
    h_{xy} + \partial_x \xi_y + \partial_y \xi_x = 
    - h_{BIC_\times} + \partial_x \xi_y + \partial_y \xi_x =
    0
    \\
    \partial_x \xi_z + \partial_x \xi_z = 0
    \\
    2 \partial_{y} \xi_{y} =
    0 
    \\
    \partial_y \xi_z + \partial_z \xi_y = 0
    \\
    h_{zz} + 2 \partial_{z} \xi_{z} = 
    - h_{BIC_\times} + 2 \partial_{z} \xi_{z} =
    0 
    \\
\end{split}
\end{equation}
Integrating $h_{BIC_+} = - 2 \partial_{t} \xi_{t}$ in $t$ results in $\xi_t$:
\begin{equation}
    \xi_t = \frac{A c}{2 \omega} ~ \big(cos(k_x x)sin(k_yy) + sin(k_xx)cos(k_yy)\big) ~ sin(\omega ct/c) ~ e^{- k_z |z|} 
    + f(x,y,z)
\end{equation}
Now integrating $h_{BIC_\times} = 2 \partial_{z} \xi_{z}$ with respect to $z$ yields the solution for $\xi_z$:
\begin{equation}
    \xi_z =\frac{A}{2 k_z} ~ \big(cos(k_x x)sin(k_yy) + sin(k_xx)cos(k_yy)\big) ~ cos(\omega ct/c) 
    ~ \big[sign(z) ( 1 - e^{- k_z |z|}) + 1\big]
    + j(t,x,y)
\end{equation}
where again the function $j(t,x,y)$ is an integration constant regarding $z$.
Employing the fourth condition \mbox{$\partial_t \xi_z + \partial_z \xi_t = 0$}:
\begin{equation}
\begin{split}
    \partial_t \big( \frac{A}{2 k_z} ~ \big(cos(k_x x)sin(k_yy) + sin(k_xx)cos(k_yy)\big) ~ cos(\omega ct/c) 
    ~ \big[sign(z) ( 1 - e^{- k_z |z|}) + 1\big] 
    + j(t,x,y) \big) \\
    + \partial_z \big( \frac{A c}{2 \omega} ~ \big(cos(k_x x)sin(k_yy) + sin(k_xx)cos(k_yy)\big) ~ sin(\omega ct/c) ~ e^{- k_z |z|} 
    + f(x,y,z) \big) 
    = 0
\end{split}
\end{equation}
Thus,
\begin{equation} \label{deltxiz_delzxit}
\begin{split}
     \frac{- A \omega}{2  k_z c} ~ \big(cos(k_x x)sin(k_yy) + sin(k_xx)cos(k_yy)\big) ~ sin(\omega ct/c) 
     ~ \big[sign(z) ( 1 - e^{- k_z |z|}) + 1\big] 
    + \partial_t \big( j(t,x,y) \big) 
    \\
    - \frac{A k_z c}{2 \omega} ~ \big(cos(k_x x)sin(k_yy) + sin(k_xx)cos(k_yy)\big) ~ sin(\omega ct/c) 
    ~ sign(z) e^{- k_z |z|} 
    + \partial_z \big( f(x,y,z) \big) 
    = 0
\end{split}
\end{equation}
By further deriving the expression with respect to 
$t$:
\begin{equation}
\begin{split}
     \frac{- A \omega^2}{2 k_z c^2} ~ \big(cos(k_x x)sin(k_yy) + sin(k_xx)cos(k_yy)\big) ~ cos(\omega ct/c) 
     ~ \big[sign(z) ( 1 - e^{- k_z |z|}) + 1\big] 
    + \partial_t^2 \big(j(t,x,y) \big) 
    \\
    - \frac{A k_z}{2} ~ \big(cos(k_x x)sin(k_yy) + sin(k_xx)cos(k_yy)\big) ~ cos(\omega ct/c) 
    ~ sign(z) e^{- k_z |z|} 
    + \partial_z \big( \partial_t (f(x,y,z)) \big) 
    = 0
    \\
    \therefore
    \partial_t^2 \big(j(t,x,y) \big) =
    \frac{A}{2} \bigg[
    \frac{\omega^2}{k_z c^2} ~ \big[sign(z) ( 1 - e^{- k_z |z|}) + 1\big]  
    + k_z ~ sign(z) e^{- k_z |z|} \bigg] \\
    \times
    ~ \big(cos(k_x x)sin(k_yy) + sin(k_xx)cos(k_yy)\big) ~ cos(\omega ct/c) 
    \\
    = \frac{A}{2} \bigg[
    \frac{\omega^2}{k_z c^2} (sign(z) + 1) 
    + \frac{sign(z) e^{- k_z |z|}}{k_z c^2} (- \omega^2 + k_z^2 c^2) \big]   \bigg] \\
    \times
    ~ \big(cos(k_x x)sin(k_yy) + sin(k_xx)cos(k_yy)\big) ~ cos(\omega ct/c) \\
\end{split}
\end{equation}
that should hold for all values of $z$, of which the left-hand side is independent (note that $sign(z) = -1$ is not true for all $z$ and that $\omega^2 \neq k_z^2 c^2$), making the equality in Eq.~\ref{deltxiz_delzxit} also false. These imply a contradiction between $h_{BIC_+} = - 2 \partial_{t} \xi_{t}$, $h_{BIC_\times} = 2 \partial_{z} \xi_{z}$ and $\partial_t \xi_z + \partial_z \xi_t = 0$ for any vector field $\xi_\mu$ and that the perturbation $h_{\mu \nu_{\times}}$ cannot be nullified by a gauge transformation. An identical conclusion can be obtained by considering other components of $\xi_\mu$. We may thus conclude that both bound metrics carry energy and are not a mere coordinate artifact.

\subsubsection{$h_{xx}, h_{yy}$ BIC - Lorenz gauge}

The trace $h_{{}_+}$ of the perturbation tensor $h_{{\mu \nu}_+}$ (Eq.~\ref{eq_BIC_+}) is:
\begin{equation} \label{trace_bic_+}
\begin{split}
    h_{{}_+} =  \eta^{\mu \alpha}h_{\alpha \mu}
    & = \eta^{tt}h_{tt} + \eta^{xx}h_{xx} + \eta^{yy}h_{yy} + \eta^{zz}h_{zz} 
    \\ & = h_{tt} - h_{xx} - h_{yy}- h_{zz} 
    \\ & = h_{{BIC}_+} - (- h_{{BIC}_+}) - h_{{BIC}_+} - (-h_{{BIC}_+}) 
    = 2 h_{{BIC}_+}
\end{split}
\end{equation}

Then, the trace-reversed perturbation tensor components are:
\begin{equation} \label{trace_reversed_bic_+}
\begin{split}
    \overline{h}_{tt} 
    & = h_{tt} - \frac{1}{2}\eta_{tt}h_{{}_+}
    = h_{{BIC}_+} - \frac{1}{2} (2 h_{{BIC}_+}) = 0
    \\
    \overline{h}_{tx}
    = \overline{h}_{xt}
    & = h_{tx} - \frac{1}{2}\eta_{tx}h_{{}_+}
    = 0
    \\
    \overline{h}_{ty}
    = \overline{h}_{yt}
    & = h_{ty} - \frac{1}{2}\eta_{ty}h_{{}_+}
    = 0
    \\
    \overline{h}_{tz}
    = \overline{h}_{zt}
    & = h_{tz} - \frac{1}{2}\eta_{tz}h_{{}_+}
    = 0
    \\
    \overline{h}_{xx}
    & = h_{xx} - \frac{1}{2}\eta_{xx}h_{{}_+}
    = - h_{{BIC}_+} + \frac{1}{2} (2 h_{{BIC}_+}) = 0
    \\
    \overline{h}_{xy}
    = \overline{h}_{yx}
    & = h_{xy} - \frac{1}{2}\eta_{xy}h_{{}_+}
    = 0
    \\
    \overline{h}_{xz}
    = \overline{h}_{zx}
    & = h_{xz} - \frac{1}{2}\eta_{xz}h_{{}_+}
    = 0
    \\
    \overline{h}_{yy}
    & = h_{yy} - \frac{1}{2}\eta_{yy}h_{{}_+}
    = h_{{BIC}_+} + \frac{1}{2} (2 h_{{BIC}_+}) 
    = 2 h_{{BIC}_+}
    \\
    \overline{h}_{yz}
    = \overline{h}_{zy}
    & = h_{yz} - \frac{1}{2}\eta_{yz}h_{{}_+}
    = 0
    \\
    \overline{h}_{zz}
    = \overline{h}_{zz}
    & = h_{zz} - \frac{1}{2}\eta_{zz}h_{{}_+}
    = - h_{{BIC}_+} + \frac{1}{2} (2 h_{{BIC}_+})
    = 0
\end{split}
\end{equation}
Note that the symmetry of the flat spacetime metric tensor and of the perturbation tensor were invoked (i.e. $\eta_{\mu \nu} = \eta_{\nu \mu}$ and $h_{\mu \nu} = h_{\nu \mu}$). Here, only the $\overline{h}_{yy}$ term is non-zero. The derivatives in the contravariant form might be calculated via $\partial^\mu = \eta^{\mu \nu} \partial_\nu$:
\begin{equation}
\begin{split}
    \partial^t & = \eta^{t \nu} \partial_{\nu}
    = \eta^{t t} \partial_{t} + \eta^{t x} \partial_{x}
    + \eta^{t y} \partial_{y} + \eta^{t z} \partial_{z}
    = \partial_t
   \\ %
    \partial^x & = \eta^{x \nu} \partial_{\nu}
    = \eta^{x t} \partial_{t} + \eta^{x x} \partial_{x}
    + \eta^{x y} \partial_{y} + \eta^{x z} \partial_{z}
    = - \partial_x
   \\ %
    \partial^y & = \eta^{y \nu} \partial_{\nu}
    = \eta^{y t} \partial_{t} + \eta^{y x} \partial_{x}
    + \eta^{y y} \partial_{y} + \eta^{y z} \partial_{z}
    = - \partial_y
   \\ %
    \partial^z & = \eta^{z \nu} \partial_{\nu}
    = \eta^{z t} \partial_{t} + \eta^{z x} \partial_{x}
    + \eta^{z y} \partial_{y} + \eta^{z z} \partial_{z}
    = - \partial_z
\end{split}
\end{equation}

We may finally check if the trace-reversed perturbation tensor fulfills the Lorenz gauge (Eq.~\ref{lor_gauge}):
\begin{equation} \label{lor_gauge_bic_+}
\begin{split}
   \partial^{\mu}\overline{h}_{t \mu}
   & = \partial^{t}\overline{h}_{tt} 
   + \partial^{x}\overline{h}_{tx}
   + \partial^{y}\overline{h}_{ty}
   + \partial^{z}\overline{h}_{tz}
   = 0
   \\
   \partial^{\mu}\overline{h}_{x \mu}
   & = \partial^{t}\overline{h}_{xt} 
   + \partial^{x}\overline{h}_{xx}
   + \partial^{y}\overline{h}_{xy}
   + \partial^{z}\overline{h}_{xz}
   = 0
   \\
   \partial^{\mu}\overline{h}_{y \mu}
   & = \partial^{t}\overline{h}_{yt} 
   + \partial^{x}\overline{h}_{yx}
   + \partial^{y}\overline{h}_{yy}
   + \partial^{z}\overline{h}_{yz}
   = - \partial_y (2 h_{{BIC}_+}) 
   = 0
   \\
   \partial^{\mu}\overline{h}_{z \mu}
   & = \partial^{t}\overline{h}_{zt} 
   + \partial^{x}\overline{h}_{zx}
   + \partial^{y}\overline{h}_{zy}
   + \partial^{z}\overline{h}_{zz}
   = 0
\end{split}
\end{equation}
given that the metric perturbation $h_{{BIC}_+}$ is not a function of $y$ (Eq.~\ref{eq_hbic_plus}). Therefore, the metric perturbation $h_{{\mu \nu}_+}$ fulfills the conditions of the Lorenz gauge. It remains to be checked if the trace-reversed perturbation tensor coefficients are solutions of the homogeneous wave equation for a suitable dispersion:
\begin{equation}
\begin{split}
    \Box \overline{h}_{{\mu \nu}_+}
    & = \bigg(
    \partial_t\partial_t - \bigtriangleup\bigg) \overline{h}_{{\mu \nu}_+}
    = \bigg(
    \partial_t \partial_t - \partial_x \partial_x - \partial_y \partial_y - \partial_z \partial_z\bigg)\overline{h}_{{\mu \nu}_+} 
    \\
    & = \bigg(
    \partial_t \partial_t - \partial_x \partial_x - \partial_y \partial_y - \partial_z \partial_z\bigg)(2 h_{{BIC}_+})
    \\
    & = \bigg(- \frac{\omega^2}{c^2} + k_x^2 - 
    k_z( k_z ~ sign^2(z) - 2\delta (z))
    \bigg)(2 h_{{BIC}_+})
\end{split}
\end{equation}

yielding a distributional expression at the localization plane $z = 0$, but a regular vacuum solution outside of it, with an identical dispersion to the one required for solving the linearized EFEs via null $R_{{\sigma \nu}_+}$ and $R_{{}_+}$, as obtained in Eq.~\ref{eq_dispersion_+}.

\subsubsection{$h_{xy}, h_{yx}$ BIC - Lorenz gauge}

Repeating the procedure above for the BIC metric associated with the $h_{xy}, h_{yx}$ coefficients, the trace of the perturbation tensor $h_{{BIC}_{\times}}$ (Eq.~\ref{eq_BIC_x}):

\begin{equation} \label{trace_bic_x}
\begin{split}
    h_{{}_\times} =  \eta^{\mu \alpha}h_{\alpha \mu}
    & = \eta^{tt}h_{tt} + \eta^{xx}h_{xx} + \eta^{yy}h_{yy} + \eta^{zz}h_{zz} 
    \\ & = h_{tt} - h_{zz} 
    \\ & = h_{{BIC}_{\times}} - (-h_{{BIC}_{\times}}) 
    = 2 h_{{BIC}_{\times}}
\end{split}
\end{equation}

The trace-reverse perturbation tensor components are:
\begin{equation} \label{trace_reversed_bic_x}
\begin{split}
    \overline{h}_{tt} 
    & = h_{tt} - \frac{1}{2}\eta_{tt}h_{{}_\times}
    = h_{{BIC}_\times} - \frac{1}{2} (2 h_{{BIC}_\times}) = 0
    \\
    \overline{h}_{tx}
    = \overline{h}_{xt}
    & = h_{tx} - \frac{1}{2}\eta_{tx}h_{{}_\times}
    = 0
    \\
    \overline{h}_{ty}
    = \overline{h}_{yt}
    & = h_{ty} - \frac{1}{2}\eta_{ty}h_{{}_\times}
    = 0
    \\
    \overline{h}_{tz}
    = \overline{h}_{zt}
    & = h_{tz} - \frac{1}{2}\eta_{tz}h_{{}_\times}
    = 0
    \\
    \overline{h}_{xx}
    & = h_{xx} - \frac{1}{2}\eta_{xx}h_{{}_\times}
    = \frac{1}{2} (2 h_{{BIC}_\times}) = h_{{BIC}_\times}
    \\
    \overline{h}_{xy}
    = \overline{h}_{yx}
    & = h_{xy} - \frac{1}{2}\eta_{xy}h_{{}_\times}
    = - h_{{BIC}_\times} 
    \\
    \overline{h}_{xz}
    = \overline{h}_{zx}
    & = h_{xz} - \frac{1}{2}\eta_{xz}h_{{}_\times}
    = 0
    \\
    \overline{h}_{yy}
    & = h_{yy} - \frac{1}{2}\eta_{yy}h_{{}_\times}
    = \frac{1}{2} (2 h_{{BIC}_\times}) 
    = h_{{BIC}_\times}
    \\
    \overline{h}_{yz}
    = \overline{h}_{zy}
    & = h_{yz} - \frac{1}{2}\eta_{yz}h_{{}_\times}
    = 0
    \\
    \overline{h}_{zz}
    = \overline{h}_{zz}
    & = h_{zz} - \frac{1}{2}\eta_{zz}h_{{}_\times}
    = - h_{{BIC}_\times} + \frac{1}{2} (2 h_{{BIC}_\times})
    = 0
\end{split}
\end{equation}

For this metric perturbation, the terms $\overline{h}_{xx}$, $\overline{h}_{xy}$, $\overline{h}_{yx}$ and $\overline{h}_{yy}$ are non-zero. We may thus check if the trace-reversed metric tensor fulfills the Lorenz gauge (Eq.~\ref{lor_gauge}):
\begin{equation} \label{lor_gauge_bic_x}
\begin{split}
   \partial^{\mu}\overline{h}_{t \mu}
   & = \partial^{t}\overline{h}_{tt} 
   + \partial^{x}\overline{h}_{tx}
   + \partial^{y}\overline{h}_{ty}
   + \partial^{z}\overline{h}_{tz}
   = 0
   \\
   \partial^{\mu}\overline{h}_{x \mu}
   & = \partial^{t}\overline{h}_{xt} 
   + \partial^{x}\overline{h}_{xx}
   + \partial^{y}\overline{h}_{xy}
   + \partial^{z}\overline{h}_{xz}
   = - \partial_x (h_{{BIC}_\times})
   - \partial_y (-h_{{BIC}_\times})
   \\
   & = (k_x - k_y) 
   ~ (sin(k_x x) sin(k_y y) - cos(k_x x) cos(k_y y)) ~ cos( \omega ct/c) ~ 
   e^{- k_z |z|} \\
   \\
   \partial^{\mu}\overline{h}_{y \mu}
   & = \partial^{t}\overline{h}_{yt} 
   + \partial^{x}\overline{h}_{yx}
   + \partial^{y}\overline{h}_{yy}
   + \partial^{z}\overline{h}_{yz}
   = - \partial_x ( - h_{{BIC}_\times})
   - \partial_y (h_{{BIC}_\times})
   \\
   & = (-k_x + k_y) 
   ~ (sin(k_x x) sin(k_y y) - cos(k_x x) cos(k_y y)) ~ cos( \omega ct/c) ~ 
   e^{- k_z |z|} \\
   \\
   \partial^{\mu}\overline{h}_{z \mu}
   & = \partial^{t}\overline{h}_{zt} 
   + \partial^{x}\overline{h}_{zx}
   + \partial^{y}\overline{h}_{zy}
   + \partial^{z}\overline{h}_{zz}
   = 0
\end{split}
\end{equation}
implying that the Lorenz gauge is fulfilled if the in-plane modulations $k_x$ and $k_y$ are equal. This is the exact condition necessary to solve the linear EFEs for this metric perturbation. Indeed, the trace-reversed perturbation tensor coefficients are also solutions of the homogeneous wave equation (for $z \neq 0$):
\begin{equation}
\begin{split}
    \Box \overline{h}_{{\mu \nu}_\times} 
    & = \bigg(
    \partial_t\partial_t - \bigtriangleup\bigg) \overline{h}_{{\mu \nu}_\times}
    = \bigg(
    \partial_t \partial_t - \partial_x \partial_x - \partial_y \partial_y - \partial_z \partial_z\bigg)\overline{h}_{{\mu \nu}_\times}
    \\
    & = \bigg(
    \partial_t \partial_t - \partial_x \partial_x - \partial_y \partial_y - \partial_z \partial_z\bigg)(\pm h_{{BIC}_\times})
    \\
    & = \bigg(- \frac{\omega^2}{c^2} + k_x^2 + k_y^2 
    - k_z( k_z ~ sign^2(z) - 2\delta (z))
    \bigg)(\pm h_{{BIC}_\times})
\end{split}
\end{equation}

again leading to a distribution at the localization plane for the non-zero spacetime indices and to a regular vacuum solution for $z \neq 0$ if using a dispersion identical to Eq.~\ref{eq_dispersion_x_SM}. Note that the homogeneous wave equation itself does not impose the condition of a square lattice ($k_x = k_y$) to the metric perturbation, which comes from Ricci tensor and scalar and the Lorenz gauge condition.

\subsection{Energy-momentum conservation}
The surface polariton-like dispersions imply vacuum solutions $T^{\mu \nu} = 0$ outside the plane $z = 0$ for both ($+,\times$) bound polarizations, making the local conservation of energy-momentum evident.
It can be demonstrated that both bound metric obey the condition for local energy and momentum conservation in linearized gravity:
\begin{equation} \label{lemconservation_SM}
    \partial_\mu T^{\mu \nu} = 
    \partial_t T^{t \nu} + \partial_x T^{x \nu} 
    + \partial_y T^{y \nu} + \partial_z T^{z \nu}
    = 0
\end{equation}
also at $z = 0$, regardless of the distributional nature of the stress-energy momentum at the localization plane. From the trace-reversed perturbation $\bar{h}_{\mu \nu_{(+,\times)}}$, we can calculate the covariant stress-energy tensor $T_{\mu \nu_{(+, \times)}}$ as:
\begin{equation}
    \Box \bar{h}_{\mu \nu_{(+,\times)}} = -\frac{16 \pi G}{c^4} T_{\mu \nu_{(+,\times)}}
\end{equation}
\subsubsection{$h_{xx}, h_{yy}$ BIC}
From Eq.~\ref{trace_reversed_bic_+}:
\begin{equation}
    \Box \bar{h}_{yy} = -\frac{16 \pi G}{c^4} T_{yy}
    \quad \therefore \quad 
    T_{yy} = \frac{c^4}{8 \pi G}  \bigg(\frac{\omega^2}{c^2} - k_x^2 + k_z \big(k_z~sign^2(z) -2\delta (z)\big) \bigg) h_{BIC_{+}}
\end{equation}
The contravariant stress-energy tensor $T^{\mu \nu}$ is:
\begin{equation}
    T^{\mu \nu} = \eta^{\mu \alpha} \eta^{\nu \beta} T_{\alpha \beta}
    = \eta^{\mu y} \eta^{\nu y} T_{yy}
\end{equation}
being non-zero only for $\mu = \nu = y$, i.e.~$T^{yy} = T_{yy}$.
Thus, the local energy-momentum conservation Eq.~\ref{lemconservation_SM} is evidently zero for $\nu = (t, x, z)$. For $\nu = y$:
\begin{equation} \label{lemconservation_+}
\begin{split}
    \partial_\mu T^{\mu \nu} = & ~
    \partial_t T^{t y} + \partial_x T^{x y} 
    + \partial_y T^{y y} + \partial_z T^{z y}
    = \partial_y T^{y y}
    \\
    = & ~ \partial_y \bigg[ \frac{c^4}{8 \pi G}  \bigg(\frac{\omega^2}{c^2} - k_x^2 + k_z \big(k_z~sign^2(z) -2\delta (z)\big) \bigg) h_{BIC_{+}} \bigg]
        \\
    = & ~ \partial_y \bigg[ \frac{c^4}{8 \pi G}  \bigg(\frac{\omega^2}{c^2} - k_x^2 + k_z \big(k_z~sign^2(z) -2\delta (z)\big) \bigg) A ~ cos(k_x x) ~ cos(\omega ct/c) ~
    e^{- k_z |z|} \bigg]
    = 0
\end{split}
\end{equation}
as the stress-energy tensor is not a function of the spacetime coordinate $y$. So the ($+$)-polarized BIC solution is divergence-free and obeys local energy-momentum conservation for all values of the $z$ coordinate. 
\subsubsection{$h_{xy}, h_{yx}$ BIC}
Similarly, to obtain $T_{\mu \nu}$ from the trace-reversed perturbation of the ($\times$)-polarized BIC (Eq.~\ref{trace_reversed_bic_x}):
\begin{equation} \label{SET_x}
\begin{split}
    \Box \bar{h}_{xx} = -\frac{16 \pi G}{c^4} T_{xx}
    \quad \therefore \quad 
    T_{xx} = \frac{c^4}{16 \pi G}  \bigg(\frac{\omega^2}{c^2} - k_x^2 - k_y^2 + k_z \big(k_z~sign^2(z) -2\delta (z)\big) \bigg) h_{{BIC}_\times}
    \\
    \Box \bar{h}_{xy} = -\frac{16 \pi G}{c^4} T_{xy}
    \quad \therefore \quad 
    T_{xy} = \frac{c^4}{16 \pi G}  \bigg(- \frac{\omega^2}{c^2} + k_x^2 + k_y^2 - k_z \big(k_z~sign^2(z) -2\delta (z)\big) \bigg) h_{{BIC}_\times}
    \\
    \Box \bar{h}_{yx} = -\frac{16 \pi G}{c^4} T_{yx}
    \quad \therefore \quad 
    T_{yx} = \frac{c^4}{16 \pi G}  \bigg(- \frac{\omega^2}{c^2} + k_x^2 + k_y^2 - k_z \big(k_z~sign^2(z) -2\delta (z)\big) \bigg) h_{{BIC}_\times}
    \\
    \Box \bar{h}_{yy} = -\frac{16 \pi G}{c^4} T_{yy}
    \quad \therefore \quad 
    T_{yy} = \frac{c^4}{16 \pi G}  \bigg(\frac{\omega^2}{c^2} - k_x^2 - k_y^2 + k_z \big(k_z~sign^2(z) -2\delta (z)\big) \bigg) h_{{BIC}_\times}
\end{split}
\end{equation}
The contravariant tensor components are:
\begin{equation}
    T^{\mu \nu} = \eta^{\mu \alpha} \eta^{\nu \beta} T_{\alpha \beta}
    = 
    \eta^{\mu x} \eta^{\nu x} T_{xx} +
    \eta^{\mu y} \eta^{\nu x} T_{yx} +
    \eta^{\mu x} \eta^{\nu y} T_{xy} +
    \eta^{\mu y} \eta^{\nu y} T_{yy}
\end{equation}
implying:
\begin{equation}
    T^{xx} = T_{xx} ~;~ 
    T^{xy} = T_{xy} ~;~ 
    T^{yx} = T_{yx} ~;~ 
    T^{yy} = T_{yy}
\end{equation}
Using Eq.~\ref{lemconservation_SM} (Note $\partial_\mu T^{\mu \nu} = 0$ for $\nu = (t,z)$):
\begin{equation} \label{lemconservation_SM_x}
\begin{split}    
    \partial_\mu T^{\mu x} = & ~ 
    \partial_x T^{xx} + \partial_y T^{yx} 
    \\
    = & ~
    \partial_x \bigg[
    \frac{c^4}{16 \pi G}  \bigg(\frac{\omega^2}{c^2} - k_x^2 - k_y^2 + k_z \big(k_z~sign^2(z) -2\delta (z)\big) \bigg) h_{{BIC}_\times}
    \bigg]
    \\
    & +
    \partial_y \bigg[
    \frac{c^4}{16 \pi G}  \bigg(- \frac{\omega^2}{c^2} + k_x^2 + k_y^2 - k_z \big(k_z~sign^2(z) -2\delta (z)\big) \bigg) h_{{BIC}_\times}
    \bigg]
    \\
    = & ~
    (-k_x + k_y) ~ \frac{Ac^4}{16 \pi G}  \bigg(\frac{\omega^2}{c^2} - k_x^2 - k_y^2 + k_z \big(k_z~sign^2(z) -2\delta (z)\bigg) \\ & \quad \times
    (cos(k_x x) cos(k_y y) - sin(k_y y) sin(k_x x)) ~ cos(\omega ct/c) ~ e^{- k_z |z|}
    \\
    \\
    \partial_\mu T^{\mu y} = & ~ 
    \partial_x T^{xy} + \partial_y T^{yy} 
    \\
    = & ~
    \partial_x \bigg[
    \frac{c^4}{16 \pi G}  \bigg(- \frac{\omega^2}{c^2} + k_x^2 + k_y^2 - k_z \big(k_z~sign^2(z) -2\delta (z)\big) \bigg) h_{{BIC}_\times}
    \bigg]
    \\
    & +
    \partial_y \bigg[
    \frac{c^4}{16 \pi G}  \bigg(\frac{\omega^2}{c^2} - k_x^2 - k_y^2 + k_z \big(k_z~sign^2(z) -2\delta (z)\big) \bigg) h_{{BIC}_\times}
    \bigg]
    \\
    = & ~
    (k_x - k_y) ~ \frac{Ac^4}{16 \pi G}  \bigg(\frac{\omega^2}{c^2} - k_x^2 - k_y^2 + k_z \big(k_z~sign^2(z) -2\delta (z)\bigg) \\ & \quad \times
    (cos(k_x x) cos(k_y y) - sin(k_y y) sin(k_x x)) ~ cos(\omega ct/c) ~ e^{- k_z |z|}
\end{split}
\end{equation}
both vanishing for a square lattice perturbation ($k_x = k_y$), thus a requirement for a divergence-free stress-energy tensor and for local energy-momentum conservation of the ($\times$)-polarized BIC. 
%
            
\subsection{Energy-momentum flux}
At first order, the planar GWs solutions of the linearized equations imply that no energy nor momentum is carried by the perturbations of the Minkowsky spacetime. Thus, second order effects are required to evaluate the energy density and also the energy and momentum flux by the propagating waves. A common gauge-invariant approach employs the Isaacson stress-energy tensor \cite{Isaacson1968}, providing a spacetime averaged effective stress tensor, suitable for high-frequency waves or whose wavelength is much smaller than the background curvature. In linearized gravity, for a solution that fulfills the Lorenz gauge the Isaacson stress-energy tensor can be written as:
\begin{equation} \label{isaacsonset}
    t_{\mu \nu}^{BIC_{(+,\times)}} = \frac{c^4}{32 \pi G} 
    \langle 
    \partial_\mu \bar{h}_{{\alpha \beta}_{(+,\times)}} ~ 
    \partial_\nu \bar{h}^{{\alpha \beta}}_{_{(+,\times)}}
    -\frac{1}{2}
    \partial_\mu \bar{h}_{_{(+,\times)}} ~ \partial_\nu \bar{h}_{_{(+,\times)}}
    \rangle
\end{equation}
in which $\langle \cdot \cdot \cdot \rangle$ refers to the average of the expression in brackets for high-frequency spacetime functions and $\bar{h}$ is the trace of the trace-reversed perturbation tensor.
\subsubsection{$h_{xx}, h_{yy}$ BIC}
For the ($+$)-polarized BIC, the relevant quantities are (Eq.~\ref{trace_reversed_bic_+}): $\bar{h}_{yy} = \bar{h}^{yy} = 2h_{BIC_{+}}$, and $\bar{h}_{_{+}} = \eta^{yy} \bar{h}_{yy} = - 2h_{BIC_{+}}$. Also, by summing in $\alpha$ and $\beta$:
\begin{equation}
    \partial_\mu \bar{h}_{{\alpha \beta}} ~ 
    \partial_\nu \bar{h}^{{\alpha \beta}}
    =
    \partial_\mu \bar{h}_{{yy}} ~ 
    \partial_\nu \bar{h}^{{yy}}
\end{equation}
So, Eq.~\ref{isaacsonset} becomes:
\begin{equation}
    t_{\mu \nu}^{BIC_+} = \frac{c^4}{32 \pi G} 
    \langle 
    \partial_\mu \bar{h}_{{yy}} ~ 
    \partial_\nu \bar{h}^{{yy}}
    -\frac{1}{2}
    \partial_\mu \bar{h}_{_{+}} ~ \partial_\nu \bar{h}_{_{+}}
    \rangle
\end{equation}
Yielding the components:
\begin{equation} \label{isaacson_tt+}
\begin{split}
        t_{tt}^{BIC_+} = & \frac{c^4}{32 \pi G} 
    \langle 
    \partial_t \bar{h}_{{yy}} ~ 
    \partial_t \bar{h}^{{yy}}
    -\frac{1}{2}
    \partial_t \bar{h}_{_{+}} ~ \partial_t \bar{h}_{_{+}}
    \rangle
    \\
        = & \frac{c^4}{32 \pi G} 
    \langle 
    \partial_t 2h_{BIC_{+}} ~ 
    \partial_t 2h_{BIC_{+}}
    -\frac{1}{2}
    \partial_t (-2h_{BIC_{+}}) ~ \partial_t (-2h_{BIC_{+}})
    \rangle
    \\
        = & \frac{c^4}{32 \pi G} 
    \langle 
    \big[
    -\frac{2A \omega}{c} ~ cos(k_x x) ~ sin(\omega ct/c) ~ e^{- k_z |z|} 
    \big]^2
    -\frac{1}{2}
    \big[
    \frac{2A \omega}{c} ~ cos(k_x x) ~ sin(\omega ct/c) ~ e^{- k_z |z|} 
    \big]^2
    \rangle
    \\
        = & \frac{A^2 \omega^2 c^2}{16 \pi G} ~ e^{- 2 k_z |z|}
    \langle 
    cos^2(k_x x) ~ sin^2(\omega ct/c) 
    \rangle
\end{split}
\end{equation}
The averaging in the Isaacson stress-energy tensor does not apply to the exponential decay as this is a monotonic function, and not one that varies periodically on the $z$ coordinate. Given the independence of the spacetime coordinate variables, the average may be written as:
\begin{equation}
\begin{split}
    \langle 
    cos^2(k_x x) ~ sin^2(\omega ct/c) 
    \rangle
    = 
        \langle 
    cos^2(k_x x)
    \rangle
        \langle 
    sin^2(\omega ct/c)
    \rangle
\end{split}
\end{equation}
The periodic functions will be integrated over their full period:
\begin{equation}
\begin{split}
        \langle 
    cos^2(k_x x)
    \rangle
    & = \frac{1}{2\pi/k_x - 0} 
        \int_{0}^{2\pi/k_x}{cos^2(k_x x) dx}
    = \frac{k_x}{2\pi} 
        \int_{0}^{2\pi/k_x}{\frac{cos(2 k_x x) - 1}{2} dx}
        \\
    & = \frac{k_x}{2\pi} 
        \bigg[ \frac{sin(2 k_x x)}{4k_x} + \frac{x}{2} \bigg]_{0}^{2\pi/k_x}
    = \frac{1}{2}
    \\ \\
        \langle 
    sin^2(\omega ct/c)
    \rangle
    & = \frac{1}{2\pi c/\omega - 0} 
    \int_{0}^{2\pi c/\omega}{sin^2(\omega ct/c) dt}
    = \frac{\omega}{2\pi c} 
    \int_{0}^{2\pi c/\omega}{\frac{1 - cos(2 \omega ct/c)}{2} dt}
    \\
    & = \frac{\omega}{2\pi c} 
    \bigg[ \frac{ct}{2} - \frac{c ~ sin(2 \omega ct/c)}{4 \omega}  \bigg]_{0}^{2\pi c/\omega}
    = \frac{1}{2}
\end{split}
\end{equation}
Thus, the energy density (or the flux of $t$-momentum in the $t$-direction) of the ($+$)-BIC is:
\begin{equation}
    t_{tt}^{BIC_+} =
    \frac{A^2 \omega^2 c^2}{64 \pi G} e^{- 2 k_z |z|}
\end{equation}
denoting the non-propagating nature of the wave and implying the localization of its energy density at $z = 0$.
\begin{equation}
\begin{split}
    t_{tx}^{BIC_+} = t_{xt}^{BIC_+} = & \frac{c^4}{32 \pi G} 
        \langle 
        \partial_t \bar{h}_{{yy}} ~ 
        \partial_x \bar{h}^{{yy}}
        -\frac{1}{2}
        \partial_t \bar{h}_{_{+}} ~ \partial_x \bar{h}_{_{+}}
        \rangle
    \\
    = & \frac{c^4}{32 \pi G} 
        \langle 
        \partial_t 2h_{BIC_{+}} ~ 
        \partial_x 2h_{BIC_{+}}
        -\frac{1}{2}
        \partial_t (-2h_{BIC_{+}}) ~ \partial_x (-2h_{BIC_{+}})
        \rangle
    \\
    = & \frac{c^4}{32 \pi G} 
        \langle 
        \frac{2A^2 k_x \omega}{c} ~ cos(k_x x) ~ sin(k_x x) ~ sin(\omega ct/c) ~ cos(\omega ct/c) ~ e^{- 2k_z |z|} 
        \rangle
    \\
    = & \frac{A^2 k_x \omega c^3}{16 \pi G} ~ e^{- 2k_z |z|}
        \langle 
        cos(k_x x) sin(k_x x) ~ sin(\omega ct/c) cos(\omega ct/c) 
        \rangle
\end{split}
\end{equation}
Noting that when averaging the periodic functions over their full period:
\begin{equation}
\begin{split}
        \langle 
    sin(k_x x) cos(k_x x)
    \rangle
    & = \frac{1}{2\pi/k_x - 0} 
        \int_{0}^{2\pi/k_x}{sin(k_x x) cos(k_x x) dx}
        \\
    & = \frac{k_x}{2\pi} 
        \bigg[ \frac{sin^2(k_x x)}{2k_x} \bigg]_{0}^{2\pi/k_x}
    = 0
    \\ \\
        \langle 
    sin(\omega ct/c) cos(\omega ct/c)
    \rangle
    & = \frac{1}{2\pi c/\omega - 0} 
    \int_{0}^{2\pi c/\omega}{sin(\omega ct/c) cos(\omega ct/c) dt}
    \\
    & = \frac{\omega}{2\pi c} 
    \bigg[ \frac{c ~ sin^2(\omega ct/c)}{2 \omega}  \bigg]_{0}^{2\pi c/\omega}
    = 0
\end{split}
\end{equation}
Thus,
\begin{equation}
    t_{tx}^{BIC_+} = t_{xt}^{BIC_+}  = 0
\end{equation}
no $t$($x$)-momentum flux/Poynting vector in the $x$($t$)-direction. Same for $y$ and $z$:
\begin{equation}
\begin{split}
    t_{ty}^{BIC_+} = t_{yt}^{BIC_+} = & \frac{c^4}{32 \pi G} 
        \langle 
        \partial_t \bar{h}_{{yy}} ~ 
        \partial_y \bar{h}^{{yy}}
        -\frac{1}{2}
        \partial_t \bar{h}_{_{+}} ~ \partial_y \bar{h}_{_{+}}
        \rangle
    = 0
\end{split}
\end{equation}
\begin{equation}
\begin{split}
    t_{tz}^{BIC_+} = t_{zt}^{BIC_+} = & \frac{c^4}{32 \pi G} 
        \langle 
        \partial_t \bar{h}_{{yy}} ~ 
        \partial_z \bar{h}^{{yy}}
        -\frac{1}{2}
        \partial_t \bar{h}_{_{+}} ~ \partial_z \bar{h}_{_{+}}
        \rangle
    \\
    = & \frac{c^4}{32 \pi G} 
        \langle 
        \partial_t 2h_{BIC_{+}} ~ 
        \partial_z 2h_{BIC_{+}}
        -\frac{1}{2}
        \partial_t (-2h_{BIC_{+}}) ~ \partial_z (-2h_{BIC_{+}})
        \rangle
    \\
    = & \frac{A^2 k_z \omega c^3}{16 \pi G} ~ sign(z)~e^{- 2k_z |z|}
        \langle 
        cos^2(k_x x) ~ sin(\omega ct/c) cos(\omega ct/c) 
        \rangle
    \\
    = & ~ 0
\end{split}
\end{equation}
\begin{equation}
\begin{split}
    t_{xx}^{BIC_+} = & \frac{c^4}{32 \pi G} 
        \langle 
        \partial_x \bar{h}_{{yy}} ~ 
        \partial_x \bar{h}^{{yy}}
        -\frac{1}{2}
        \partial_x \bar{h}_{_{+}} ~ \partial_x \bar{h}_{_{+}}
        \rangle
    \\
    = & \frac{c^4}{32 \pi G} 
        \langle 
        \big[
        \partial_x 2h_{BIC_{+}}
        \big]^2
        -\frac{1}{2}
        \big[
        \partial_x (-2h_{BIC_{+}})
        \big]^2
        \rangle
    \\
    = & \frac{A^2 k_x^2 c^4}{16 \pi G} ~e^{- 2k_z |z|}
        \langle 
        sin^2(k_x x) ~ cos(\omega ct/c) 
        \rangle
\end{split}
\end{equation}
thus, the $x$-momentum transfer across $x$ boundaries is:
\begin{equation}
    t_{xx}^{BIC_+} =
    \frac{A^2 k_x^2 c^4}{64 \pi G} ~e^{- 2k_z |z|}
\end{equation}
\begin{equation}
\begin{split}
    t_{xy}^{BIC_+} = t_{yx}^{BIC_+} = & \frac{c^4}{32 \pi G} 
        \langle 
        \partial_x \bar{h}_{{yy}} ~ 
        \partial_y \bar{h}^{{yy}}
        -\frac{1}{2}
        \partial_x \bar{h}_{_{+}} ~ \partial_y \bar{h}_{_{+}}
        \rangle
    = 0
\end{split}
\end{equation}
\begin{equation}
\begin{split}
    t_{xz}^{BIC_+} = t_{zx}^{BIC_+} = & \frac{c^4}{32 \pi G} 
        \langle 
        \partial_x \bar{h}_{{yy}} ~ 
        \partial_z \bar{h}^{{yy}}
        -\frac{1}{2}
        \partial_x \bar{h}_{_{+}} ~ \partial_z \bar{h}_{_{+}}
        \rangle
    \\
    = & \frac{c^4}{32 \pi G} 
        \langle 
        \partial_x 2h_{BIC_{+}} ~ 
        \partial_z 2h_{BIC_{+}}
        -\frac{1}{2}
        \partial_x (-2h_{BIC_{+}}) ~ \partial_z (-2h_{BIC_{+}})
        \rangle
    \\
    = & \frac{A^2 k_x k_z c^4}{16 \pi G} ~ sign(z)~e^{- 2k_z |z|}
        \langle 
        sin(k_x x) cos(k_x x) ~ cos^2(\omega ct/c) 
        \rangle
    \\
    = & ~ 0
\end{split}
\end{equation}
\begin{equation}
\begin{split}
    t_{yy}^{BIC_+} = & \frac{c^4}{32 \pi G} 
        \langle 
        \partial_y \bar{h}_{{yy}} ~ 
        \partial_y \bar{h}^{{yy}}
        -\frac{1}{2}
        \partial_y \bar{h}_{_{+}} ~ \partial_y \bar{h}_{_{+}}
        \rangle
    = 0
\end{split}
\end{equation}
\begin{equation}
\begin{split}
    t_{yz}^{BIC_+} = t_{zy}^{BIC_+} = & \frac{c^4}{32 \pi G} 
        \langle 
        \partial_y \bar{h}_{{yy}} ~ 
        \partial_z \bar{h}^{{yy}}
        -\frac{1}{2}
        \partial_y \bar{h}_{_{+}} ~ \partial_z \bar{h}_{_{+}}
        \rangle
    = 0
\end{split}
\end{equation}
\begin{equation} \label{isaacson_zz+}
\begin{split}
    t_{zz}^{BIC_+} = & \frac{c^4}{32 \pi G} 
        \langle 
        \partial_z \bar{h}_{{yy}} ~ 
        \partial_z \bar{h}^{{yy}}
        -\frac{1}{2}
        \partial_z \bar{h}_{_{+}} ~ \partial_z \bar{h}_{_{+}}
        \rangle
    \\
    = & \frac{c^4}{32 \pi G} 
        \langle 
        \big[
        \partial_z 2h_{BIC_{+}}
        \big]^2
        -\frac{1}{2}
        \big[
        \partial_z (-2h_{BIC_{+}})
        \big]^2
        \rangle
    \\
    = & \frac{A^2 k_z^2 c^4}{16 \pi G} ~ sign^2(z)~e^{- 2k_z |z|}
        \langle 
        cos^2(k_x x) ~ cos^2(\omega ct/c) 
        \rangle
    \\
    = & \frac{A^2 k_z^2 c^4}{64 \pi G} ~ sign^2(z)~e^{- 2k_z |z|}
\end{split}
\end{equation}
Implying a discontinuity at $z = 0$, due to the $sign^2(z)$ function becoming zero. Thus, interestingly, the $z$-momentum transfer across $z$ boundaries, usually associated with the radiation pressure of a planar GW propagating in the $z$-direction gets nullified at the localization plane.
Summarizing in a matrix representation:
\begin{equation} \label{isaacsonset_+}
    t_{\mu \nu}^{BIC_{+}} = e^{- 2k_z |z|}
    \begin{pmatrix}
			\frac{A^2 \omega^2 c^2}{64 \pi G} 	& 0 		& 0 		& 0 \\ 
			0 	& \frac{A^2 k_x^2 c^4}{64 \pi G}  	& 0 		& 0 \\ 
			0 	& 0 		& 0 	& 0 \\ 
			0 	& 0 		& 0 		& \frac{A^2 k_z^2 c^4}{64 \pi G} sign^2(z)
    \end{pmatrix}
\end{equation}
\subsubsection{$h_{xy}, h_{yx}$ BIC}
From the components of the trace-reversed perturbation of the ($\times$)-polarized BIC (Eq.~\ref{trace_reversed_bic_x}), we may calculate $\bar{h}_{_{\times}}$:
\begin{equation}
    \bar{h}_{_{\times}} = \eta^{\mu \nu} \bar{h}_{\mu \nu_{\times}}
    = \eta^{xx} \bar{h}_{xx} + \eta^{yy}\bar{h}_{yy} 
    = - 2 h_{BIC_{\times}}
\end{equation}
While the contravariant components are identical to the covariant ones:
\begin{equation}
    \bar{h}^{xx} = \bar{h}_{xx} ~;~
    \bar{h}^{xy} = \bar{h}_{xy} ~;~
    \bar{h}^{yx} = \bar{h}_{yx} ~;~
    \bar{h}^{yy} = \bar{h}_{yy} ~;~
\end{equation}
Noting that for the first term of the Isaacson tensor:
\begin{equation}
\begin{split}
    \partial_\mu \bar{h}_{\alpha \beta} ~ \partial_\nu \bar{h}^{\alpha \beta}
        = & ~ \partial_\mu \bar{h}_{xx} ~ \partial_\nu \bar{h}^{xx}
        + \partial_\mu \bar{h}_{yx} ~ \partial_\nu \bar{h}^{yx}
        + \partial_\mu \bar{h}_{xy} ~ \partial_\nu \bar{h}^{xy}
        + \partial_\mu \bar{h}_{yy} ~ \partial_\nu \bar{h}^{yy}
    \\
    = & ~ 4 \partial_\mu (h_{BIC_{\times}}) ~ \partial_\nu (h_{BIC_{\times}})
\end{split}
\end{equation}
The components of $t_{\mu \nu}^{BIC_{\times}}$ are:
\begin{equation} \label{isaacson_ttx}
\begin{split}
    t_{tt}^{BIC_{\times}} = & \frac{c^4}{32 \pi G} 
        \langle 
        4 \partial_t (h_{BIC_{\times}}) ~ \partial_t (h_{BIC_{\times}})
        -\frac{1}{2}
        \partial_t \bar{h}_{_{\times}} ~ \partial_t \bar{h}_{_{\times}}
        \rangle
    \\
    = & \frac{A^2 \omega^2 c^2}{16 \pi G} ~ e^{- 2 k_z |z|}
        \langle 
        \big( cos(k_x x)sin(k_y y) + sin(k_x x)cos(k_y y) \big )^2
        ~ sin^2(\omega ct/c) 
        \rangle
\end{split}
\end{equation}
in which the average:
\begin{equation}
    \langle 
        \big( 
        cos(k_x x)sin(k_y y) + sin(k_x x)cos(k_y y) 
        \big)^2
        \rangle
        = 
        \langle 
        sin^2(k_x x + k_y y) 
        \rangle
        = 
        \langle 
        \big(1 - cos(2k_x x + 2k_y y)\big)/2
        \rangle
\end{equation}
\begin{equation}
\begin{split}
        \langle 
    sin^2(k_x x + k_y y)
    \rangle
    & = \frac{1}{2\pi/k_x - 0} 
        \frac{1}{2\pi/k_y - 0} 
        \int_{0}^{2\pi/k_y}{dy}
        \int_{0}^{2\pi/k_x}{\frac{1 - cos(2k_x x + 2k_y y)}{2} dx}
        \\
    & = \frac{k_x k_y}{4\pi^2} 
        \bigg[ \frac{xy}{2} + \frac{cos(2 k_x x + 2k_y y)}{8k_x k_y} \bigg]_{0,0}^{2\pi/k_x, 2\pi/k_y}
    = \frac{1}{2}
\end{split}
\end{equation}
resulting for $t_{tt}^{BIC_{\times}}$:
\begin{equation}
    t_{tt}^{BIC_{\times}} =
    \frac{A^2 \omega^2 c^2}{64 \pi G} e^{- 2 k_z |z|}
\end{equation}
having an energy density identical to that of the ($+$)-polarized BIC. For the remaining components:
\begin{equation}
\begin{split}
    t_{tx}^{BIC_{\times}} = t_{xt}^{BIC_{\times}} = & \frac{c^4}{32 \pi G} 
        \langle 
        4 \partial_t (h_{BIC_{\times}}) ~ \partial_x (h_{BIC_{\times}})
        -\frac{1}{2}
        \partial_x \bar{h}_{_{\times}} ~ \partial_t \bar{h}_{_{\times}}
        \rangle
    \\
    = & \frac{A^2 k_x \omega c^3}{16 \pi G} ~ e^{- 2 k_z |z|}
        \langle 
        \big( cos(k_x x)sin(k_y y) + sin(k_x x)cos(k_y y) \big)
        \\
        & \times
        \big( sin(k_x x)sin(k_y y) - cos(k_x x)cos(k_y y) \big)
        \\
        & \times sin(\omega ct/c) cos(\omega ct/c) 
        \rangle
\end{split}
\end{equation}
where the average:
\begin{equation}
\begin{split}
        \langle 
            \big( cos(k_x x)sin(k_y y) + sin(k_x x)cos(k_y y) \big)
            \times
            \big( sin(k_x x)sin(k_y y) - cos(k_x x)cos(k_y y) \big)
            \rangle
        \\
        =         
        \langle 
            sin(k_x x + k_y y)
            ~
            cos(k_x x + k_y y)
            \rangle
        = 0    
\end{split}
\end{equation}
Thus
\begin{equation}
    t_{tx}^{BIC_{\times}} = t_{xt}^{BIC_{\times}} = 0
\end{equation}
\begin{equation}
\begin{split}
    t_{ty}^{BIC_{\times}} = t_{yt}^{BIC_{\times}} = & \frac{c^4}{32 \pi G} 
        \langle 
        4 \partial_t (h_{BIC_{\times}}) ~ \partial_y (h_{BIC_{\times}})
        -\frac{1}{2}
        \partial_t \bar{h}_{_{\times}} ~ \partial_y \bar{h}_{_{\times}}
        \rangle
    \\
    = & \frac{A^2 k_y \omega c^3}{16 \pi G}  
        ~ e^{- 2 k_z |z|}
        \langle 
        \big( cos(k_x x)sin(k_y y) + sin(k_x x)cos(k_y y) \big)
        \\
        & \times
        \big( sin(k_x x)sin(k_y y) - cos(k_x x)cos(k_y y) \big)
        \\
        & \times sin(\omega ct/c) cos(\omega ct/c)
        \rangle
        \\ = & ~ 0
\end{split}
\end{equation}
\begin{equation}
\begin{split}
    t_{tz}^{BIC_{\times}} = t_{zt}^{BIC_{\times}} = & \frac{c^4}{32 \pi G} 
        \langle 
        4 \partial_t (h_{BIC_{\times}}) ~ \partial_z (h_{BIC_{\times}})
        -\frac{1}{2}
        \partial_t \bar{h}_{_{\times}} ~ \partial_z \bar{h}_{_{\times}}
        \rangle
    \\
    = & \frac{A^2 k_z \omega c^3}{16 \pi G}  
        ~ sign(z) ~ e^{- 2 k_z |z|}
        \langle 
        \big( cos(k_x x)sin(k_y y) + sin(k_x x)cos(k_y y) \big)^2
        \\
        & \times sin(\omega ct/c) cos(\omega ct/c)
        \rangle
        \\ = & ~ 0
\end{split}
\end{equation}
Like the previous case, there is no momentum density ($t_{tk}^{BIC_{\times}} = t_{kt}^{BIC_{\times}} = 0$) for the ($\times$)-polarized BIC.
\begin{equation}
\begin{split}
    t_{xx}^{BIC_{\times}} = & \frac{c^4}{32 \pi G} 
        \langle 
        4 \partial_x (h_{BIC_{\times}}) ~ \partial_x (h_{BIC_{\times}})
        -\frac{1}{2}
        \partial_x \bar{h}_{_{\times}} ~ \partial_x \bar{h}_{_{\times}}
        \rangle
    \\
    = & \frac{A^2 k_x^2 c^4}{16 \pi G}  
        ~ e^{- 2 k_z |z|}
        \langle 
        \big( sin(k_x x)sin(k_y y) - cos(k_x x)cos(k_y y) \big)^2
        \\
        & \times cos^2(\omega ct/c)
        \rangle
    \\
    = & \frac{A^2 k_x^2 c^4}{16 \pi G}  
        ~ e^{- 2 k_z |z|}
        \langle 
        cos^2(k_x x + k_y y) ~ cos^2(\omega ct/c)
        \rangle
    \\
    = & \frac{A^2 k_x^2 c^4}{64 \pi G}  
        ~ e^{- 2 k_z |z|}
\end{split}
\end{equation}
\begin{equation}
\begin{split}
    t_{xy}^{BIC_{\times}} = t_{yx}^{BIC_{\times}} = & \frac{c^4}{32 \pi G} 
        \langle 
        4 \partial_x (h_{BIC_{\times}}) ~ \partial_y (h_{BIC_{\times}})
        -\frac{1}{2}
        \partial_x \bar{h}_{_{\times}} ~ \partial_y \bar{h}_{_{\times}}
        \rangle
    \\
    = & \frac{A^2 k_x k_y c^4}{16 \pi G}  
        ~ e^{- 2 k_z |z|}
        \langle 
        \big( sin(k_x x)sin(k_y y) - cos(k_x x)cos(k_y y) \big)^2
        \\
        & \times cos^2(\omega ct/c)
        \rangle
    \\
    = & \frac{A^2 k_x k_y c^4}{64 \pi G}  
        ~ e^{- 2 k_z |z|}
\end{split}
\end{equation}
These components imply a localized $x(y)$-momentum flow in the $y(x)$ direction, which are absent for the $(+)$-polarized BIC.
\begin{equation}
\begin{split}
    t_{xz}^{BIC_{\times}} = t_{zx}^{BIC_{\times}} = & \frac{c^4}{32 \pi G} 
        \langle 
        4 \partial_x (h_{BIC_{\times}}) ~ \partial_z (h_{BIC_{\times}})
        -\frac{1}{2}
        \partial_x \bar{h}_{_{\times}} ~ \partial_z \bar{h}_{_{\times}}
        \rangle
    \\
    = & \frac{A^2 k_x k_z c^4}{16 \pi G}  
        ~ sign(z) ~ e^{- 2 k_z |z|}
        \langle 
        \big( sin(k_x x)sin(k_y y) - cos(k_x x)cos(k_y y) \big)
        \\
        & \times
        \big( cos(k_x x)sin(k_y y) + sin(k_x x)cos(k_y y) \big)
        \\
        & \times cos^2(\omega ct/c)
        \rangle
        \\ = & ~ 0
\end{split}
\end{equation}
\begin{equation}
\begin{split}
    t_{yy}^{BIC_{\times}} = & \frac{c^4}{32 \pi G} 
        \langle 
        4 \partial_y (h_{BIC_{\times}}) ~ \partial_y (h_{BIC_{\times}})
        -\frac{1}{2}
        \partial_y \bar{h}_{_{\times}} ~ \partial_y \bar{h}_{_{\times}}
        \rangle
    \\
    = & \frac{A^2 k_y^2 c^4}{16 \pi G}  
        ~ e^{- 2 k_z |z|}
        \langle 
        \big( sin(k_x x)sin(k_y y) - cos(k_x x)cos(k_y y) \big)^2
        \\
        & \times cos^2(\omega ct/c)
        \rangle
    \\
    = & \frac{A^2 k_y^2 c^4}{16 \pi G}  
        ~ e^{- 2 k_z |z|}
        \langle 
        cos^2(k_x x + k_y y) ~ cos^2(\omega ct/c)
        \rangle
    \\
    = & \frac{A^2 k_y^2 c^4}{64 \pi G}  
        ~ e^{- 2 k_z |z|}
\end{split}
\end{equation}
having an identical structure to that of $t_{xx}^{BIC_{\times}}$.
\begin{equation}
\begin{split}
    t_{yz}^{BIC_{\times}} = t_{yx}^{BIC_{\times}} = & \frac{c^4}{32 \pi G} 
        \langle 
        4 \partial_y (h_{BIC_{\times}}) ~ \partial_z (h_{BIC_{\times}})
        -\frac{1}{2}
        \partial_y \bar{h}_{_{\times}} ~ \partial_z \bar{h}_{_{\times}}
        \rangle
    \\
    = & \frac{A^2 k_y k_z c^4}{16 \pi G}  
        ~ sign(z) ~ e^{- 2 k_z |z|}
        \langle 
        \big( sin(k_x x)sin(k_y y) - cos(k_x x)cos(k_y y) \big)
        \\
        & \times
        \big( cos(k_x x)sin(k_y y) + sin(k_x x)cos(k_y y) \big)
        \\
        & \times cos^2(\omega ct/c)
        \rangle
        \\ = & ~ 0
\end{split}
\end{equation}
Lastly,
\begin{equation} \label{isaacson_zzx}
\begin{split}
    t_{zz}^{BIC_{\times}} = & \frac{c^4}{32 \pi G} 
        \langle 
        4 \partial_z (h_{BIC_{\times}}) ~ \partial_z (h_{BIC_{\times}})
        -\frac{1}{2}
        \partial_z \bar{h}_{_{\times}} ~ \partial_z \bar{h}_{_{\times}}
        \rangle
    \\
    = & \frac{A^2 k_z^2 c^4}{16 \pi G} 
        ~ sign^2(z) ~ e^{- 2 k_z |z|} 
        \langle 
        \big( sin(k_x x)cos(k_y y) - sin(k_x x)cos(k_y y) \big)^2
        \\
        & \times cos^2(\omega ct/c)
        \rangle
    \\
    = & \frac{A^2 k_z^2 c^4}{64 \pi G} 
        ~ sign^2(z) ~ e^{- 2 k_z |z|}
\end{split}
\end{equation}
As for the ($+$)-polarized BIC, the radiation pressure shows a discontinuity, becomes null at $z = 0$ and also averages to zero away from the localization plane. The full Isaacson stress-energy tensor in a matrix representation is:
\begin{equation} \label{isaacsonset_x}
    t_{\mu \nu}^{BIC_{\times}} = e^{- 2 k_z |z|}
    \begin{pmatrix}
			\frac{A^2 \omega^2 c^2}{64 \pi G} 	& 0 		& 0 		& 0 \\ 
			0 	& \frac{A^2 k_x^2 c^4}{64 \pi G}  	& \frac{A^2 k_x k_y c^4}{64 \pi G} 		& 0 \\ 
			0 	& \frac{A^2 k_x k_y c^4}{64 \pi G} 		& \frac{A^2 k_y^2 c^4}{64 \pi G} 	& 0 \\ 
			0 	& 0 		& 0 		& \frac{A^2 k_z^2 c^4}{64 \pi G} sign^2(z)
    \end{pmatrix}
\end{equation}
Despite having different non-zero terms ($xy, yx, yy$) when compared with the ($+$)-BIC Isaacson tensor, again all terms decay exponentially to zero away from the source. Thus, as for the ($+$)-polarized BIC, the localized mode has a null $t_{\mu \nu}^{BIC_{\times}}$ at this limit, differently from a planar propagating GW.


\subsection{BIC mode energy}
The BIC mode energy ($E_{BIC_{(+,\times)}}$) can be calculated by integrating the energy density of each polarization ($t_{tt}^{BIC_{(+,\times)}}$) over the spatial spacetime coordinates in their respective unit cells:
\begin{equation} \label{modeenergy}
    E_{BIC_{(+,\times)}} = \int_{Unit~cell}{t_{tt}^{BIC_{(+,\times)}} d^3x}
\end{equation}
Noting that the energy density of both ($+$) and ($\times$) modes are identical (Eq.~\ref{isaacsonset_+} and \ref{isaacsonset_x}), i.e.:
\begin{equation}
    t_{tt}^{BIC_+} = t_{tt}^{BIC_\times} = \frac{A^2 \omega^2 c^2}{64 \pi G}
    ~ e^{- 2 k_z |z|}
\end{equation}
While the strain modulation in $x$ and in $y$ clearly implies a finite unit cell for the ($\times$)-BIC, the orientation of the axis leads to an infinite unit cell in the $y$-direction for the ($+$)-BIC mode (Fig. 1 of the main manuscript), which would lead to an infinite energy per unit cell when integrated. However, this is exactly the same situation as for the 1D photonic BIC \cite{Berte2023, Berte2025}, whose translation invariance in its non-periodic and non-confined direction implies an infinite unit cell for the idealized BIC mode, that, endowed with a finite electric field (with energy density $u \propto |E|^2$), would also lead to an infinite energy per unit cell. Thus, to obtain a sensible finite result for the idealized BIC, the unit cell size in the invariant direction must be made finite. To maintain the equivalence of modes observed throughout the manuscript, the unit cell for the ($+$)-BIC might be arbitrarily defined within the interval $[0,2 \pi/k_y]$, such that Eq.~\ref{modeenergy} becomes:
\begin{equation}
    E_{BIC_{(+,\times)}} = \int_{Unit~cell}{t_{tt}^{BIC_{(+,\times)}} d^3x}
    = \int_{0}^{2 \pi/k_x}{dx} 
    \int_0^{2 \pi/k_y}{dy}
    \int_{- \infty}^{\infty}{\frac{A^2 \omega^2 c^2}{64 \pi G}
    ~ e^{- 2 k_z |z|} dz}
\end{equation}
being
\begin{equation}
    \int_{- \infty}^{\infty}{~ e^{- 2 k_z |z|} dz}
    = 
    \bigg[ \frac{sign(z)}{2 k_z}(1 - e^{- 2 k_z |z|}) + \frac{1}{2 k_z}\bigg]_{- \infty}^{\infty}
    = \frac{1}{k_z}
\end{equation}
Implying an identical finite energy per unit cell for both BIC polarizations:
\begin{equation}
    E_{BIC_{(+,\times)}} = \frac{A^2 \omega^2 c^2 \pi}{16 G k_x k_y k_z}
\end{equation}

\end{document}